\documentclass{iucrjournals}
\hypersetup{pdftitle={Bond-valence theory is a general electrostatic valence rule},
  pdfauthor={M. L. Whittaker},
  pdfsubject={Bond valence, Pauling's second rule, Gauss's law, and electron-density topology}}

\usepackage{amsmath,amssymb,mathtools,bm}
\usepackage[section]{placeins}
\usepackage{longtable}
\providecommand{\onecolumngrid}{}
\providecommand{\twocolumngrid}{}
\renewcommand{\addcontentsline}[3]{}

\newcommand{\dd}{\mathrm{d}}

\newsavebox{\fitwidthbox}
\newcommand{\fitwidth}[1]{%
  \sbox{\fitwidthbox}{#1}%
  \ifdim\wd\fitwidthbox>\textwidth
    \resizebox{\textwidth}{!}{\usebox{\fitwidthbox}}%
  \else
    \usebox{\fitwidthbox}%
  \fi}
\newif\ifiucrj\iucrjtrue
\begin{document}
\nolinenumbers\singlespacing

\title{Bond-valence theory is a general electrostatic valence rule}

\author[a]{M.~L.~Whittaker\IUCrOrcidlink{0000-0002-9724-3409}\IUCrCemaillink{mwhittaker@lbl.gov}}
\affil[a]{Lawrence Berkeley National Laboratory, Berkeley, CA 94720, USA}

\maketitle

\begin{synopsis}
Defining the bond valence as the electrostatic flux through the face that two atomic
basins of the electron density share makes the valence-sum rule, taken over all faces of
a basin, Gauss's law, and shows the bond valence to be Pauling's strength raised to a
distance exponent. The electron density fixes the softness $B$ while $R_0$ is set by the
bond lengths through the sum rule, so neither is a transferable constant of a
cation--anion pair.
\end{synopsis}

\begin{abstract}
Pauling's electrostatic valence rule assigns to each bond of a cation of
charge $z$ with $n$ anion neighbors the strength $z/n$. Bond-valence
theory replaces the equal split of valence among bonds by $s=\exp[(R_0-d)/B]$ in the bond
length $d$ and takes the sum rule, that the valences at each atom add to
its formal charge, as its first axiom. Here the bond valence is defined in
the electron density. Atomic basins of a first-principles density are
bounded by faces, one per bond critical point, and the electric flux of
that density through a face is a charge associated with the bond. The
fluxes leaving a cation sum to its basin charge by Gauss's law, so on
this surface the sum rule is a theorem. Across 3680 oxides, halides,
chalcogenides and pnictides the flux follows the bond-valence law site by site,
with a median deviation of ten percent. The softness
$B$ varies nearly an order of magnitude from stiff sulfides to softer arsenides. 
In all cases it is set by the electrostatic
flux at atomic basin faces. $R_0$ then follows from the bond lengths and their differences. 
Anion valences equal those of bonded cations to within 5\% 
for most of the oxide, fluoride, sulfide and nitride sites. 
Transferability of $(R_0,B)$ between compounds fails in 96 of 106 families, suggesting
that the softness is a local site property defined by the basin charge 
rather than a transferable constant of a given cation--anion pair.
Eliminating $B$ by the sum rule gives $s=(z/n)^{(R_0-d)/(R_0-d_B)}$,
Pauling's strength raised to a distance exponent at the softness-weighted 
mean bond length $d_B$. Thus, bond-valence theory is Pauling's rule
with a distance exponent, and the electron density supplies the surface
on which the two are equivalent.
 \end{abstract}


\section{Introduction}\label{sec:intro}

Pauling's second rule~\cite{pauling1929} is the electrostatic valence principle,
which states that in a stable ionic
structure the electrostatic bond strength $z/n$ of a cation of charge $z$
with $n$ anion neighbors, summed over the cations bonded to an anion,
equals the anion's charge. The bond-valence model~\cite{brown2009,brown2016}
extends the rule to inequivalent bonds by assigning to every cation--anion
contact of length $d$ a valence
\begin{equation}
  s = \exp\!\left[\frac{R_0 - d}{B}\right],
  \label{eq:bvlaw}
\end{equation}
with a pair-specific length $R_0$ and softness $B$ fitted to crystal
structures~\cite{brown1985,brese1991,adams2001,gagne2015}. A
power law $s=(d/R_1)^{-N}$, with its own length $R_1$, has alternatively been
proposed~\cite{brown1973,brown1976}.
Both forms require that the valences meeting at each atom sum to its formal
oxidation state $z$. The charge of the atom after the ionic approximation of its
heteronuclear bonds~\cite{karen2016} provides an integer $z$.
The bond-valence sum is the standard means of assigning an
oxidation state
to a site from its structure~\cite{brown2009,jablonka2021}, a use whose
known limits are the choice of which contacts count as bonds, the
multiplicity of parameter sets, and a form that sees bond lengths
only~\cite{jablonka2021}.
The charge that a partition of the electron density assigns to an atom 
is a different, method-dependent number that need not change by one electron
when the oxidation state does~\cite{raebiger2008,walsh2017,walsh2018}.

The bond-valence model predicts bond lengths, detects distortion and locates ions,
yet its relation to the electrostatic valence principle has remained informal.
Pauling's rule contains no bond length and the exponential law contains
no coordination number. The valence carried by a bond via Eq.~\eqref{eq:bvlaw} 
is defined in ``valence units,'' and its normalization to $z$ has not 
been given an operational definition in the electron density.
Pauling's rule assumes that every bond of a coordination
polyhedron carries the same strength. Tested with equal strengths
across the known oxides, it holds closely only on undistorted
polyhedra~\cite{george2020}, which is the limitation the exponential law
was designed to address. Donnay and Allmann, who introduced the term
\emph{bond valence} for a strength read from a measured bond length,
called $z/n$ the ideal bond valence and offered their curve as a
quantitative application of Pauling's principle~\cite{donnay1970}. Brown
and Shannon at first kept Pauling's own name for the
quantity~\cite{brown1973}, and the two names are retained ``to avoid
confusion with Pauling bond-strengths''~\cite{gibbs2022}. The strength
has been called an approximation to the valence~\cite{brown2009} and
identified with the valence at the polyhedron's mean bond
length~\cite{donnay1970,brown1973,hawthorne1994}, but no relation has
been written that carries one into the other. 

Brown proposed that the bond valence be understood as the electrostatic flux linking a
cation to an anion~\cite{preiser1999}, and bond-valence theory states
the valence-sum rule as the first of its three axioms, with Gauss's flux
theorem as its justification~\cite{gibbs2022}. Gauss's theorem, however,
fixes only the total flux leaving an ion. How that flux divides among
the bonds depends on the charge distribution to which the theorem is
applied, and must be computed from the electron density itself. 
The quantum theory of atoms in molecules (QTAIM) partitions any electron
density $\rho(\mathbf r)$ exactly into atomic basins bounded by
zero-flux surfaces of $\nabla\rho$, with one bond critical point on each
interatomic face~\cite{bader1990}, enabling basin charges to be
computed~\cite{henkelman2006,yu2011}. A basin charge, however,
is one number per atom and not per bond, so it does not directly give the bond valence. 
The bond length, Pauling's bond strength and $\rho_b$, the electron density at the bond critical point,
are tied by power laws in oxide molecules and crystals~\cite{gibbs1998,gibbs2003,gibbs2014,gibbs2022}. 
A correlation of $\rho_b$ with bond length does not, however, explain why the
strengths meeting at an atom sum to its charge.

This paper gives the bond valence an operational definition in the
electron density and shows that, on it, bond-valence theory and
Pauling's rule are one theory. QTAIM supplies both the faces between
atoms, one per bond point, and the electrostatic field of the atomic basins.
The electric flux of the electron density through a face is a charge
attached to that bond, and the fluxes leaving a cation basin sum to its
basin charge $q$ by Gauss's law. Counting $q$ instead of the formal charge $z$
makes the valence-sum rule an identity, so that the first axiom of
bond-valence theory becomes a theorem, and the normalization to $z$,
otherwise a convention, becomes the ratio of two charges, one measured
and one `ideal'. The definition is tested on a corpus of 3680 oxides,
fluorides, chlorides, sulfides, selenides, tellurides, nitrides,
phosphides and arsenides from Materials Project electron density distributions, 
on which the flux valence follows Eq.~\eqref{eq:bvlaw} site by site, 
with a median scatter of 10\%. The
electron density fixes the softness, which is read from how a distorted
site distributes its flux. It is not generally a property of a cation--anion pair,
although in a minority of cases it does transfer among bond pair types. 
The bond lengths fix $R_0$ through the valence-sum rule and the ratio $q/z$ 
fixes the unit of the valence. Eliminating the
softness between the valence-sum rule and the electrostatic valence rule 
gives Pauling's strength raised to a distance exponent, which
reproduces the measured fluxes of the corpus more closely than the
exponential law at the same parameters.

Section~\ref{sec:theory} states the definition and the identities that
follow from it. Sec.~\ref{sec:methods} describes the electron densities,
the faces and the fluxes. Sec.~\ref{sec:results} presents the basin
charges, the valence-sum rule at the anion, the bond-valence law and its softness, 
the Pauling form, and the test of transferability. Sec.~\ref{sec:discussion}
returns to the questions of this introduction, what Gauss's theorem
leaves to the electron density, what the flux adds to a Bader charge and
to the bond-point density, how the formal charge and the basin charge are
related, and what standing the conventional parameters have, and states
the limits of the present electron densities.
\section{Flux valence and Pauling's rule}\label{sec:theory}

\subsection{Faces and fluxes of the electron density}

Let $\rho(\mathbf r)$ be the electron density of a periodic crystal and
$\{Z_i, \mathbf R_i\}$ its nuclei. The index $i$ labels a cation and $j$ an anion.
The gradient field $\nabla\rho$
partitions the crystal unit cell into atomic basins $\Omega_i$, each containing one
nucleus, separated by zero-flux surfaces $\nabla\rho\cdot\hat n = 0$.
Two basins that share a surface patch $F_{ij}$ share, on that patch, a
$(3,-1)$ critical point of $\rho$ (a bond point) whose two steepest-ascent
paths end at the two nuclei. In a Morse-regular electron density every patch carries
exactly one bond point and every bond point lies on exactly one patch. The
patches are the two-cells dual to the bonds of the Morse complex.

The same electron density and nuclei define the charge density
$\varrho(\mathbf r) = \sum_i Z_i \delta(\mathbf r - \mathbf R_i) - \rho(\mathbf r)$
and, through Poisson's equation, the electrostatic field $\mathbf E$ of
the crystal. We define the \emph{face flux}
\begin{equation}
  \Phi_{ij} = \frac{1}{4\pi}\oint_{F_{ij}} \mathbf E\cdot \hat n_{ij}\,\dd S ,
  \label{eq:flux}
\end{equation}
oriented from $i$ into $j$. Hartree atomic units are used throughout, in
which Gauss's law reads $\oint\mathbf E\cdot\hat n\,\dd S=4\pi Q$ and
every charge, $Z_i$, $q_i$, $\Phi_{ij}$ and the formal charge $z_i$
alike, is a number of elementary charges, so that $\Phi_{ij}$ is a
charge in electrons and the ratios $z_i/q_i$ and $\Phi_{ij}/q_i$ formed
below are pure numbers. Gauss's law applied to $\Omega_i$ gives, for
every atom,
\begin{equation}
  \sum_{j} \Phi_{ij} = q_i \equiv Z_i - \int_{\Omega_i}\rho\,\dd^3 r ,
  \label{eq:gauss}
\end{equation}
where $q_i$ is the net (Bader) charge of the basin. Equations~\eqref{eq:flux}
and~\eqref{eq:gauss} hold for any pair of basins, cation--cation and
anion--anion pairs included, and Eq.~\eqref{eq:gauss} is exact and holds
face by face for whatever partition is used. The content
of this paper is that, on the \emph{bond-point faces} of the electron density, the
individual terms $\Phi_{ij}$ are the bond valences.

The construction is illustrated on a two-dimensional model density in SI
Fig.~\ref{fig:basins} and the \hyperref[app:notes]{SI notes}. The field
crosses a cation--anion face only near its bond point, and the
field-tangent part of a face carries no flux. In the crystals the
anion--anion faces and contacts carry a net flux that is reported with
the results. By Eq.~\eqref{eq:gauss} the faces of an undistorted site
share its charge $q_i$ equally, Pauling's share, and the unequal shares of
a distorted site are the within-site law from which the softness $B$ is
measured below.

\subsection{Definition of the flux valence}

Define, for a cation $i$ of formal charge $z_i$ and basin charge $q_i$, the
\emph{flux valence} of each of its faces
\begin{equation}
  s_{ij} = \Phi_{ij}\,\frac{z_i}{q_i},
  \label{eq:s}
\end{equation}
the face flux rescaled by the ratio $z_i/q_i$ of the formal charge to
the basin charge. It is a charge in the same unit as $\Phi_{ij}$, the
elementary charge, which is the valence unit of bond-valence theory, since
the valences at an atom sum to an integer number of elementary charges,
and in atomic units that number is what enters the logarithm of
Eq.~\eqref{eq:law}. Then (i) by
Eq.~\eqref{eq:gauss}, $\sum_j s_{ij} = z_i$ identically, so the
valence-sum rule is Gauss's law with the basin charge counted as the formal
charge. The sum runs over every face and contact of the basin,
including those a cation shares with other cations, which carry a net
flux $L_i$ that is small on most sites and not on all, and over the
cation--anion faces alone it is
$z_i(1-L_i/q_i)$. The coordination number
$n_i$ of Pauling's rule is the number of those faces that carry a bond
valence. A face whose flux valence is below a fraction $\epsilon$ of the
largest at the site changes the site's sum by less than $\epsilon$ of one
bond, below the precision to which $B$ is defined, and by
Eq.~\eqref{eq:law} lies more than $B\ln(1/\epsilon)$ beyond the shortest
bond, outside any coordination polyhedron. We use $\epsilon=0.05$. The faces this
excludes, and the flux they carry, are reported with the results.
(ii) Empirically, on the corpus,
\begin{equation}
  s_{ij} = \exp\!\left[\frac{R_0 - d_{ij}}{B}\right],
  \label{eq:law}
\end{equation}
within every site, with a softness $B$ that is measured independently of
$R_0$ from the response of the flux to distortion (the within-site form
derived below). Whether one pair $(R_0,B)$ then serves a family across
hosts and coordination numbers is a separate statement, tested in
Sec.~\ref{sec:transfer}.
$R_0$ at fixed $B$ is set by the bond lengths through
the sum rule, as derived below, up to a term that depends only on how
unequally a distorted site shares its flux, so it is not a prediction of
the electron density beyond what $B$ already carries. (iii) The screening ratio
$q_i/z_i$ is one number per site, so that the flux itself,
$\Phi_{ij} = (q_i/z_i)\, s_{ij}$, obeys the same law at every site with
a prefactor that is the site's own. Across hosts it is not a constant, and
its distribution over the corpus, a band for each cation species with
each anion, is reported with the results.

\subsection{\texorpdfstring{Formal charge and basin charge}{Formal charge and basin charge}}\label{sec:qz}

Equation~\eqref{eq:s} is the only place in the construction where a formal
charge appears, and the two charges it relates are different kinds of
quantity. The basin charge $q_i$ is measured. By Eq.~\eqref{eq:gauss} it is
the whole flux the cation emits, and no choice in the construction can
change it. The formal charge $z_i$ is an integer number of elementary
charges assigned by the oxidation-state rules, a label, and is not a
charge that the electron density carries anywhere. The two
differ by a deficit $z_i-q_i$, which is electron density that has stayed inside the cation
basin instead of crossing into an anion.

The ratio $q_i/z_i$ is one number per
site, not one per bond, so it multiplies every face of a site by the same
factor. It therefore fixes only the unit in which the valence is quoted and
cancels from everything that depends on $d$. Substituting
$\Phi=(q/z)\,s$ in Eq.~\eqref{eq:law} leaves $B$ unchanged and shifts $R_0$
by $B\ln(q/z)$. This is why the conventional
normalization to formal charges works. A fit to structures cannot see $q$,
and does not need to. It recovers the same softness and an $R_0$ displaced
by $B\ln(q/z)$, which varies little across a species' band with a given
anion (Sec.~\ref{sec:results}).

\subsection{From bond valence to Pauling's rule}

Two identities follow without further input. Let a cation site $i$ have
$n$ cation--anion faces, one per bond point, with bond lengths $d_{ij}$,
$j=1,\dots,n$. This $n$ is the coordination number used throughout. It is
read off the topology of the electron density, not chosen by a distance cutoff, and
the faces a cation basin shares with anions beyond its coordination
polyhedron are typically negligible. Equation~\eqref{eq:law}
and the valence-sum rule then give
\begin{equation}
  R_0 = B\ln\!\Big[\frac{z_i}{\sum_j e^{-d_{ij}/B}}\Big]
      \;\simeq\; \langle d\rangle + B\ln\frac{z_i}{n},
  \label{eq:r0site}
\end{equation}
the second form holding for a regular polyhedron,
so that the coordination dependence of the valence is Pauling's term
$B\ln(z/n)$. With the family's own $B$ the mean bond length shifts between
coordination numbers by exactly $B\ln(n_2/n_1)$ and $R_0$ is then the
same in both coordinations, while a fit at a common softness must let $R_0$ carry the
difference, $(B_{\rm family}-B)\ln(n/z)$. Conversely, within one polyhedron $R_0$ cancels:
\begin{equation}
  \ln\frac{s_{ij}\,n}{z_i} = -\frac{d_{ij} - \langle d\rangle}{B} + O\big((d_{ij}-\langle d\rangle)^2\big),
  \label{eq:distortion}
\end{equation}
which measures $B$ from the response of the flux to bond-length distortion
alone. Equation~\eqref{eq:distortion} is the form in which $B$ is extracted
below. It is also the reason $B$ cannot be read off a single undistorted
structure, where $R_0$ and $B$ enter only through the combination that
fixes the sum rule. Reading $B$ from distortion has a precedent in the
distortion theorem, under which $B=\sigma^2/2\Delta R_a$ has been proposed as a way
to determine the softness~\cite{urusov2003}, with $\sigma^2$ the variance
of a site's bond lengths and $\Delta R_a$ the rise of its mean bond length
above the regular-site length at fixed valence sum, and $\Delta R_a$
itself has been adopted as the measure of a distortion~\cite{brown2006}.
Equation~\eqref{eq:distortion} is the per-bond form of the same idea. It
needs no regular-site length because the valence of each face is
measured.

The same degeneracy can be turned around. The exact sum rule on
a site of any shape reads $R_0 = d_B + B\ln(z_i/n)$, with
\begin{equation}
  d_B = -B\ln\Big[\frac1n\sum_k e^{-d_{ik}/B}\Big]
  \label{eq:dB}
\end{equation}
the softness-weighted mean bond length of the site, so that eliminating
$B$ gives
\begin{equation}
  s_{ij} = \Big(\frac{z_i}{n}\Big)^{\displaystyle\frac{R_0-d_{ij}}{R_0-d_B}},
  \label{eq:pauling}
\end{equation}
exactly, for every polyhedron, with $\sum_j s_{ij}=z_i$ identically.
This is Pauling's electrostatic valence $z/n$~\cite{pauling1929} raised
to a distance exponent. The exponent is the bond's own deviation from the
unit-valence length $R_0$ in units of the site's mean deviation, so that
$s=z_i/n$ at $d_{ij}=d_B$ (Pauling's rule), $s=1$ at $d_{ij}=R_0$, and the
interpolation between them is log-linear. Written with the site's own $R_0$,
the one its sum rule defines, the form is the law itself. Written with a
family $R_0$ it is a prediction from $z$, $n$ and the site's bond lengths,
whose residual is the offset of the site's $R_0$ from the family's, and
whose closure is exact only up to that offset. The one subtlety is that $d_B$
is defined with $B$. It depends on $B$ only through the spread of the
site's bond lengths. With $\langle d\rangle$ the arithmetic mean of the
site's bond lengths, $\sigma^2$ their variance and $\kappa_3$ their third
cumulant, the cumulant expansion of Eq.~\eqref{eq:dB} is
\begin{equation}
  d_B = \langle d\rangle - \frac{\sigma^2}{2B} + \frac{\kappa_3}{6B^2} - \cdots.
  \label{eq:cumulant}
\end{equation}
Equation~\eqref{eq:cumulant} is the distortion theorem in its
quantitative form. The length $d_B=R_0-B\ln(z_i/n)$ is the one a regular site of
the same valence sum would have, so $\langle d\rangle-d_B$ is the
increase $\Delta R_a$ of the mean bond length that is the measure
of distortion, written through the valences as
$(B/n)\sum_j\ln(z_i/ns_{ij})$ \cite{brown2006}, and the leading term
$\sigma^2/2B$ with its mode-dependent third-moment correction is the
expansion of the theorem in the moments of the bond
lengths~\cite{urusov2003}. For a regular site,
$\sigma=0$, $d_B=\langle d\rangle$ exactly and Eq.~\eqref{eq:pauling}
involves no $B$ at all. For a distorted site the
$B$-dependence enters the exponent at relative order
$\sigma^2/[2B(R_0-\langle d\rangle)]$, and the sum rule removes it order by
order. To first order $d_B=\langle d\rangle$, which is the
regular-polyhedron form, with exponent $(R_0-d_{ij})/(R_0-\langle d\rangle)$.
To second order the variance term, with the sum-rule value of $B$
substituted, lowers $d_B$ below the arithmetic mean, so that the short
bonds count for more (\hyperref[app:notes]{notes of the supporting information, SI}). The exact
$d_B$ at the family $B$ is the working form and the arithmetic mean its
first-order approximation, the gap between them being small on most
sites and not on all (Sec.~\ref{sec:results}). Equation~\eqref{eq:pauling} contains no softness
parameter. It embeds the sum-rule value $B=(R_0-d_B)/\ln(z_i/n)$ of each
site, and it is degenerate when $z_i=n$ (the tetrahedral Si of the
silicates), where the base is unity and every bond has unit valence
whatever its length, and there the distortion softness of
Eq.~\eqref{eq:distortion} is the only source of a distance dependence.
Eliminating $R_0$ instead of $B$ gives the dual form,
\begin{equation}
  s_{ij} = z_i\,\frac{e^{-d_{ij}/B}}{\sum_k e^{-d_{ik}/B}},
  \label{eq:softmax}
\end{equation}
a Boltzmann weighting of the site's bond lengths with $B$ as the
temperature, exact in closure for any polyhedron, free of $R_0$, and the
form that Eq.~\eqref{eq:distortion} probes. The two forms are
complementary, Eq.~\eqref{eq:pauling} exact at the unit-valence point and
Eq.~\eqref{eq:softmax} exact in closure, and they coincide when $d_B$ is
used. Equation~\eqref{eq:pauling} is tested below face by face
with the exact $d_B$ and with the arithmetic mean.

Nothing in Eqs.~\eqref{eq:r0site}--\eqref{eq:softmax} depends on the
exponential form. For any law that is log-linear in a variable $u(d)$,
$s=\exp[(u_0-u)/B_u]$, the sum rule eliminates $B_u$ in the same way and
gives $s=(z_i/n)^{(u_0-u)/(u_0-u_B)}$, with $u_B$ the corresponding
weighted mean. The exponential law is the case $u=d$ and the power law
$s=(d/R_1)^{-N}$ of Brown and Shannon~\cite{brown1973} the case
$u=\ln d$ with $B_u=1/N$, and within one site
Eq.~\eqref{eq:distortion} cannot tell the two apart (\hyperref[app:notes]{SI notes}). We keep the exponential because
it is linear in the bond length, so that $B$ is a length and $R_0$ an
intercept. The identities above use only that the law is log-linear in
$d$. The physical basis of that form, as the flux of a screened Coulomb
field between ions, is derived in Ref.~\cite{whittaker_prm} and is not
repeated here.

The equivalence is now explicit. Pauling's rule is the statement
$s=z/n$ with the sum rule at the anion, the bond-valence model is
Eq.~\eqref{eq:law} with the sum rule at every atom, and
Eq.~\eqref{eq:pauling} is both, the exponential law with its softness
eliminated by the sum rule, which reduces to Pauling's rule on a regular
polyhedron and to the exponential law at any fixed $B$. What the electron density
adds is that the elimination holds bond by bond on a measured flux, with
$B$ measurable on its own from distortion, $R_0$ fixed by the bond lengths,
and $q/z$ a unit fixed site by site.
\section{Electron densities, faces and fluxes}\label{sec:methods}

\paragraph{Electron densities.}
Each host's density is the self-consistent PBE pseudo-valence density that
the Materials Project~\cite{mp2013,mp2025} stores at its relaxed geometry,
completed by analytic frozen Slater-type cores carrying the electrons each
pseudopotential freezes, so that grid and core electrons sum to the
nuclear charge and Eq.~\eqref{eq:gauss} closes without rescaling. The
stored grid is the VASP CHGCAR output, which VASP and the Materials Project
call a charge density. It is the valence electron density and carries no
nuclear charge. The
faces lie far from the cores, so the fluxes and the law are expected to be
insensitive to the core representation. The pipeline, the core completion
and its all-electron reference are the makeAIM
construction~\cite{makeaim1}, set out in the \hyperref[app:notes]{SI notes}.

\paragraph{Faces and fluxes.}
The gradient field $\nabla\rho$ partitions each cell into atomic basins
whose shared boundaries carry the $(3,-1)$ critical points of $\rho$. A
boundary patch at a bond point is a \emph{face}. A patch where two basins
meet without one, at a ring or cage point, is a \emph{contact}, kept with
its flux but carrying no bond valence, and each is classified by the
formal charges of its basins as cation--anion, anion--anion or
cation--cation. The electrostatic field follows from $\varrho$ by a
reciprocal-space Poisson solve, and the flux through every face and
contact is integrated on two voxel grids ($0.13$ and $0.087$~bohr) and
extrapolated to zero spacing, closing on each basin to
$10^{-12}$~electrons (\hyperref[app:notes]{SI notes}).

\paragraph{The corpus.}
The corpus is every ordered, non-magnetic, experimentally known Materials
Project compound of at most 40 sites, within $0.05$~eV of the convex
hull, computed with PBE and no Hubbard term and free of the open-shell
$3d$ elements, the lanthanides other than La and Lu, and the actinides,
whose anion is one of O, F, Cl, S, Se, Te, N, P or As. Of the 126\,315
entries with a stored density, 3933 meet the selection and 3680
yield a complete face table at both resolutions
(\ifiucrj Fig.~\ref{fig:flow}\else SI Fig.~\ref{fig:flow}\fi,
\hyperref[app:notes]{SI notes}).

\newcommand{\captionflow}{How the corpus was defined (Sec.~\ref{sec:methods}), from the Materials Project entries with a stored electron density to the families of the transferability test. Band widths are proportional to the counts, on a finer scale to the right of the pale dotted band, which marks the change of scale and is not a flow.}
\ifiucrj
\begin{figure*}[t]
\includegraphics[width=\textwidth]{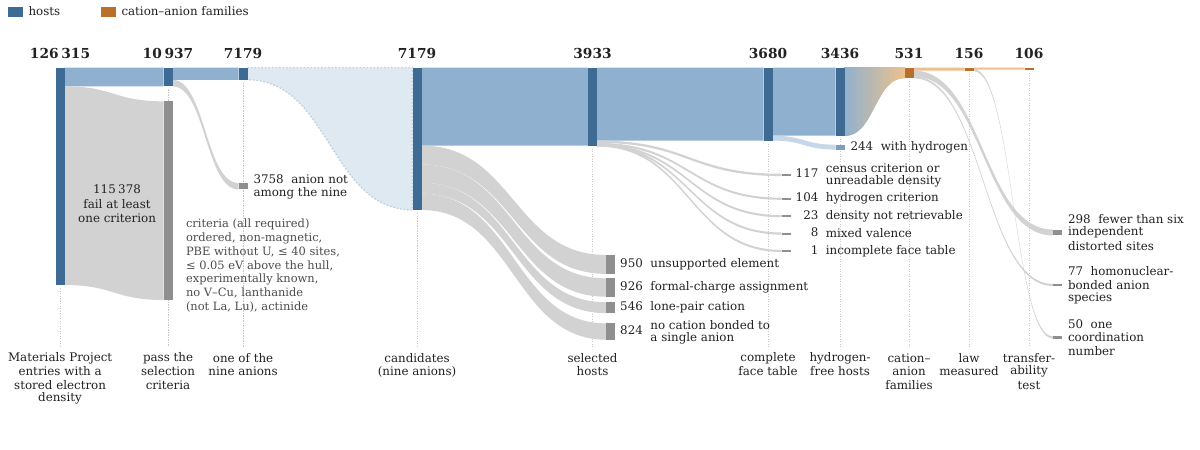}
\caption{\captionflow}
\label{fig:flow}
\end{figure*}
\fi

\begin{figure*}[p]
\includegraphics[width=\textwidth,height=0.74\textheight,keepaspectratio]{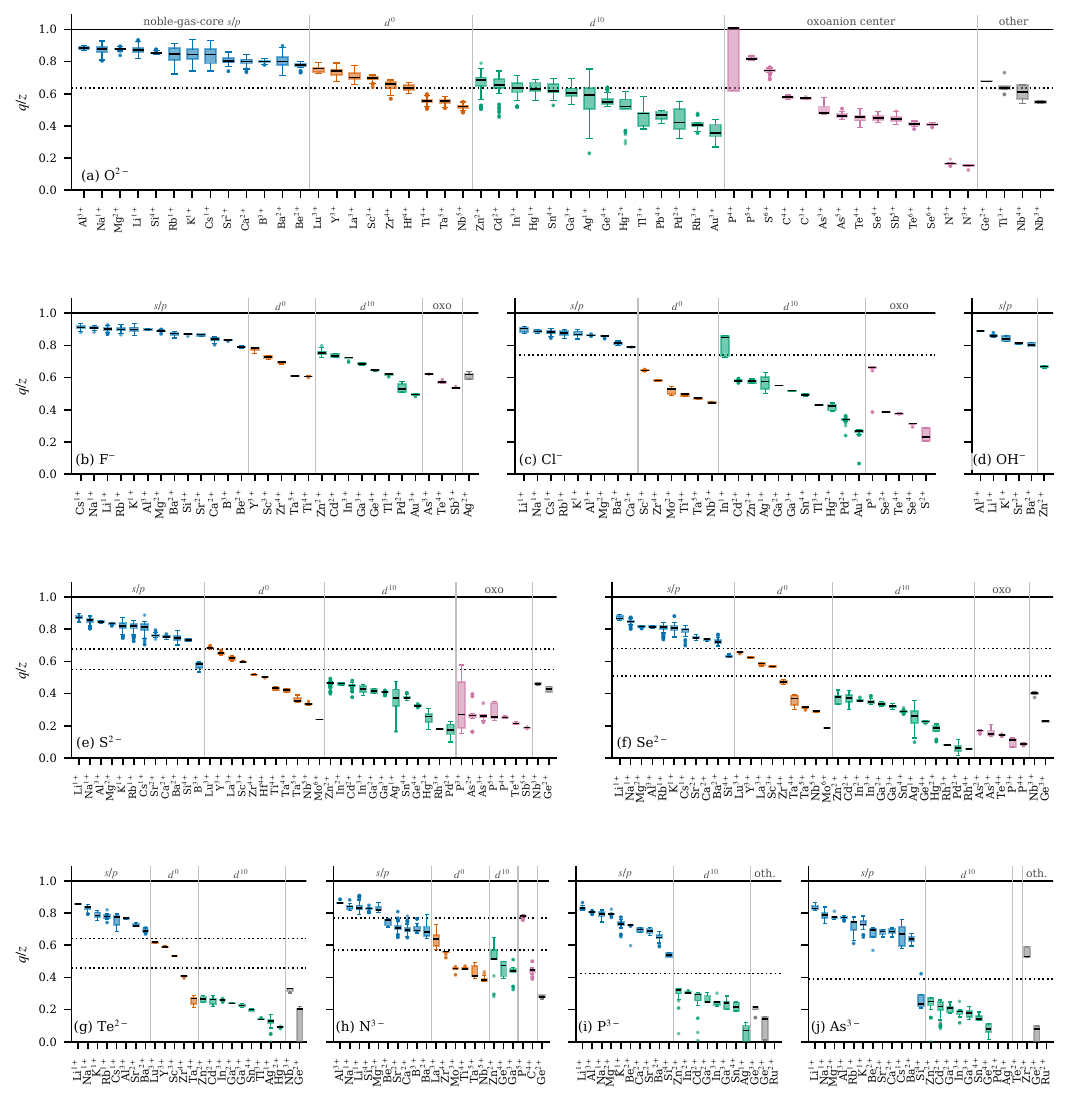}
\caption{The screening ratio $q/z$ of the cation sites bonded to a single anion species (hydrogen-free hosts, OH$^{-}$ denotes oxygen basins with one merged hydrogen), for species with at least eight sites, as box-and-whisker plots grouped by cation class and ordered by median within a class. The solid line is $q/z=1$ and the dotted lines the valleys of the cation distribution with that anion (SI Fig.~\ref{fig:qzhist} and SI text). The fluorides have none, and none is placed for OH$^{-}$.}
\label{fig:charges}
\end{figure*}

\paragraph{Formal charges and families.}
Formal charges are assigned per site by a bond-valence analysis of each
structure~\cite{ong2013,okeeffe1991}, the assignment checked against the
$q/z$ bands of Fig.~\ref{fig:charges}. A family is a cation species with
an anion species. The law, its softness and the Pauling form are measured
on every family with at least six crystallographically independent
distorted sites and a within-site bond-length range of at least
$0.05$~\AA{}, which 156 families have, and the test of transferability,
which needs in addition sites at two coordination numbers, on 106 of them
(SI Table~\ref{tab:families}, \hyperref[app:notes]{SI notes}). A
hydrogen-bearing host is used only when each of its hydrogens ascends to
an anion, into which it is merged. The 244 such hosts are held out of the
hydrogen-free set of 3436 on which numbers are quoted unless stated
otherwise, and adding them changes no family's verdict in the
transferability test (SI Table~\ref{tab:hsens},
\hyperref[app:notes]{SI notes}).

\paragraph{Coordination number.}
The coordination number $n$ of a cation site is the number of
cation--anion faces of its basin, one per bond point, and it involves no
distance cutoff.

\paragraph{Family metrics.}
The within-site softness $B$ is the slope of Eq.~\eqref{eq:distortion}
pooled over the faces of a family's distorted sites, with a 68\% bootstrap
interval over its independent sites. The family $R_0=\langle d+B\ln
s\rangle$ follows from the sum rule, Eq.~\eqref{eq:r0site}, at that $B$.
The coordination-branch softness $B^{*}$ is the slope of the site mean
bond length against $-\ln(z/n)$ over the family's sites with three or more
bonded faces, $D=B-B^{*}$, and the family passes the transferability test
when $|D|\le\max(0.03~\text{\AA},\text{half-width})$. The per-site and
per-coordination softnesses, the host-to-host spread $\sigma_n$, the
two-parameter fit and its variance decomposition, and the bootstrap,
held-out and reproducibility conventions are defined in the
\hyperref[app:notes]{SI notes}.
\section{Results}\label{sec:results}

The corpus comprises 169\,238 bonded cation--anion faces on 3680
hosts, 160\,300 of them on the 3436 hydrogen-free hosts on which the
numbers below are quoted. The hydrogen-free hosts carry 29\,123 cation
sites, the 27\,971 of Table~\ref{tab:byanion}, bonded to a single one
of the nine anions, 436 bonded to a single homonuclear-bonded anion
species or to Sb$^{3-}$, 715 bonded to more than one anion species
(Sec.~\ref{sec:methods}) and one whose cation--anion faces all lie
below the 5\% threshold, the Rh site of Na$_3$Rh(NO$_2$)$_6$, whose
ligands are the nitrite nitrogens and so count as cations.
Table~\ref{tab:byanion} summarizes the corpus anion by
anion. The law is shown family by family in SI Fig.~\ref{fig:law} for the
families of the transferability test, and every family is tabulated in SI
Tables~\ref{tab:families} and~\ref{tab:transfer}.

\begin{figure}[t]
\centering
\includegraphics[width=\columnwidth,height=0.5\textheight,keepaspectratio]{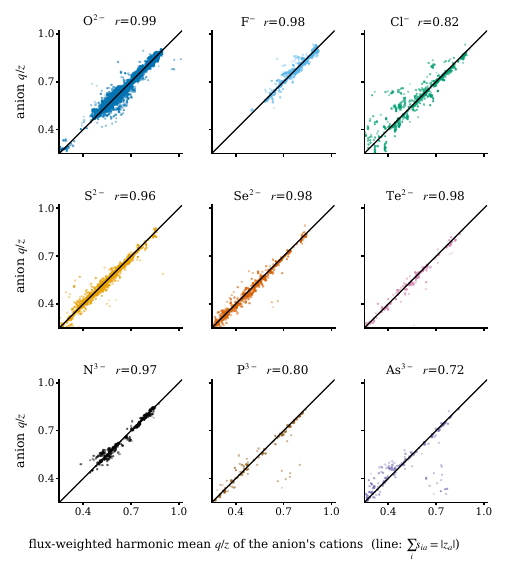}
\caption{The valence-sum rule at the anion (hydrogen-free hosts), as $q/z$ of every anion site against the flux-weighted harmonic mean $q/z$ of its bonded cations, one panel per anion species, with the Pearson $r$. The line is $\sum_i s_{ia}=|z_a|$, Pauling's rule on the flux (Eq.~\eqref{eq:anionsum}).}
\label{fig:anionrule}
\end{figure}

\begin{table*}[t]
\caption{The corpus by anion (hydrogen-free hosts). For each anion, the hosts with a bonded face to it, the cation sites bonded to it alone and their bonded faces, the cation--anion families present (the families of the homonuclear-bonded anion species and the antimonide family are not listed), the number on which the law is measured, the number in the transferability test, and over the latter the range of the within-site softness $B$. The host column sums to more than the number of hosts because a host can carry more than one anion.}
\label{tab:byanion}
\centering\footnotesize\setlength{\tabcolsep}{4pt}
\fitwidth{%
\begin{tabular}{lrrrrrrc}
\toprule
anion & hosts & sites & faces & families & law & in test & $B$ (\AA) \\
\midrule
F$^{-}$ & 271 & 1374 & 10\,881 & 42 & 12 & 8 & 0.44--0.55 \\
Cl$^{-}$ & 240 & 1241 & 8853 & 62 & 16 & 8 & 0.47--0.75 \\
O$^{2-}$ & 1593 & 14\,626 & 81\,931 & 61 & 44 & 35 & 0.33--0.76 \\
S$^{2-}$ & 510 & 4377 & 23\,942 & 64 & 28 & 21 & 0.26--0.85 \\
Se$^{2-}$ & 283 & 2155 & 11\,738 & 59 & 20 & 13 & 0.45--0.83 \\
Te$^{2-}$ & 121 & 634 & 3596 & 42 & 7 & 3 & 0.58--0.77 \\
N$^{3-}$ & 181 & 1572 & 7479 & 39 & 11 & 7 & 0.43--0.96 \\
P$^{3-}$ & 111 & 729 & 3392 & 40 & 7 & 4 & 0.65--0.85 \\
As$^{3-}$ & 164 & 1263 & 5874 & 44 & 11 & 7 & 0.63--0.94 \\
\midrule
all & 3474 & 27\,971 & 157\,686 & 453 & 156 & 106 & \\
\bottomrule
\end{tabular}
}
\end{table*}

\subsection{Basin charges and the screening ratio}

The basin charge $q$ of a cation is by Eq.~\eqref{eq:gauss} the total
flux leaving it. Figure~\ref{fig:charges} gives the screening ratio
$q/z$ of every cation site of the corpus, species by species and anion
by anion. The findings are four.

\emph{Each cation species occupies a band of $q/z$ with each anion.}
With oxygen the noble-gas-core $s$- and $p$-block cations form a sharp
band that begins at $q/z=0.72$, with 8689 of their 8692 sites
above it, while the $d^0$ and $d^{10}$ cations spread from $0.45$ to
$0.77$ with 84\% and 96\% of their sites below $0.72$ (SI
Fig.~\ref{fig:qzhist}). The deepest valley of the cation distribution,
at $0.64$, lies inside that spread. It separates Ti, Nb, Ta and Ge from
a middle band of Zn, Cd, In, Ga, Sn, Zr, Hf, Sc, Y and La ($0.64$--$0.72$),
which is resolved from the $s$/$p$ band only at a finer kernel (SI
text). The centers of the polyatomic oxoanions (C, N, P, S, Se, As)
form a class of their own that straddles the $0.72$ edge. With the
other anions the bands are fewer or more. The chlorides, phosphides and
arsenides have one valley ($0.74$, $0.42$ and $0.39$), with the $d^{10}$
cations below it and the noble-gas-core cations above it, and the
fluorides have none. The sulfides, selenides, tellurides and nitrides
have two. A middle band of the hard trivalent cations Sc, Y, La and Lu
(B as well with S, Si with Se) opens between the $d^{10}$ band and the
alkali and alkaline-earth band with S, Se and Te ($0.55$--$0.68$,
$0.51$--$0.68$ and $0.46$--$0.64$), and with N it is Ca, Sr, Ba, Be, B
and La that sit apart ($0.57$--$0.77$), a tenth below where Ca, Sr and
Ba sit with oxygen, while Li, Mg, Al and Si stay in the top band.

\emph{The band belongs to the cation--anion pair, not to the cation.}
For the alkali and alkaline-earth cations $q/z$ falls from the
fluorides through the chlorides and oxides to the sulfides, selenides,
tellurides and pnictides (SI Fig.~\ref{fig:anions_h3}). Within a pair
it rises with the coordination number for every anion but fluoride
(K$^{+}$ from $0.79$ to $0.87$ with O, $0.77$ to $0.84$ with S and
$0.86$ to $0.90$ with Cl, while with F it stays at $0.90$--$0.91$).

\emph{Within a species the ratio is nearly fixed, and the $d$ cations
are the most screened.} The cations with a closed $1s^2$ (Li, Be) or
$2s^22p^6$ (Mg, Al, Si) outer shell carry, with oxygen, a fraction of
their formal charge at medians of $0.87$ (Li$^{+}$), $0.78$
(Be$^{2+}$), $0.88$ (Mg$^{2+}$ and Al$^{3+}$) and $0.86$ (Si$^{4+}$),
with interquartile ranges of $0.01$--$0.03$, so the host moves the
deficit $z-q$ of these cations by a few percent. The $d^0$ cations are
more screened, Ti$^{4+}$ at $0.56$ and Zr$^{4+}$ at $0.66$, as are the
$d^{10}$ cations (Zn$^{2+}$ at $0.69$, Ge$^{4+}$ at $0.55$), and the
anions bonded to them receive correspondingly less flux. What lets
$q/z$ be factored out of the valence as a unit (Sec.~\ref{sec:qz}) is
that it is one number per site, not that it is nearly one number per
species.

\emph{The anion's ratio is set by its cations.} On the anion side the
screening ratio follows the flux-weighted harmonic mean of its
cations' (Fig.~\ref{fig:anionrule}), which is the valence-sum rule at
the anion and is examined next. The anion's total flux closes exactly
to its charge, and the part that does not arrive through faces from
its cations is carried by the anion--anion contacts of the same
partition (SI Fig.~\ref{fig:anion}). The amounts are small throughout,
a root-mean-square of $0.008$--$0.065$~e by anion and no contact above
$0.1$ electrons, so an anion's screening ratio is set by its cations
to within a few percent (\hyperref[app:notes]{SI notes}).

\subsection{The valence-sum rule at the anion and at the cation}

\begin{table}[t]
\caption{The valence-sum rule at the anion, Eq.~\eqref{eq:anionsum} (hydrogen-free hosts, bonded faces). Per anion species, the number of sites, the median, the 10th--90th percentile range and the share within 5\% of $S_a/|z_a|$, and the robust spread $\delta$ (half the 16th--84th percentile interval) with its two parts, the spread of the screening mismatch $(q_a/z_a)\langle z/q\rangle_\Phi$ and of the leak $L_a/|z_a|$. The homonuclear-bonded species are listed beneath the monatomic ones.}
\label{tab:anionsum}
\centering\footnotesize\setlength{\tabcolsep}{3pt}
\begin{tabular}{lrcccccc}
\toprule
anion & sites & median & 10th--90th & $\leq5\%$ & $\delta$ & $\delta_{\rm scr}$ & $\delta_{\rm leak}$ \\
\midrule
F$^{-}$ & 2813 & 1.000 & 0.948--1.035 & 84\% & 0.028 & 0.013 & 0.029 \\
Cl$^{-}$ & 2714 & 1.000 & 0.879--1.135 & 57\% & 0.073 & 0.081 & 0.065 \\
O$^{2-}$ & 20\,878 & 0.995 & 0.945--1.039 & 82\% & 0.033 & 0.021 & 0.025 \\
S$^{2-}$ & 5020 & 0.998 & 0.951--1.037 & 84\% & 0.031 & 0.026 & 0.019 \\
Se$^{2-}$ & 2372 & 1.000 & 0.935--1.071 & 70\% & 0.047 & 0.047 & 0.028 \\
Te$^{2-}$ & 678 & 1.000 & 0.930--1.056 & 74\% & 0.036 & 0.029 & 0.019 \\
N$^{3-}$ & 1399 & 0.999 & 0.953--1.036 & 83\% & 0.027 & 0.017 & 0.015 \\
P$^{3-}$ & 486 & 1.000 & 0.942--1.080 & 74\% & 0.036 & 0.034 & 0.008 \\
As$^{3-}$ & 767 & 1.001 & 0.882--1.179 & 61\% & 0.071 & 0.076 & 0.019 \\
\bottomrule
\end{tabular}
 \end{table}

\emph{The rule at the anion.} Pauling defined the strength of an
electrostatic valence bond at the cation, $s=z/n$ for a cation of charge
$z$ at the center of a polyhedron of $n$ anions, and stated his rule at
the anion. In a stable structure the charge of each anion tends to be
compensated by the sum of the strengths of the bonds reaching it from the
cations at the centers of the polyhedra of which it forms a corner,
$|z_a|=\sum_i s_{ia}$~\cite{pauling1929}. In its modern form the rule is
the valence-sum rule, that the bond valences meeting at any atom, cation
or anion, sum to that atom's valence~\cite{brown2009}. On the flux the
sum over all faces and contacts of a cation is an identity
(Eq.~\eqref{eq:s}), and the sum over its bonded faces departs from it
only by the cation--cation leak reported below. The sum at the anion
is the one that can fail, and it is the test of the rule. Let
$S_a=\sum_i s_{ia}$ be the flux valences reaching anion $a$ through its
bonded faces. Applying Eq.~\eqref{eq:gauss} to the anion basin gives,
exactly,
\begin{equation}
  \frac{S_a}{|z_a|}
  = \frac{q_a}{z_a}\Big\langle\frac{z}{q}\Big\rangle_{\!\Phi}
  + \frac{L_a}{|z_a|}\Big\langle\frac{z}{q}\Big\rangle_{\!\Phi},
  \label{eq:anionsum}
\end{equation}
where $\langle z/q\rangle_\Phi$ is the flux-weighted mean of $z_i/q_i$
over the anion's bonded cations, the inverse of the flux-weighted
harmonic mean of their screening ratios, and $L_a$ is the flux the anion
passes on through its other faces and contacts, the anion--anion
contacts and the faces below the 5\% threshold. The rule $S_a=|z_a|$ therefore holds when the
anion's screening ratio is the harmonic mean of its cations' and the
leak is small, and neither is a theorem. On the corpus both hold
(Table~\ref{tab:anionsum}, Fig.~\ref{fig:anionrule}). The vertical
distance of a site from the line of Fig.~\ref{fig:anionrule} is its
valence-sum deviation $S_a/|z_a|-1$ times its cations' mean screening
ratio, less the leak $L_a/|z_a|$ through the anion's other faces and
contacts. For the $20\,878$
oxygen sites the median of $S_a/|z_a|$ is $0.995$, the 10th and 90th
percentiles are $0.945$ and $1.039$, and 82\% of the sites lie within
5\% of the rule. Fluoride, sulfide and nitride are as tight
($0.948$--$1.035$, $0.951$--$1.037$ and $0.953$--$1.036$, 83--84\%
within 5\%), telluride, phosphide and selenide somewhat wider (74\%,
74\% and 70\%), and chloride and arsenide the widest ($0.879$--$1.135$
and $0.882$--$1.179$, 57\% and 61\%). The spread divides about equally
between its two terms for oxygen ($\delta=0.033$, with $0.021$ from the
screening mismatch and $0.025$ from the leak, added in quadrature), and
chloride is the
one anion for which both terms are large ($0.081$ and $0.065$). The
homonuclear-bonded species, whose formal charge is an assigned share of
an anion--anion bond, scatter by $0.06$--$0.43$. There the rule has no
fixed charge to close on, and the flux says so. For the oxides,
fluorides, sulfides and nitrides, then, the valence-sum rule holds at
the anion to within 5\% for four fifths of the sites and to within 10\%
for 94--97\% of them, with no parameter fitted to make it so, and for
the chlorides, phosphides and arsenides to within 10\% for 72--83\%. The
deviations are the anion-side analogue of the bond-valence-sum
residuals of the empirical model, and here they have two identified
sources, the spread of the screening ratio among the cations meeting at
the anion and the flux that leaves the anion through faces that carry
no bond valence.

\emph{The sum rule at the cation.} Gauss's law closes on every face and
contact of a cation basin, and not all of them are cation--anion faces.
Four fifths of the $29\,123$ cation sites of the hydrogen-free hosts
share at least one contact with another cation, 12\% a face at a bond
point, and the $65\,251$ such contacts carry $0.7\%$ of the total cation
flux. The bond-point faces below the 5\% threshold of
Sec.~\ref{sec:theory} are $0.3\%$ of all bond-point faces, lie
$1.4$--$1.9$~\AA{} (tenth percentile to median) beyond the shortest bond
of their site and carry $10^{-4}$ of the cation flux, so counting them or
not changes the sum at the cation only at that order. Let $L_i$ be the net flux a cation passes to other cations
through them. The sum of
the bonded flux valences at the cation is then $z_i(1-L_i/q_i)$. The
ratio $L_i/q_i$ is small on most sites and large on a few. Over the
$20\,086$ sites of the 106 families it is $0.05\%$ at the median,
$2.7\%$ at the 90th percentile and $22\%$ at the 99th, and it exceeds
$5\%$ on $1311$ sites in 50 of the families. The large values belong to the $d^{10}$ and $d^{8}$ cations Ag$^{+}$,
Hg$^{2+}$, Pd$^{2+}$ and Au$^{3+}$, whose low basin charge makes them
conduits for the flux of their neighbors, and to the alkali and
alkaline-earth cations of the pnictides (\hyperref[app:notes]{SI notes}). Because $q_i$ is one number per site,
$L_i/q_i$ multiplies every bonded face of the site by the same factor,
as $q/z$ does (Sec.~\ref{sec:qz}). The within-site law and $B$ are
untouched, and the site's $R_0$ shifts by $B\ln(1-L_i/q_i)$, which is
part of the host-to-host spread $\sigma_n$ of the soft families
reported with the test of transferability.

\subsection{The face flux obeys the bond-valence law}

The softness $B$ is extracted first, independently of $R_0$, through
Eq.~\eqref{eq:distortion}, from the faces of every distorted site of a
family in the hosts where it occurs. On the 156 families with six or more independent distorted sites
(SI Table~\ref{tab:families}) it is determined to a 68\% bootstrap
half-width of $0.03$~\AA{} at the median ($0.02$--$0.05$~\AA{}
interquartile) and, over the 147 families determined to better than
$0.15$~\AA{}, runs from $0.33$~\AA{} (Si--O) to $0.76$~\AA{} (Tl--O)
with O, $0.38$--$0.61$~\AA{} with F, $0.44$--$1.02$~\AA{} with Cl,
$0.15$~\AA{} (P$^{5+}$--S) to $0.85$~\AA{} with S, $0.45$--$0.83$~\AA{}
with Se, $0.58$--$0.77$~\AA{} with Te and $0.43$--$1.21$~\AA{} with the
pnictides, stiff for the small, highly charged cations and soft for the
large ones, softer with the softer anions, as in the ``soft''
parametrizations~\cite{adams2001}, and nowhere a universal $0.37$~\AA{}.
The nine families determined less well are those whose polyhedra are
nearly regular or which have two or three hosts (Zn--S, half-width
$0.6$~\AA{}, and Mo$^{2+}$--Cl, $1.0$~\AA{}, among them).
The crystallographically independent sites of one family scatter about
the family value by $0.08$--$0.11$~\AA{} for Si--O, Ge--O, Nb--O and
Ti--O and more widely for the alkali and alkaline-earth families and for
most families of the softer anions (SI Fig.~\ref{fig:bsites}), the
family value being their pooled slope. Within a family the per-site $B$
depends on coordination number, rising with coordination for the compact
cations and falling for the soft ones, with no sign common to all
families (SI Fig.~\ref{fig:bcn}, SI Table~\ref{tab:bcn}, \hyperref[app:notes]{SI notes}).

With $B$ held at the family's own value, SI Table~\ref{tab:families} fits
$\ln s$ against $d$ with one free parameter $R_0$ per family.
Individual faces scatter about the law by $1.4$--$46$\% in $s$
(quartiles $5.8$, $10.2$ and $16.2$\%), the compact cations at the low
end and the soft ones at the high. Because $\sum_j s_{ij}=z_i$ on every
site by construction, the sum rule fixes the site mean of $s$ at $z/n$,
and the site mean of $\ln s$ that $R_0=\langle d+B\ln s\rangle$ needs
differs from $\ln(z/n)$ by the Jensen gap $G=-\langle\ln(sn/z)\rangle\geq0$,
which vanishes when the faces carry equal valences and grows with how
unequally a distorted site shares its flux. $R_0$ is therefore
$\langle d\rangle + B\ln(z/n) - BG$ averaged over the family's sites
(Eq.~\eqref{eq:r0site}), bond lengths and coordination numbers plus a
gap that carries the same within-site information as $B$. A null model
measures the gap. Assigning Pauling's equal split $z/n$ to every face of
a site, at the family $B$, gives an $R_0$ within $0.005$~\AA{} of the
flux value in 35 of the 156 families and within $0.015$~\AA{} in 73
(B--O, Si--O, Al--O and P--O differ by $0.001$--$0.003$~\AA{}),
$0.019$~\AA{} at the median. The rest, mostly soft cations, differ by up
to $0.34$~\AA{} (As$^{2+}$--S), the gap $BG$ of their widely distorted
sites (SI Table~\ref{tab:families}). The
electron density enters where the sum rule is silent, in how a
distorted site distributes its flux over bonds of unequal length. There
the exponential law has $0.3$--$0.6$ (interquartile range) of the
scatter of the equal split and less than the equal split in 147 of the
156 families (Si--O $4.1$\% against $7.0$\%, P--O $2.7$\% against $11$\%,
S--O $2.9$\% against $16$\%, Al--O $4.5$\% against $12$\%, Mg--O $6.5$\%
against $13$\%, Ti--O $6.1$\% against $22$\%, B--O $2.6$\% against
$6.9$\%). The nine exceptions are families whose polyhedra are nearly
regular, where the equal split is already within $2$--$8$\% (Zn--S
$1.9$\%, Mg--Se $4.7$\%) and the within-site softness is barely
defined. That distribution is what the softness measures.

\begin{figure*}[t]
\includegraphics[width=\textwidth]{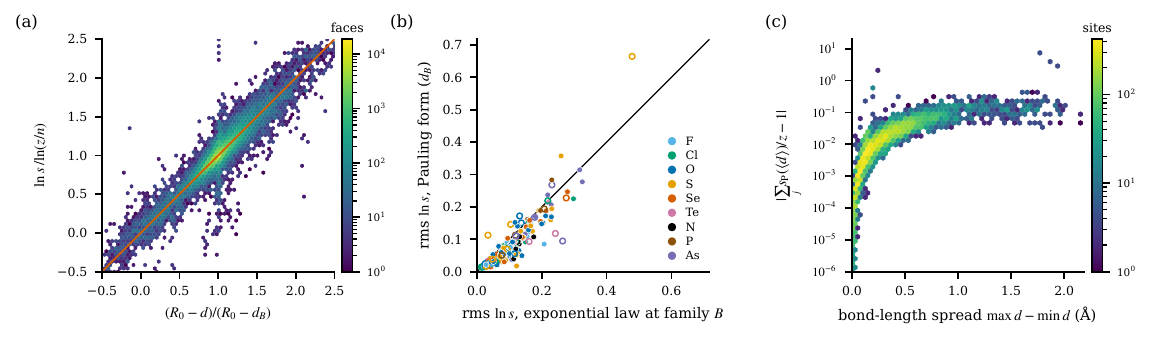}
\caption{Pauling's rule with a distance exponent, Eq.~\eqref{eq:pauling}, with the family $R_0$ and $B$ of SI Table~\ref{tab:families}, on every bonded cation--anion face of a site with $z\neq n$ (hydrogen-free hosts). (a)~The measured $\ln s/\ln(z/n)$ against the exponent $(R_0-d)/(R_0-d_B)$, as a density of faces on a logarithmic color scale, and the line is Eq.~\eqref{eq:pauling}. (b)~Per family, the rms of $\ln s$ about Eq.~\eqref{eq:pauling} against the rms about the exponential law at the same $R_0$ and $B$, colored by anion, filled for the families of the transferability test and open for the others. (c)~The closure error per site of the form with the arithmetic mean $\langle d\rangle$ in place of $d_B$, against the spread of the site's bond lengths, as a density of sites.}
\label{fig:pauling}
\end{figure*}

\emph{Pauling's rule with a distance exponent.} Equation~\eqref{eq:pauling}
is Eq.~\eqref{eq:law} with the softness eliminated by the sum rule, and
on a site whose faces follow the law at the site's own $R_0$ it is an
identity. Its content as a prediction is that a family $R_0$ and the
site's own bond lengths suffice. With $R_0$ and $B$ of
SI Table~\ref{tab:families}, the latter entering only through $d_B$, it
assigns to every face a valence from $z$, $n$ and the bond lengths
alone. Figure~\ref{fig:pauling} tests it on the 126\,453 faces of the
21\,688 sites with $z\neq n$ in the 154 families of
SI Table~\ref{tab:paulingcorpus} (two of the 156 have $z=n$ at every
site). The measured $\ln s/\ln(z/n)$ follows
the exponent $(R_0-d)/(R_0-d_B)$ on one line for all families
(Fig.~\ref{fig:pauling}(a)), with an rms of $0.11$ in $\ln s$ pooled
over the corpus, $0.021$ for Si--O, $0.031$ for Al--O, $0.037$ for Mg--O
and $0.038$ for Ti--O, $0.13$ and $0.15$ for Na--O and K--O. It lies
closer to the flux than the exponential law at the same family $R_0$
and $B$ ($0.13$ pooled) in 137 of the 154 families
(Fig.~\ref{fig:pauling}(b)), because it imposes the sum rule on each
site, which the exponential law at a family $R_0$ does not, and it has a
third of the scatter of the equal split ($0.30$ pooled, quartiles of the ratio $0.23$--$0.46$),
which it beats in every family. The form with the arithmetic mean in
place of $d_B$, the regular-polyhedron form, has an rms of $0.16$ and
does not close. Its closure error grows with the spread of the site's
bond lengths (Fig.~\ref{fig:pauling}(c)), which is what the cumulant
correction of Eq.~\eqref{eq:cumulant} removes. On the 21\,688 sites
the gap $\langle d\rangle-d_B$ at the family $B$ is $0.004$~\AA{} at the
median and $0.04$~\AA{} at the 90th percentile, but it exceeds
$0.05$~\AA{} on 7\% of the sites and reaches $0.49$~\AA{} on the most
distorted. The per-family maximum is
$9\%$ at the median and reaches $750\%$ on one site where
$R_0-\langle d\rangle$ passes through zero. With a family $R_0$ the $d_B$ form
closes to $0.2\%$ on average and $20\%$ at worst, the residual being the
offset of the site's own $R_0$ from the family's
(next subsection). The 11\,394 faces of the 3031 sites with
$z=n$, the Si and Zn tetrahedra among them, are the degenerate case. The
form assigns unit valence to every face, and the measured valences
scatter about it by $0.16$.

\subsection{Transferability across hosts}\label{sec:transfer}

Bond-valence theory rests on the premise that $R_0$ and $B$ are
properties of the cation--anion pair, carried unchanged from one
compound to the next~\cite{brown2009,brese1991}. On the electron density
that premise is two statements. First, that the softness read from how a
distorted site shares its flux, the tangent of the family's flux--length
curve, equals the softness $B^{*}$ read from the shift of the mean bond
length between coordination numbers, its secant, so $B=B^{*}$. Second,
that one $R_0$ serves every host, so that regular sites at the same
coordination number have the same mean bond length, since the sum rule
fixes $R_0=\langle d\rangle+B\ln(z/n)$. Of the 156 families on which the
law is measured, a family enters the test when its sites occupy at least
two coordination numbers with three or more hosts each, and it passes
when $|D|\leq\max(0.03~\text{\AA},\text{half-width})$, $D=B-B^{*}$. 106
families meet the
requirement, their distribution over the anions and the 50 families
excluded given in the \hyperref[app:notes]{SI notes}. SI
Fig.~\ref{fig:transfer} gives the result on the density partition, SI
Table~\ref{tab:transfer} every family, and SI
Table~\ref{tab:partitions} $B$ on six geometric partitions of the same
densities beside it.

\emph{What the test finds.} The equality fails in 96 of the 106
families and holds within the resolution of the test in ten (Si--O,
Al--O, Na--F, Na--Cl, Ba--Cl, Al--S, Li--S, Ba--Se, Na--Te and Li--As).
Ten agreements among 106 tests with intervals of this width, following no
order in cation or anion, are what the scatter of the test allows and are
not evidence of transferability. Where the test fails the within-site
response is almost always the softer one. The within-site $B$ exceeds
$B^{*}$ in 92 of the 96 failing families, by $0.04$--$1.45$~\AA{} (median
$0.28$~\AA{}), and only four fail the other way. Which of the two
statements each family satisfies, and how the within-site scatter of
$\ln s$ follows the same order, are set out family by family in the
\hyperref[app:notes]{SI notes}.

\emph{The failure is in the density, not in the partition.} $B^{*}$ is
computed from bond lengths and coordination numbers alone, so it is the
same number for every partition of a cell, while $B$ is the only
quantity in $D$ that depends on where the faces are. Moved across seven
partitions, from a cation's own zero-flux surface to a plane halfway to
its anions and from the eroded to the dilated density basins, $B$ stays
above $B^{*}$ in almost every failing family (SI
Table~\ref{tab:partitions}, \hyperref[app:notes]{SI notes}), so the
failure is in how the density distributes a site's flux over its bonds,
relative to how its bond lengths change with coordination number, and not
in the surface through which the flux is counted. The relaxed geometries
and a few unrepresentative distorted sites, two further explanations, are
excluded in the \hyperref[app:notes]{SI notes}.

\emph{What the two failures are.} Read through Sec.~\ref{sec:theory}, a
within-site $B$ above $B^{*}$ means a bond's flux responds to the site's
own distortion more softly than the regular bond length responds to a
change of coordination number, two derivatives of one density that the
single-valued law assumes equal. Where in addition $R_0$ varies between
hosts, the valence of a bond of given length depends on the host. The
mechanism and the empirical parameters it explains are set out in the
\hyperref[app:notes]{SI notes}. Neither failure touches the sum rule or
the identity of Sec.~\ref{sec:theory} between the exponential law and
Pauling's rule, which holds site by site with the site's own $R_0$. What
they show is that the transferability the empirical model assumes is not
a property of the electron density. The softness is a property of the
site, not of the cation--anion pair, and where the two softnesses
coincide, for a few compact cations, the test cannot distinguish
coincidence from chance. What does transfer is measured directly in SI
Fig.~\ref{fig:transfercomp} and SI Tables~\ref{tab:transferselect}
and~\ref{tab:transfercomp}, which split the scatter of the family law
into its host, site and within-site parts and give the host-to-host
spread of $R_0$ and the reproducibility of the within-site $B$.

\emph{What the failure depends on.} The gap $D$ is not predicted by any
cation descriptor, and the coordination branch is not curved in a
universal way, so the two failures are distinct and the failure is a
within-site softness that does not follow the branch rather than a bending
of the law (SI Figs.~\ref{fig:branch} and~\ref{fig:branchall},
\hyperref[app:notes]{SI notes}).

\emph{Across the anions.} Across every anion the failure keeps its sign,
the within-site softness staying above the coordination branch, and the
families that pass follow no order in cation or anion (SI
Figs.~\ref{fig:anions_h1} and~\ref{fig:anions_h2}, SI
Table~\ref{tab:anions}, \hyperref[app:notes]{SI notes}).
\section{Discussion}\label{sec:discussion}

\paragraph{What Gauss's theorem fixes and what the electron density adds.}
Bond-valence theory takes the valence-sum rule as the first of its three
axioms and glosses it as a corollary of Gauss's flux
theorem~\cite{gibbs2022}. Gauss's theorem constrains the
total flux through any closed surface whatever and says nothing about how
that total divides or which parts of the surface are bonds. Without a
surface the corollary can only be asserted. The electron density supplies
it. A basin boundary fixed by $\nabla\rho$ divides into faces in
one-to-one correspondence with the bond points and carries a flux
computed from the same density. On it $\sum_j\Phi_{ij}=q_i$ is Gauss's law and $\sum_j s_{ij}=z_i$ follows
by the choice of unit, so the rule is not a hypothesis that structures
happen to obey but holds for any electron density to the accuracy of the
Poisson solve. How the total divides is then measured rather than
assumed. On 160\,300 bonded faces the division follows the bond length
through Eq.~\eqref{eq:law}, with a face-to-face scatter of 1.4--46\% in $s$
by family (median 10\%) and no free charge, while the parts of the
surface that are not cation--anion faces carry little, no anion--anion
contact above $0.1$~e (Sec.~\ref{sec:results}). The
step that counts the basin charge as the formal charge, and the relation
of Eq.~\eqref{eq:flux} to Brown's point-charge identification of the
valence with the flux~\cite{preiser1999}, have physical readings taken up
in the \hyperref[app:notes]{SI notes}.

\paragraph{Bader charges, bond points and the bond-point density.}
QTAIM and the present construction share the partition. The basins, their
zero-flux faces and the bond points are Bader's~\cite{bader1990}, and the
charge $q_i$ in Eq.~\eqref{eq:gauss} is the Bader
charge~\cite{henkelman2006,yu2011}. The difference is what is attached to
a face. A Bader charge is one number per atom. Gauss's law distributes it
over the atom's faces, and that distribution is the new information,
telling two cations of equal basin charge apart by fluxes that follow
their bond lengths (Eq.~\eqref{eq:distortion}). The bond-point density
$\rho_b$, on which a connection with Pauling's bond strength has been
built~\cite{gibbs1998,gibbs2003,gibbs2014,gibbs2022}, is a value at one
point rather than a charge, so the bond-point densities at an atom are
not constrained to sum to anything, whereas the flux is an integral over
the face that inherits the basin charge by Gauss's law. A correlation of
$\rho_b$ with bond length can witness the valence but cannot close it. A
bond point is not, however, what makes a bond. The
anion--anion faces of the topology are genuine faces with genuine bond
points, yet together with the other contacts they carry a root-mean-square
net flux of at most $0.07$~e per anion (Sec.~\ref{sec:results}), so the
flux, not the existence of a face, decides what is a valence bond, and it
decides in favor of the cation--anion faces. How the flux sits on a face,
and the relation of the construction to the bond-point and bond-order
routes to the valence, are taken up in the \hyperref[app:notes]{SI
notes}.

\paragraph{The formal charge and the basin charge.}
QTAIM reports that Bader charges are smaller than formal charges and
leaves the warning that an oxidation state is not an atomic charge as a
warning~\cite{raebiger2008,walsh2017,walsh2018}. Here the gap is given a
role. The physical valence carried by the faces is $\Phi$, in electrons,
with $\sum\Phi=q$, and $z$ enters only through $s=\Phi z/q$. The ratio
$q/z$ is one number per site rather than one per bond, so it factors out
of the valence entirely. It sets the unit in which $s$ is quoted, leaves
$B$, the law and the sum rule untouched, and displaces $R_0$ by
$B\ln(q/z)$, $-0.05$ to $-0.06$~\AA{} for the closed-shell cations with
oxygen (Sec.~\ref{sec:qz}), so that fits which normalize to $z$ absorb
the screening into $R_0$. Across hosts $q/z$ is a band for each cation
species and anion (Fig.~\ref{fig:charges}). For the five elements the
corpus carries in two oxidation states bonded to oxygen the median basin
charge rises by only $0.31$--$0.43$ electrons per unit of $z$ (Ti from
$1.91$ to $2.22$ between $3+$ and $4+$), a third to a half of the step,
so the deficit $z-q$ grows with $z$ and $q/z$ falls, the charge
self-regulation found for transition-metal ions in
insulators~\cite{raebiger2008}, here at a larger fraction of the step,
and the reason the formal charge is a label and the basin charge a
per-site unit of the model, not a per-bond correction. The
lower band of the $d^0$ and $d^{10}$ cations is $d$--$p$ mixing counted to
the cation side of the zero-flux surface, which leaves the law and the
Pauling form untouched, and the bands, set by formal charges that are
themselves bond-valence assignments, are a check a bond-length fit cannot
make (\hyperref[app:notes]{SI notes}).

\paragraph{The conventional parametrizations.}
The bond-valence model postulates Eq.~\eqref{eq:bvlaw} and the sum rule
and fits $R_0$ and $B$ to structures under the sum-rule
constraint~\cite{brown1985,brese1991,adams2001,gagne2015}. The flux gives
the two a different standing. With $s=\Phi z/q$ the sum rule holds on
every site, so $R_0$ at fixed $B$ is the sum rule acting on the bond
lengths, Eq.~\eqref{eq:r0site}, which the equal split at the family $B$
reproduces to $0.019$~\AA{} at the median (Sec.~\ref{sec:results}). $R_0$
is a property of the structures, which the electron density neither tests
nor predicts. $B$ is
where the electron density speaks. It is $0.33$--$0.76$~\AA{} with oxygen
and up to $1.2$~\AA{} with the softer anions, not the universal
$0.37$~\AA{} of the standard parametrization~\cite{brown1985,gibbs2022},
in the direction of the soft-bond-valence sets~\cite{adams2001} and the
joint $(R_0,B)$ fits~\cite{sidey2009}. The model's third element, that
$R_0$ and $B$ carry from one compound to the
next~\cite{brown2009,brese1991}, is testable on the electron density, and
the test fails in most families (Sec.~\ref{sec:transfer}). The
within-site softness exceeds the coordination-branch softness by
$0.19$--$0.40$~\AA{} (interquartile range), so the coordination
dependence of the valence is carried by $R_0$ in any single-softness fit.
The density does
not remove the failure. It measures it, and shows that the sum rule and
the Pauling form do not depend on it. What the data do not decide is the
functional form. Equation~\eqref{eq:bvlaw} remains empirical, and the electron
density does not select between its exponential and power-law
expressions, the same law linearized in $d$ and in $\ln d$
(\hyperref[app:notes]{SI notes}). Two recent descriptions from the
empirical tables~\cite{whittaker_ammin,whittaker_prm} contain the class
line $R_0=\bar R+B\ln(z/n)$, $B^{*}$ and the Boltzmann weights as
identities of the exponential form. Their screening-length reading of the
softness, the loop rule and their standing are taken up in the
\hyperref[app:notes]{SI notes}.

\paragraph{Pauling's rule and the bond valence.}
Pauling's rule contains no bond length and the exponential law no
coordination number, and the two have been kept apart by name, a strength
computed from an observed bond length being called a bond valence
expressly ``to avoid confusion with Pauling
bond-strengths''~\cite{gibbs2022}.
Equation~\eqref{eq:pauling} removes the distinction. With the softness
eliminated by the sum rule, the bond valence is Pauling's strength raised
to a distance exponent, exactly and for any polyhedron. The strength is
the valence at the site's mean bond length, the valence is the strength
at any other length, and the distance and coordination dependences are
one expression, which on the flux holds bond by bond
(Fig.~\ref{fig:pauling}, SI Table~\ref{tab:paulingcorpus}). The nearest
earlier form is the curve with which Donnay and Allmann introduced the
term~\cite{donnay1970}, whose exponent is a softness read from a tabulated
cutoff rather than from the flux (\hyperref[app:notes]{SI notes}). The
equal-split rule, exact on a regular polyhedron and only there, holds to
within $0.01$ valence units on undistorted oxide polyhedra and only
approximately on distorted ones~\cite{george2020}, while the flux form
(Table~\ref{tab:anionsum}) holds within 5\% for 82\% of the oxygen sites
and within 10\% for 97\% of them. The exponential law carries on every
distorted site the departure the equal split cannot, to the accuracy of
the leak and the screening mismatch (\hyperref[app:notes]{SI notes}).

\paragraph{Limits of the present electron densities.}
The fluxes are computed from PBE pseudo-valence densities with frozen
Slater cores at the project's relaxed geometries, so the basin charges,
and with them $q/z$, are those of these densities, $R_0$ is quoted at
the relaxed geometry, which enters $B$ and $B^{*}$ alike
(Sec.~\ref{sec:transfer}), and the weak faces that the frozen-core
completion can manufacture in flat channels, carrying at most a few
hundredths of an electron~\cite{makeaim1}, are read as errors of the
density rather than as bonds.

\paragraph{Beyond the scope of this work.}
The corpus excludes the cations with partly filled $3d$ and $4f$ shells
(Sec.~\ref{sec:methods}), for which PBE without a Hubbard term closes
gaps that are open and overstates the $d$--$p$ mixing that sets the
screening, so their basin charges, whether their faces obey the law and
with what softness await densities from DFT$+U$, hybrid functionals or
many-body methods, and what sets the band of $q/z$ of each cation
species with each anion is not explained here.
\section{Conclusions}\label{sec:conclusions}

For the oxides, fluorides, chlorides, sulfides, selenides, tellurides,
nitrides, phosphides and arsenides of a 3680-host corpus, the faces of
the electron density's atomic basins are its bond-point set, and the
electric flux of the same electron density through those faces is the
bond valence. This is an operational definition of the valence, and on
it the valence-sum rule, until now the first axiom of the theory, is a
theorem, Gauss's law with basin charges. The coordination number of
Pauling's rule is the number of faces of the cation basin that carry a
bond valence, read from the topology of the density and not from a
distance cutoff.

The bond-valence--bond-length law is the flux--distance law of the
electron density, and on 160\,300 bonded faces of the 3436 hydrogen-free
hosts it holds on every site, with a face-to-face scatter of 1.4--46\% in
$s$ by family (median 10\%). Of its two parameters
the electron density fixes one. The softness $B$ is read from how a
distorted site distributes its flux over bonds of unequal length,
independently of $R_0$. It runs from $0.33$ to $0.76$~\AA{} with oxygen
and up to $1.2$~\AA{} with the softer anions across the 156 families on
which it is measured, stiff for the small, highly charged cations and
soft for the large ones and with the softer anions. With it the law has $0.3$--$0.6$ of the scatter
of an equal split (interquartile range) and less than the equal split
in 147 of the 156 families. $R_0$ is not an electron-density prediction.
At fixed $B$ the sum rule sets it from the bond lengths, up to a gap
that grows with how unequally a distorted site shares its flux, and an
equal split $z/n$ at the family $B$ reproduces it to $0.015$~\AA{} in
73 families and $0.019$~\AA{} at the median, the larger gaps belonging
to the widely distorted sites of the soft cations.

The normalization of the valence to the formal charge is explained. The
basin charge $q$ is measured and the oxidation state $z$ is an integer
label, and the two enter only through the screening ratio $q/z$, one
number per site that fixes the unit of the valence, leaves $B$
untouched and is absorbed into $R_0$ by any fit that normalizes to $z$,
so the empirical model never had to measure it. Across hosts
$q/z$ is a band for each cation species and anion, $0.78$--$0.88$ with
oxygen for the closed-shell cations, lower with the softer anions, and a
lower band for the $d^0$ and $d^{10}$ cations (Ti$^{4+}$ at $0.56$),
whose formally empty or filled $d$ shells mix with the anion $p$ states.
Between the oxidation states of one element the basin charge rises by
only $0.31$--$0.43$ electrons per unit of $z$. The label changes by an
electron and the density self-regulates, so the formal charge
is a label and the basin charge is not. At the anion the screening ratio
is the flux-weighted harmonic mean of its cations', which is Pauling's
rule stated at the anion. With no parameter fitted it holds to within
5\% for 82--84\% of the oxide, fluoride, sulfide and nitride sites,
70--74\% of the selenide, telluride and phosphide sites and 57--61\% of
the chloride and arsenide sites. The anion--anion contacts of the
partition carry little. No anion--anion contact of the corpus carries
more than $0.1$ electrons, and their net flux per anion is at most
$0.07$ electrons root-mean-square. A cation
shares contacts or faces with other cations in four fifths of the
sites, but the net flux through them is below 3\% of its
charge on nine sites in ten and, where it is larger, for the $d^{10}$
and $d^{8}$ cations and the soft cations of the pnictides, it rescales
the site's bonded valences by a common factor and leaves $B$ untouched.

With these, bond-valence theory and Pauling's second rule are one
statement. Eliminating the softness by the sum rule gives
$s=(z/n)^{(R_0-d)/(R_0-d_B)}$, Pauling's electrostatic valence at the
softness-weighted mean bond length $d_B$ raised to a distance exponent,
unit valence at $R_0$, no softness parameter, and exact closure on every
polyhedron at the site's own $R_0$. With a family $R_0$ it is a
prediction from $z$, $n$ and the site's bond lengths alone, and on the
$126\,453$ faces of the corpus with $z\neq n$ it reproduces the flux to
an rms of $0.11$ in $\ln s$, closer than the exponential law at the
same parameters ($0.13$) in 137 of the 154 families and a third of the
equal split ($0.30$), which it beats in every family. The coordination
dependence of the valence is Pauling's $B\ln(z/n)$, which a fit at a
common softness must load onto $R_0$. The bond valence was given its
name to keep it distinct from Pauling's bond strength. On the electron
density the two are one quantity, the strength being the valence at
the site's mean bond length and the valence the strength raised to a
distance exponent. Pauling's rule is the softness-free form of
bond-valence theory, bond-valence theory is Pauling's rule with a
distance exponent, and the electron density carries both.

What the electron density does not support is the transferability of
$R_0$ and $B$ between hosts. That premise requires the within-site
softness to equal the coordination-branch softness, the shift of the
mean bond length between coordination numbers. Tested on the 106
families, the equality fails in 96, the within-site softness exceeding
the coordination-branch softness in 92 of them by $0.19$--$0.40$~\AA{}
(interquartile range), on every partition of the density tried for the
oxides and fluorides. In 65
the bond length at fixed valence also varies between hosts, by
$0.05$--$0.33$ in $\ln s$. The ten families in which the two agree,
Si--O, Al--O and Al--S among them, do so within the resolution of the
test and in no order of cation or anion. The coordination-specific and
compound-specific parameters of the empirical tables are the trace of
that failure. The sum rule and the Pauling form do not depend on it.
They hold site by site, with the site's own $R_0$, in every host.

These conclusions are drawn from PBE pseudo-valence densities with
frozen cores at relaxed geometries, on a corpus from which the cations
with partly filled $3d$ and $4f$ shells are excluded. The basin charges,
and with them $q/z$, are those of these densities. Those cations await
densities that treat them correctly.

\begin{acknowledgements}
Claude (Anthropic) was used to develop analysis code and to assist in editing the text.
All text, figures, code and data were reviewed and verified by the author, who takes full
responsibility for the content.
\end{acknowledgements}

\begin{funding}
This work was supported by the U.S. Department of Energy, Office of Science, Basic Energy
Sciences, Materials Sciences and Engineering Division.
\end{funding}

\ConflictsOfInterest{The author declares no conflicts of interest.}

\DataAvailability{The certified critical-point censuses, face fluxes, basin charges and
every analysis table and figure of this article, with the scripts that regenerate them,
are available at \url{https://github.com/MWhitta/bond-valence-electrostatics}. The
electron densities, the voxel arrays of the basin partition and the pooled face tables
from which those results are computed are deposited at Zenodo
(\url{https://doi.org/10.5281/zenodo.XXXXXXX}).}

\newcommand{\SItitlepage}{%
  \clearpage
  \setcounter{figure}{0}\setcounter{table}{0}\setcounter{equation}{0}%
  \renewcommand{\thefigure}{S\arabic{figure}}%
  \renewcommand{\thetable}{S\arabic{table}}%
  \renewcommand{\theequation}{S\arabic{equation}}%
  \renewcommand{\theHfigure}{S.\arabic{figure}}%
  \renewcommand{\theHtable}{S.\arabic{table}}%
  \renewcommand{\theHequation}{S.\arabic{equation}}%
  \begin{center}
    {\Large\bfseries Supporting information for:\\[0.4em]
     Bond-valence theory is a general electrostatic valence rule\par}
    \vspace{1em}
    {\large M.~L.~Whittaker\IUCrOrcidlink{0000-0002-9724-3409}\par}
    \vspace{0.3em}
    Lawrence Berkeley National Laboratory, Berkeley, CA 94720, USA\par
    \href{mailto:mwhittaker@lbl.gov}{mwhittaker@lbl.gov}\par
  \end{center}
  \vspace{1em}}
\ifdefined\SItitlepage\SItitlepage\fi
\appendix
\section{Supporting information}\label{app:si}

\subsection{Supporting notes}\label{app:notes}

\begin{figure*}[t]
\includegraphics[width=\textwidth]{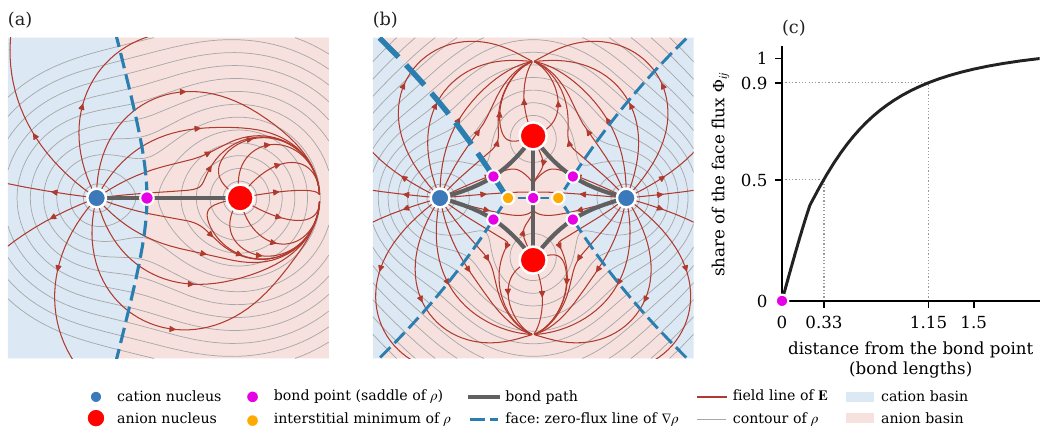}
\caption{Basins, faces and flux on a two-dimensional model density (a sum of exponential ion densities on point nuclei), with contours of $\rho$ in gray, cation basins blue, anion basins rose, bond points magenta, interstitial minima orange, faces dashed, field lines of $\mathbf E$ red (one per unit of nuclear charge). (a)~One cation and one anion. (b)~Two cations sharing two anions, with the upper-left cation--anion face drawn heavier. (c)~The share of the flux $\Phi_{ij}$ through that face lying within a given distance of its bond point along the face, in units of the bond length.}
\label{fig:basins}
\end{figure*}

\emph{The construction on a model density.} In Fig.~\ref{fig:basins}(a)
the uphill flow of $\nabla\rho$ divides the plane into a cation basin and
an anion basin. The bond point is the saddle between them, its uphill
eigendirection traces the bond path and its downhill eigendirection
generates the face $F_{ij}$. The field lines of $\mathbf E$ leave the
cation, cross the face into the anion and end at the zeros of $\mathbf E$
inside the anion's electron excess, where the anion nucleus's own lines
also end. In panel (b), two cations sharing two anions, each cation basin
is bounded by its two bond-point faces, which meet at an interstitial
minimum of $\rho$, and the mirror line between the minima is an
anion--anion face with its own bond point. Away from its bond point the
cation--anion face runs parallel to $\mathbf E$ and carries no flux, and
the anion--anion face is tangent to $\mathbf E$ by symmetry and carries
none at all. Panel (c) accumulates the flux of Eq.~\eqref{eq:flux} along
one cation--anion face. Half of the face's flux crosses within $0.33$
bond lengths of the bond point and nine tenths within $1.15$. In the
crystals the symmetry that empties the anion--anion face of the model is
absent (anion--anion contacts, below).

\emph{Electron densities and the pipeline.} The frozen Slater-type cores
carry exactly the $Z-z_{\rm val}$ electrons that each pseudopotential
freezes, semicore pseudopotentials being completed by partial shells.
Geometry and electrostatics both use the periodic quintic spline of the
stored grid, smoothed by $0.03$~\AA{} and projected onto the space group
(a grid not commensurate with the space-group operations is resampled),
plus the analytic cores. The basin charges integrate over the cores, so
the $q/z$ reported here are those of these densities. The field and the
partition are built by the makeAIM construction~\cite{makeaim1}, which
also carries the all-electron and r$^2$SCAN densities against which the
pseudo-valence results are judged.

\emph{Faces, contacts and fluxes.} The $(3,-1)$ critical points of the
density at $0.03$~\AA{} smoothing define the candidate faces. The cell is
sampled on voxel grids of spacing $0.13$ and $0.087$~bohr, every voxel
center is assigned to a nucleus by steepest-ascent integration of
$\nabla\rho$, and the basin boundaries are resolved by bisection along
every voxel edge, which recovers the sheets of cation basin, $10^{-9}$ to
$10^{-3}$~bohr thick, that separate adjacent anion basins and that a voxel
labeling alone misses. Two basins that share voxel faces after the
bisection form a contact, a face when a bond point lies on it, and the
flux through it is the flux through those voxel faces. The net
anion--anion flux of an anion is the sum of the signed fluxes through its
anion--anion contacts, by Eq.~\eqref{eq:gauss} the part of its charge not
received through faces from its cations. The Fourier coefficients of
$\varrho$ (spline, symmetry projection, analytic core transforms, nuclei
as normalized Gaussians of width $0.12$~bohr) are solved in reciprocal
space at $20.8$ modes per bohr, the electric flux through every voxel
face and the charge in every voxel are integrated analytically, and the
fluxes summed over a basin equal its voxel charge to $10^{-12}$~electrons,
the construction check. The flux through a face is integrated over a
staircase of voxel faces with an error of first order in the spacing, so
every face flux is computed on both grids and the value used is
Richardson's extrapolation to zero spacing,
$\Phi_0=\Phi(h_2)+[\Phi(h_2)-\Phi(h_1)]/(h_1/h_2-1)$ with $h_1=0.13$ and
$h_2=0.087$~bohr, whose correction is the face's uncertainty. For
Table~\ref{tab:partitions} the same voxel fluxes are re-summed over six
geometric partitions of each cell, a power diagram whose planar faces lie
at the Voronoi midplanes ($t=0$), halfway to the bond points ($t=0.5$),
through the bond points on average ($t=1$) and beyond them ($t=1.5$),
its weights calibrated per anion species from the bond points of each
host (the anion radius from Shannon where one exists, otherwise from the
bond points), and the density basins eroded and dilated by one voxel. The
power diagrams, designed for the oxide and fluoride families, are
reported for those families and the eroded and dilated basins for all.

\emph{Corpus selection.} Non-magnetic means no magnetic ordering and a
total magnetization of at most $0.01\,\mu_B$, and the excluded open-shell
elements are V, Cr, Mn, Fe, Co, Ni and Cu. Of the $126\,315$ Materials
Project entries with a stored electron density, $10\,937$ satisfy the
criteria and $7179$ of those have one of the nine anions, $3880$ an
oxide or fluoride, $3495$ a chloride, chalcogenide or pnictide and
$196$ both. The selection kept the hosts whose elements the core
construction supports, without the heavy lone-pair cations Sn$^{2+}$,
Pb$^{2+}$, Bi$^{3+}$, Sb$^{3+}$ and Tl$^{+}$, with a valid formal-charge
assignment and at least one cation site bonded to a single anion,
$3933$ hosts in all, of which $3680$ yielded a complete face table
at both resolutions. Of the $253$ excluded, $117$ failed a census
criterion at one resolution or had an unreadable stored density, $104$
failed the hydrogen criterion, $23$ had no retrievable density, eight were
mixed-valent and one had no complete face table.

\emph{Formal charges and family membership.} Formal charges are assigned
per site by the bond-valence analyzer of pymatgen~\cite{ong2013}, a
maximum a posteriori assignment of integer oxidation states from
bond-valence sums with the parameters of O'Keeffe and
Brese~\cite{okeeffe1991} and priors from the ICSD. Where it returns no
charge-neutral assignment, the oxidation-state guesses of the composition
and then the species listed by the Materials Project are tried in that
order. A host is excluded when the analyzer assigns two oxidation states
to one element (mixed valence), when no assignment has every hydrogen at
$+1$, or when any assigned state is not a common or ICSD-recorded
oxidation state of its element, negative states being accepted only for
O, F, Cl, S, Se, Te, N, P, As, Sb and C. A family is a cation species and
an anion species, each an element in a formal oxidation state. In the
persulfides, polyphosphides, polyarsenides, polyselenides and
polytellurides the anions bonded to each other (S$^{-}$ of S$_2^{2-}$,
P$^{-}$, As$^{-}$, As$^{2-}$, Se$^{-}$, Te$^{-}$ and the like) are anion
species of their own. The hydrogen-free hosts carry 531 cation--anion
families, 453 with the nine anions and 78 with the homonuclear-bonded
anion species and Sb$^{3-}$. Of the 453, 156 have the six independent
distorted sites and the within-site bond-length range of at least
$0.05$~\AA{} on which the law is measured (no family of the other 78
does), and 106 of the 156 meet the transferability requirement of a
second coordination number. A cation site enters a family only if all of
its cation--anion faces are to the family's anion species, which excludes
790 sites in 151 hosts (715 sites in 132 of the hydrogen-free hosts), and
the 5\% threshold of Sec.~\ref{sec:theory} drops the long split-shell
contacts that would otherwise flatten a cation's coordination branch.

\emph{Family metrics and uncertainties.} The per-site $B$ is the slope of
Eq.~\eqref{eq:distortion} on one site, and the per-coordination $B$ the
slope pooled over the distorted sites at one coordination number, reported
where at least five independent sites populate it. The slope of the
per-site $B$ against coordination number is a least-squares fit with each
site's weight capped at the 90th percentile. $B^{*}$ carries a bootstrap
interval over hosts, and the host-to-host spread $\sigma_n$ is the
standard deviation over hosts of the mean bond length of the sites at
fixed coordination number, pooled over the coordination numbers with
three or more hosts and expressed in $\ln s$ through $B$. The
two-parameter fit is a least-squares fit of $\ln s$ against $d$ over
every face of the family, its residual split by a one-way random-effects
decomposition into host, site-within-host and face-within-site
components, all standard deviations in $\ln s$. The held-out ratio is the
rms on held-out hosts over the rms in sample for a two-parameter fit on
half of the hosts (30 random splits, families with at least ten hosts).
The reproducibility of the per-site $B$ at fixed coordination number is
assessed where at least eight sites with four or more bonded faces and a
bond-length range of at least $0.05$~\AA{} populate a family's
coordination number, as the robust spread of the per-site $B$
(interquartile range over $1.349$) against the median standard error of a
single site's slope.

\emph{The power law.} The power law $s=(d/R_1)^{-N}$ of Brown and
Shannon~\cite{brown1973}, which descends from the curve of Donnay and
Allmann~\cite{donnay1970} and was tabulated for cation--oxygen pairs by
Brown and Wu~\cite{brown1976}, is the case $u=\ln d$ with $B_u=1/N$, and $d_B$
is then the power mean of order $-N$ and Eq.~\eqref{eq:softmax} reads
$s=(z_i/n)(d_N/d)^{N}$. The two laws are one weight linearized in $d$ and
in $\ln d$. They agree to 1\% for $0.5<s<2$ when
$N=R_1/B$~\cite{zachariasen1978,whittaker_prm}, and within one site, where
$\ln d$ varies by $\Delta d/\langle d\rangle$, Eq.~\eqref{eq:distortion}
cannot tell them apart. The distortion softness $B$ is the exponent
$N=\langle d\rangle/B$.

\emph{The softness-weighted mean to second order.} Substituting the
sum-rule value $B=(R_0-\langle d\rangle)/\ln(z_i/n)$ into the variance
term of Eq.~\eqref{eq:cumulant}, the denominator of the exponent of
Eq.~\eqref{eq:pauling} becomes
$R_0-\langle d\rangle+\sigma^2\ln(z_i/n)/[2(R_0-\langle d\rangle)]$, again
free of $B$. The quantities $\ln(z_i/n)$ and $R_0-\langle d\rangle$ have
the same sign, so the correction is positive, $d_B$ lies below the
arithmetic mean and the short bonds count for more. On the corpus the
second-order term reproduces the exact gap $\langle d\rangle-d_B$ to
$0.005$~\AA{} on nine sites in ten and misses it by up to $0.23$~\AA{} on
the rest, the widely distorted sites of As$^{2+}$--S, Se$^{4+}$--O and
Te$^{4+}$--O, where the gap reaches $0.49$~\AA{}.

\emph{Cation--cation contacts.} With $L_i$ the net flux a cation passes to
other cations through its cation--cation faces and contacts
(Sec.~\ref{sec:results}), the families of the small, highly charged
cations with oxygen (B, Be, Mg, Al, Si, Ti, Ge, Zr, Nb, Hf, Ta) have no
site with $L_i/q_i$ above $2\%$. The large values belong to Ag$^{+}$,
Hg$^{2+}$, Pd$^{2+}$ and Au$^{3+}$ (in Na$_3$AgO$_2$ the silver basin
passes to its oxygens as much flux as it receives from the sodium through
Ag--Na contacts, $L_i/q_i=-1.0$), to the alkali and alkaline-earth
cations of the pnictides (K--As up to $0.37$, Ba--As $0.47$, Ca--P
$0.50$) and, more weakly, to the alkali cations of the oxides and
chalcogenides (Na--O, K--O and Cs--O up to $0.22$).

\begin{figure*}[t]
\includegraphics[width=\textwidth]{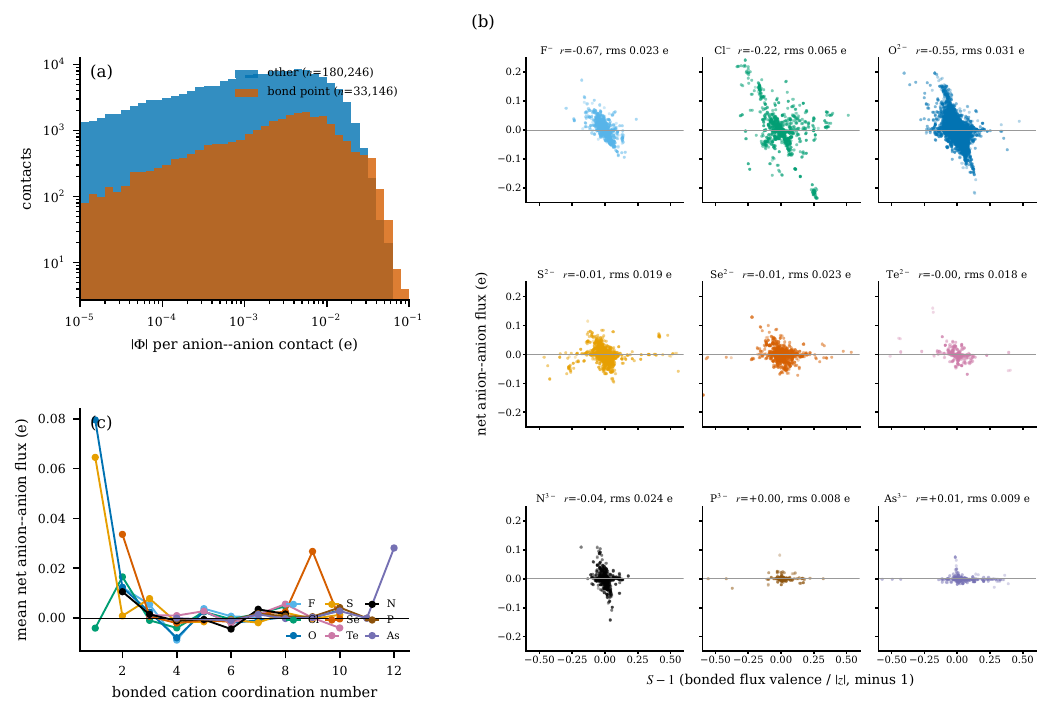}
\caption{Anion--anion contacts of the partition (hydrogen-free hosts). (a)~The magnitude of the flux through each contact, faces at a bond point against the rest (logarithmic axes). (b)~The net anion--anion flux received per anion against $S-1$, with $S$ the bonded flux-valence sum $\sum s$ over $|z|$, one panel per anion, with the Pearson $r$ and the root-mean-square net flux. (c)~The mean net flux by the anion's bonded cation coordination number, per anion.}
\label{fig:anion}
\end{figure*}

\emph{Anion--anion contacts.} Of the 213\,392 anion--anion contacts of
the hydrogen-free hosts 16\% are faces at a bond critical point, there
are about 11 per anion, their median separation is $3.5$~\AA{}, and no
contact carries $0.1$~e or more (Fig.~\ref{fig:anion}(a)). An over-bonded
oxide, fluoride or chloride anion ($S>1$) gives flux to its neighbors
and an under-bonded one receives it, the net flux being anticorrelated
with the anion's bonded flux-valence sum ($r=-0.55$ for oxygen, $-0.67$
for fluorine and $-0.22$ for chlorine), while for the chalcogenides and
pnictides it is not ($|r|\leq0.04$) (Fig.~\ref{fig:anion}(b)). Summed
over all of an anion's faces and contacts the flux closes to its charge
to $1.5\times10^{-12}$~e (Eq.~\eqref{eq:gauss}).

\emph{Flux tubes and their walls.} Each cation--anion face of a
sheet-bearing crystal has a compact part near the bond point and a sheet
part, where the cation basin thins to $10^{-9}$--$10^{-3}$~bohr between
two anion basins, and the sheet part carries at most a few percent of the
face's flux. An anion basin is the terminus of the flux tubes of its
cations. The flux enters through the bond-point faces, where $\mathbf E$
crosses the surface (Fig.~\ref{fig:basins}(c)), and the anion--anion
contacts, direct or through a cation sheet, are the walls of the tubes,
along which $\mathbf E$ runs nearly tangent to the surface. A contact
that intrudes on a cation--anion face narrows it, but the flux through a
face follows its bond length to the scatter of the law
(Sec.~\ref{sec:results}) whatever the contacts do to its area. Their
effect on a bond's valence is carried by the bond lengths they help set,
which is what the empirical model assumes in fitting $R_0$ and $B$ to
bond lengths alone~\cite{brown2016}.

\emph{The per-site softness.} The per-site $B$ of the independent
distorted sites of Mg--O and Al--O scatters about the family value by
$0.15$ and $0.20$~\AA{} (Fig.~\ref{fig:bsites}). Within a family it
depends on coordination number (Fig.~\ref{fig:bcn}, Table~\ref{tab:bcn}).
For the compact cations it rises with coordination (Ge--O from $0.45$ at
four-fold to $0.55$~\AA{} at six-fold, Mg--O $0.42$ to $0.52$, Zn--O
$0.45$ to $0.56$, Al--O $0.40$ to $0.44$ through a $0.35$~\AA{} minimum
at five-fold) and for the soft cations it falls (Na--O from $0.59$ at
four-fold to $0.36$~\AA{} at ten-fold, and K--O, Ba--O, Sr--O, Ca--O
and, more weakly, Nb--O). Of the 59 families with two or more
coordination numbers populated by five sites, the slope of the per-site
$B$ against coordination number is positive in 19 and negative in 40,
twelve of them with an interval that includes zero.

\emph{The transferability test, family by family.} Of the 106 families
that meet the requirement of Sec.~\ref{sec:transfer}, eight are with F,
eight with Cl, 35 with O, 21 with S, 13 with Se, three with Te, seven
with N, four with P and seven with As. The other 50, the phosphates,
sulfates and tetrahedral zinc sulfides among them, occupy one
coordination number, and no family of the homonuclear-bonded anion
species has six independent distorted sites. Of the ten families that
hold within the resolution of the test, in five $|D|\leq0.03$~\AA{}
outright (Si--O, Al--O, Na--F, Al--S and Li--S) and in the other five
only within a bootstrap half-width of $0.06$--$0.12$~\AA{}. For Si--O,
Al--O, Al--S and Ba--Cl both statements hold. The two softnesses agree
and the mean bond length of a regular site at fixed coordination number
is the same across hosts to $0.03$--$0.04$ in $\ln s$. For Na--F, Na--Cl,
Na--Te, Ba--Se, Li--S and Li--As the softnesses agree but
$\sigma_n=0.07$--$0.12$, so no single $R_0$ serves their hosts. Of the 96
families that fail, 31 satisfy the second statement and not the first.
Their $R_0$ is the same across hosts to within $0.05$ in $\ln s$
($\sigma_n=0.01$--$0.05$), but the within-site softness exceeds the
coordination-branch softness, and among them are B--O, Mg--O, Zr--O,
Ti--O, Nb--O, Ge--O and Y--O, the $d^0$ and lanthanide sulfides, and
Mg--S and Mg--Se. The other 65 satisfy neither, with
$\sigma_n=0.05$--$0.33$ in $\ln s$, so no single $R_0$ serves the family
even at one coordination number. The four families that fail the other
way, with $B^{*}$ above the within-site $B$, are B--S, K--Cl, Li--Cl and
Li--Se. The within-site scatter of $\ln s$ about the law follows the
order of the two statements, $0.04$--$0.08$ where both hold,
$0.10$--$0.13$ where only the softnesses agree, $0.03$--$0.18$ where only
$R_0$ is shared and $0.05$--$0.33$ where neither holds.

\emph{The partition comparison.} For Si--O and Al--O $B$ on the
density surface equals $B^{*}$ while on the Voronoi surface it is
$0.13$--$0.19$~\AA{} larger ($0.48$ and $0.59$~\AA{} against $B^{*}$ of
$0.35$ and $0.40$), decreasing monotonically as the planar faces move
toward and past the bond points, and Na--F behaves the same way ($0.57$ on
the Voronoi surface against $0.46$), so the passing families are
partition-specific (Table~\ref{tab:partitions}). Across the 40 failing
oxide and fluoride families, $B$ on the seven partitions spans a median
of $0.09$~\AA{} and exceeds $B^{*}$ on every partition in all 40 (K--O
$0.54$--$0.59$~\AA{} against $B^{*}=0.25$~\AA{}, Zr--O $0.51$--$0.56$
against $0.34$). With every anion the eroded and dilated basins move $B$
by a median of $0.03$~\AA{} and leave it above $B^{*}$ in 91 of the 96
failing families. Two further explanations of the within-site excess are
excluded. The corpus geometries are relaxed, but both softnesses are
measured on the same structures, so the relaxation enters $B$ and $B^{*}$
alike. The within-site $B$ might be carried by unrepresentative
distorted sites, but the per-site values scatter about the family value
(Fig.~\ref{fig:bsites}) by no more than $D$ for the failing oxide
families, and pooling the distorted sites of one coordination number at
a time gives, in 55 of the 59 families with two or more populated
coordination numbers, a within-site $B$ above $B^{*}$ at every
coordination number, by $0.01$--$0.84$~\AA{} at the closest. In three
(Al--O, Li--S and Na--F) $B^{*}$ lies inside the range of the
per-coordination values and in one (Si--O) above it.

\begin{figure*}[t]
\includegraphics[width=\textwidth]{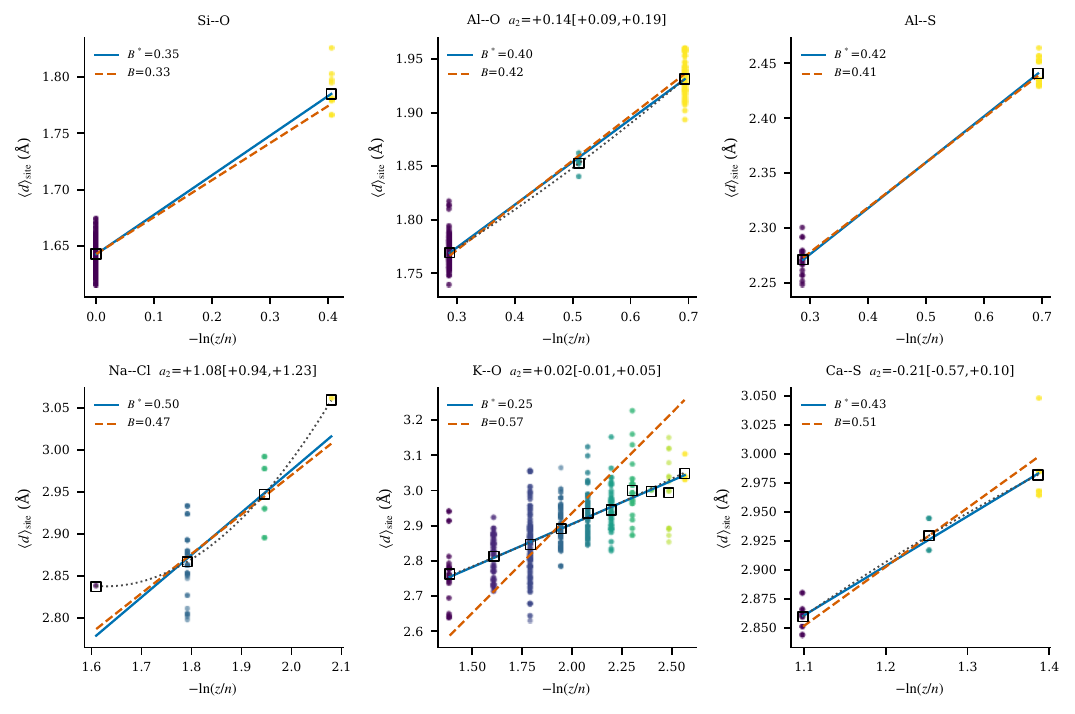}
\caption{The coordination branch for six families, as the mean bond length of every cation site with three or more bonded faces (colored by coordination number) against $-\ln(z/n)$, with the mean at each coordination number (squares). The solid line is the least-squares secant, of slope $B^{*}$, the dashed line has the slope of the family's within-site $B$, and the dotted curve is a quadratic fit with coefficient $a_2$, given where three or more coordination numbers support it.}
\label{fig:branch}
\end{figure*}

\emph{What the failure depends on.} Figure~\ref{fig:branch} shows the
coordination branch for six families and Fig.~\ref{fig:branchall} for
every family. Across the 106 families, $D$ grows only weakly with the
cation's Shannon radius, Spearman $\rho=+0.19$ with a 68\% bootstrap
interval of $[+0.09,+0.29]$, and not with its formal charge or field
strength $z/r$ ($|\rho|\leq0.06$). $B^{*}$ is not predicted by any cation
descriptor ($|\rho|\leq0.25$ against radius, charge, field strength and
electronegativity). The within-site $B$ has no dependence on
coordination number common to the families (Fig.~\ref{fig:bcn}).
$D$ is not ranked by $\sigma_n$ ($\rho=+0.26$), which itself falls with
the formal charge ($\rho=-0.59$) and the field strength ($-0.64$), so the
two failures of Sec.~\ref{sec:transfer} are distinct. The curvature of
the branch is family-specific in sign ($a_2$ in Fig.~\ref{fig:branch}),
so the failure is a within-site softness that does not follow the
branch, not a universal bending of the law. One mechanism is visible at the
level of single faces. The residual of $\ln s$ from the family law falls
with the valence the anion receives from its other cations, negatively in
90 of the 106 families (median $\rho=-0.27$), so that part of the host
offset of a bond is carried by the competition at its anion. The
coordination-specific shift of $R_0$ that a fit at one softness must then
carry, $(B-B^{*})\ln(n_2/n_1)$, is $0.06$~\AA{} for Zr--O between six- and
eight-fold coordination. It is the coordination-specific $R_0$ of the
empirical parameter sets~\cite{gagne2015} seen from the density, and the
larger softness, compound-specific parameters and cutoff of the fitted
sets~\cite{adams2001} are the trace of the host-to-host variation of
$R_0$.

\begin{figure*}[t]
\includegraphics[width=\textwidth]{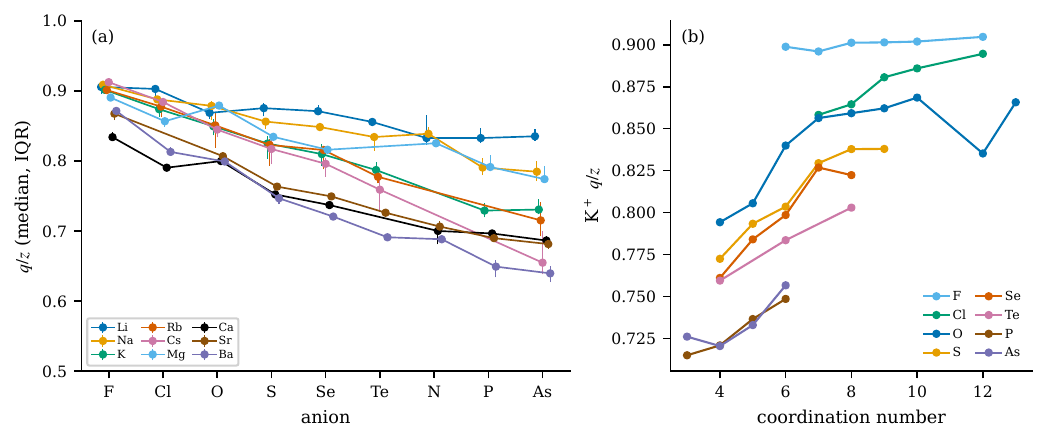}
\caption{The screening ratio $q/z$ across the anions. (a)~$q/z$ by anion for the alkali and alkaline-earth cations present with at least three anions, as the median over sites with the interquartile range (sites weighted within each coordination number). (b)~$q/z$ against coordination number for K$^{+}$ with each anion.}
\label{fig:anions_h3}
\end{figure*}

\begin{figure*}[t]
\includegraphics[width=\textwidth]{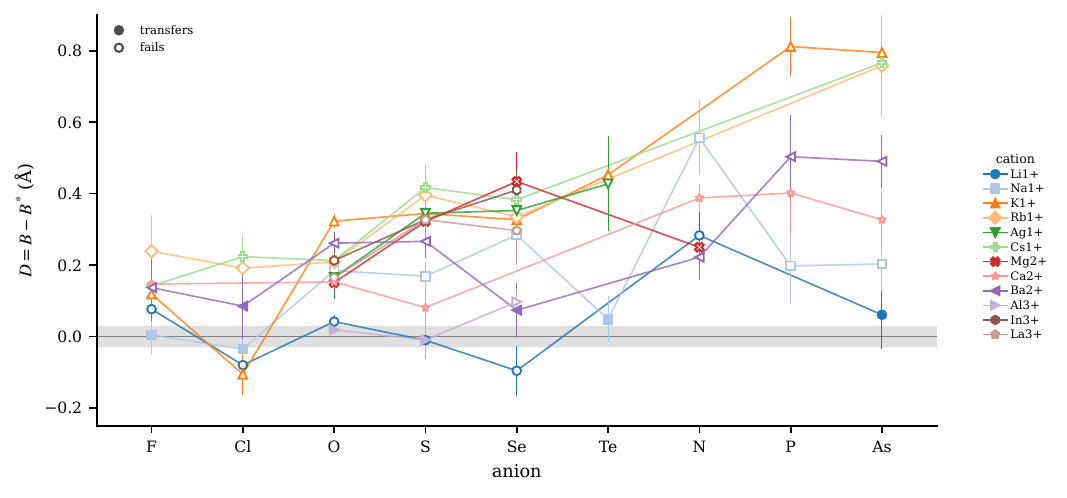}
\caption{The softness gap $D=B-B^{*}$ for the twelve cations with eligible families in at least three anions, one line per cation through its families, with 68\% intervals. Filled marks pass, open marks fail, and the band is $\pm0.03$~\AA{}.}
\label{fig:anions_h1}
\end{figure*}

\begin{figure*}[t]
\includegraphics[width=\textwidth]{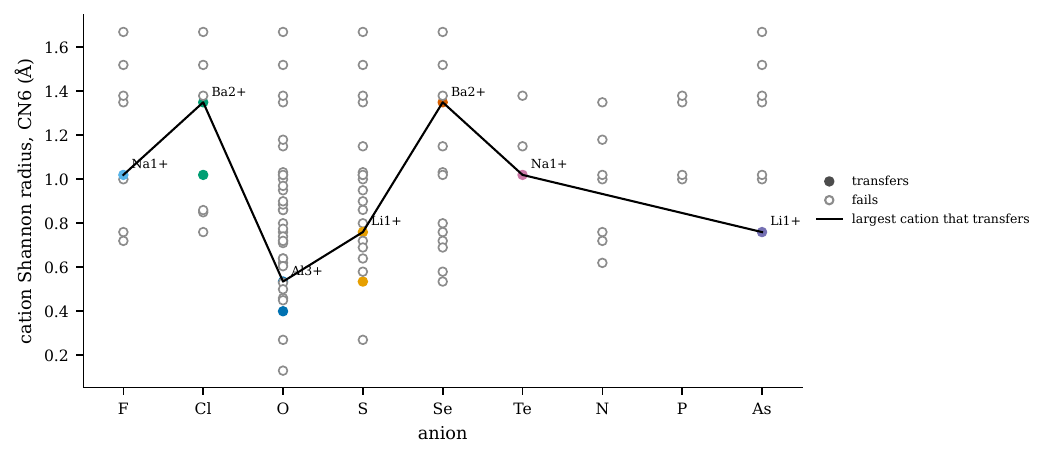}
\caption{For each anion, the cations of its eligible families at their Shannon six-fold radius. Filled marks pass, open marks fail, and the line connects the largest cation that passes with each anion.}
\label{fig:anions_h2}
\end{figure*}

\emph{Across the anions.} Of the twelve cations with eligible families
in at least three anions, eight fail with every anion, and four pass
with some anions and not with others (Fig.~\ref{fig:anions_h1},
Table~\ref{tab:anions}). Al$^{3+}$ passes with O and S and not with Se,
Ba$^{2+}$ with Cl and Se and not with F, O, S, N, P or As, Li$^{+}$ with
S and As and not with F, Cl, O, Se or N, and Na$^{+}$ with F, Cl and Te
and not with O, S, Se, N, P or As. The largest cation that passes ranges
from Al$^{3+}$ ($0.54$~\AA{}) with O to Ba$^{2+}$ ($1.35$~\AA{}) with Cl
and Se, and none passes with N or P (Fig.~\ref{fig:anions_h2}), as
expected of agreements within the resolution of the test. What does not
depend on the anion is the sign of the failure, the within-site softness
exceeding the coordination-branch softness with every anion
(Table~\ref{tab:transfer}). The screening ratio of the alkali and
alkaline-earth cations, by contrast, is ordered by the anion, F above Cl
for all eight cations present with both, Cl above O for six of eight, O
above S for eight of nine, S above Se for all nine and Se above Te for
all seven (Fig.~\ref{fig:anions_h3}).

\emph{Hydrogen sensitivity.} A host qualifies for the hydrogen-bearing
set only when none of its hydrogens is a maximum of the pseudo-valence
density, each instead ascending to the anion it is merged into. Added to
the corpus with each hydrogen merged into its bonded anion, the 244
hydrogen-bearing hosts held out of the hydrogen-free set
(Sec.~\ref{sec:methods}) enter 35 families with 1111 sites and move the
within-site $B$ by at most $0.033$~\AA{} (Be--O), and $B^{*}$ and $D$ by
at most $0.15$ and $0.17$~\AA{} in N--O, the nitrates, whose coordination
branch is not defined ($B^{*}<0$). The median move is zero, and the same
ten families pass with and without them (Table~\ref{tab:hsens}). The
hydrogen-host faces sit $0.029\pm0.004$ below the no-hydrogen family law
in $\ln s$, by the same amount for faces to hydrogen-merged anions as for
faces to plain ones ($-0.029$ each), so the shift is a host offset and
not a hydroxyl effect.

\begin{figure*}[t]
\includegraphics[width=\textwidth]{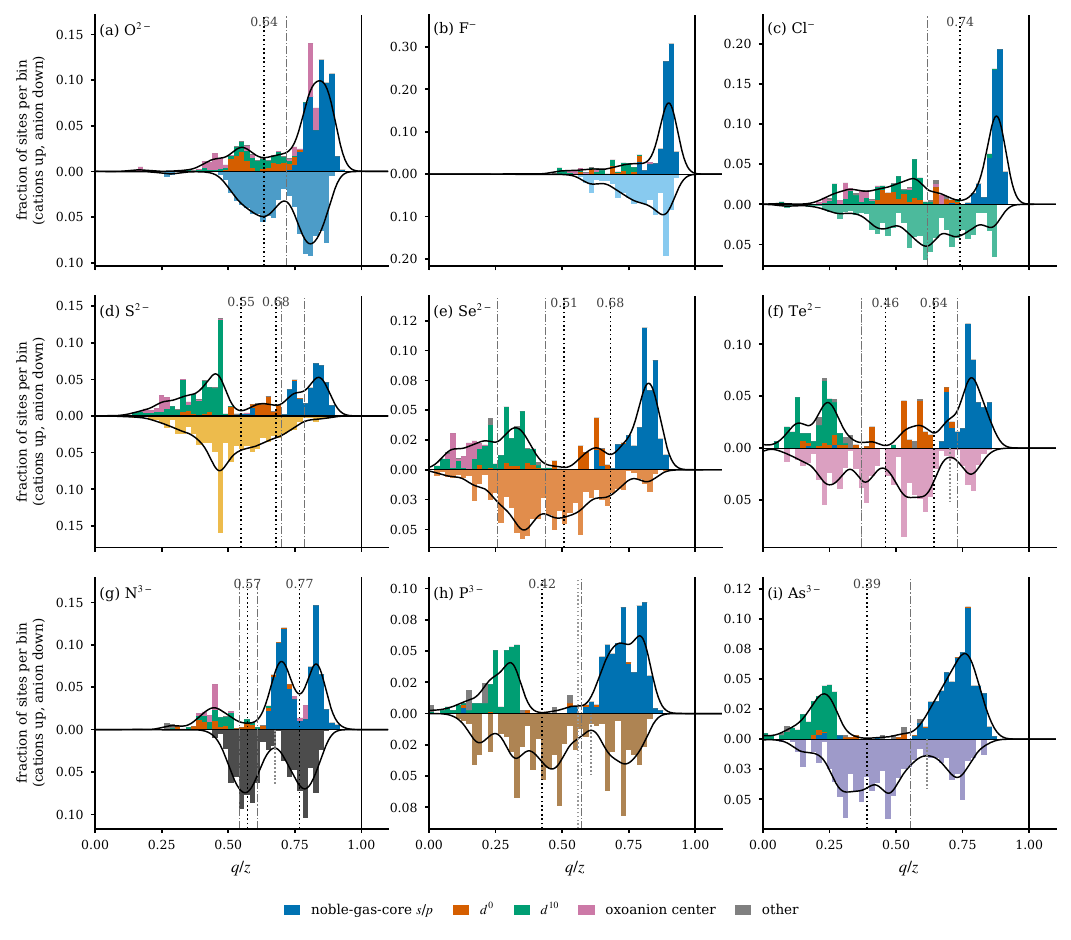}
\caption{The screening-ratio distributions behind the bands of Fig.~\ref{fig:charges}, one panel per anion. Above the axis, the cation sites of Fig.~\ref{fig:charges} stacked by cation class, as the fraction of those sites per bin of $0.02$, with the kernel density estimate (kernel $0.03$) on which the valleys are placed (notes above). Below the axis, mirrored, the anion's own $q/|z|$ (hydrogen-free hosts) as the fraction of the anion sites per bin, with its estimate at the same kernel. Dotted lines with a value are the accepted valleys of the cation distribution, gray dotted lines the candidate minima that fail the depth or the resampling criterion, gray dash-dot lines the valleys accepted only at a kernel of $0.015$ (notes above), and the short gray marks in the lower half the minima of the anion's own distribution. The solid line is $q/z=1$.}
\label{fig:qzhist}
\end{figure*}

\emph{The screening-ratio bands.} The valleys of Figs.~\ref{fig:charges}
and \ref{fig:qzhist} are placed on the cation distribution of each anion
by one rule. A Gaussian kernel density estimate is taken with a kernel of
$0.03$ in $q/z$, about $1.4$ times the median spread of one species with
one anion ($0.022$ with O and with N), so that a species counts as one
object and not as several. Its modes are the local maxima above a tenth
of the maximum. A valley is a local minimum between two modes, with at
least 5\% of the sites on each side, whose density is at most $0.7$ of
the lower of the two side maxima and which recurs, within $0.05$, in at
least 95\% of 200 resamples of the hosts (hosts, not sites, being the
independent units). The rule gives one valley with O ($0.64$), Cl
($0.74$), P ($0.42$) and As ($0.39$), none with F, and two with S
($0.55$, $0.68$), Se ($0.51$, $0.68$), Te ($0.46$, $0.64$) and N
($0.57$, $0.77$). The second valleys of S and N recur in 96\% and 97\% of
the resamples, close to the criterion. The anion's own distribution
(lower half of Fig.~\ref{fig:qzhist}) has its minima elsewhere ($0.71$
with O). Since the anion's ratio is a mean over the cations of its host
(Fig.~\ref{fig:anionrule}), its minima reflect which cations the corpus
combines in a host, and it is shown for comparison and not used. At a
kernel of $0.015$, below the species spread, further valleys pass the
same depth and resampling criteria (gray dash-dot lines in
Fig.~\ref{fig:qzhist}). They are reported, not used. With oxygen the one
at $0.72$ is the lower edge of the noble-gas-core band, while the valley
of the rule, $0.64$, separates the 1083 sites of Zn, La, Zr, Sc, Cd, In,
Ga, Y, Sn and Hf between $0.64$ and $0.72$ from the more screened Ti, Nb,
Ta and Ge. The valley of the rule is the deepest gap of the cation
distribution, not the edge of the $s$/$p$ band, which lies a tenth
higher. With every other anion that has a valley, at least 97\% of the
noble-gas-core sites lie above the lowest one and at least 91\% of the
$d^{10}$ sites below it (all of them with S, Se, P and As).

\emph{The Donnay--Allmann curve.} The curve with which Donnay and Allmann
introduced the term \cite{donnay1970} was offered as an approximation to
Pauling's logarithmic law~\cite{pauling1947}. For bonds shorter than the
polyhedron's own mean length $\bar L$, $s=(z/n)(\bar L/L)^{p}$, continued
for the longer bonds by a straight line that reaches zero at a tabulated
cutoff $L_{\max}$, with $p=\bar L/(L_{\max}-\bar L)$ fixed by matching the
slopes of the two segments at $\bar L$. In the variable $\ln d$ the power
segment is the dual form Eq.~\eqref{eq:softmax} of the power law,
$s=(z/n)(d_N/d)^{N}$, with the arithmetic mean in place of the power mean
$d_N$ and $p$ for $N$, and to first order in
$(d-\langle d\rangle)/\langle d\rangle$ it is the exponential law with
$B=\bar L/p=L_{\max}-\bar L$, so that their exponent is a softness read
from the cutoff rather than from the flux. On a regular polyhedron the
two coincide exactly. On a distorted one the arithmetic mean leaves a
closure error of second order in the spread, $p(p+1)\sigma^2/2\bar L^2$,
which the power mean removes as Eq.~\eqref{eq:cumulant} removes it for
the exponential law. What the sum rule adds is therefore the weighted
mean, with which closure is exact, and the elimination of the exponent
itself (Eq.~\eqref{eq:pauling}) in place of a slope condition at a cutoff.

\emph{Global and screened bond valence.} The data-driven global bond
valence of Ref.~\cite{whittaker_ammin} and the Yukawa-screening
derivation of the bond-valence rule of Ref.~\cite{whittaker_prm} both
normalize to the formal charge and neither uses an electron density.
Three of their elements are identities of the exponential form. The class
line $R_0 = \bar R + B\ln(z/n)$, derived in Ref.~\cite{whittaker_prm} in
the uniform-bond limit of a screened flux, is Eq.~\eqref{eq:r0site}, which
follows from the sum rule alone and which the electron density obeys on
every site. The coordination-branch softness $B^{*}$ of
Ref.~\cite{whittaker_prm} is the secant that Sec.~\ref{sec:transfer}
compares with the within-site softness, and the coordination dependence
of $R_0$ that both references report at a common softness is
$(B_{\rm family}-B)\ln(n/z)$, of which their pole law
$\beta = dB/dR_0 = 1/\ln(z/n)$ is the derivative. The Boltzmann weights,
shell centroid and entropy of Ref.~\cite{whittaker_prm} are
Eq.~\eqref{eq:softmax} and its moments, which a zero-temperature
electrostatic flux carries verbatim, so the thermodynamic language is an
interpretation and not an additional mechanism. Two points separate the
descriptions. Reference~\cite{whittaker_prm} obtains the softness from a
screening length. The present data measure $B$ but do not trace it to a
length, and the screening they contain is the in-basin deficit $z-q$ of
Sec.~\ref{sec:qz}, a fraction of $z$ fixed site by site, which neither
decays with distance nor enters the distance dependence of the valence.
And the small softnesses that joint fits of $(R_0,B)$ per structure
report for the alkali--oxygen pairs~\cite{whittaker_ammin,whittaker_prm}
are a matter of definition. A joint fit under $\sum s=z$ is degenerate
whenever the shells are nearly regular ($R_0$ and $B$ trade off along
$\bar R+B\ln(z/n)$), so the fitted $B$ reports the correlation of
$\bar R$ with $n$ across structures, whereas Eq.~\eqref{eq:distortion}
reads the same equation along $d$ within one site, where $R_0$ cancels.
What those descriptions contribute and the electron density does not,
the formation-potential threshold of Ref.~\cite{whittaker_ammin} and the
covalent-radius reading of $R_0$, are statements about energies and about
thousands of structures. The flux valence is a statement about one
electron density at a time, and the two are complementary.

\emph{The screening of the $d^0$ and $d^{10}$ cations.} For Ti$^{4+}$
with oxygen ($q/z=0.56$) the deficit is $1.8$ electrons that the formal
count assigns to oxygen and the density leaves in the titanium basin. It
is within $0.3$ electrons of the Bader charge of Ti in TiO$_2$ that the
literature reports, about $2.5$ electrons~\cite{walsh2017,walsh2018}, so
it is a property of the Bader partition of a $d^0$ oxide and not of the
present construction, and it is what the formally empty $3d$ shell does
in an oxide. The Ti $3d$ orbitals mix with O $2p$ in the filled valence
band, and the part of that density on the titanium side of the zero-flux
surface is counted to titanium. The $3d$ shell is compact, so the mixing
is stronger for Ti than for Zr with its more extended $4d$ shell ($0.56$
against $0.66$), and the filled, polarizable $d^{10}$ shells of Ge and Zn
are screened by a similar amount ($0.55$ and $0.69$). The
self-regulation of Sec.~\ref{sec:discussion} holds for every element the
corpus carries in two oxidation states bonded to oxygen (species with at
least 20 sites). Ti rises by a third of the step between $3+$ and $4+$,
Hg from $0.63$ to $1.04$ electrons between $1+$ and $2+$, As from $1.44$
to $2.31$, Se from $1.79$ to $2.44$ and Te from $1.82$ to $2.48$ between
their two states, against the tenth of an electron or less by which a
change of oxidation state moves the local charge of a transition-metal
ion in an insulator~\cite{raebiger2008}. The screening does not touch the
valence. Ti--O obeys the law and the Pauling form as well as the
closed-shell families (rms $0.038$ in $\ln s$, against $0.061$ for the
exponential law and $0.22$ for the equal split,
Table~\ref{tab:paulingcorpus}), because the screening is one number per
site that cancels from everything that depends on $d$
(Sec.~\ref{sec:qz}). The one caveat is the functional. PBE
over-delocalizes $d$ states, so the $3d$--$2p$ mixing, and with it the
deficit, is if anything overstated. A Hubbard term or a hybrid functional
would be expected to raise $q/z$ for the $d^0$ cations and to leave the
law untouched, and neither expectation is tested here.

\emph{The labels.} Assigning formal charges by bond-valence sums
(Sec.~\ref{sec:methods}) has three documented limits. The contacts
counted as bonds must first be chosen, the parameter sets are many, and
the sum sees bond lengths only, so that it errs where the oxidation state
is carried by the shape of the coordination polyhedron rather than by its
bond lengths, and some of the oxidation states recorded in the structural
databases are themselves wrong~\cite{jablonka2021}. The flux answers the
first two limits and not the third. The coordination number is the face
count of the partition, no parameter table enters, and the law is a
property of the electron density, but $z$ is taken as input and not
inferred. What the label touches is bounded. The law, $B$, the sum rule
on $\Phi$ and the closure at the anion contain no $z$. Only $s$ and $q/z$
do, and a site whose label is wrong by the ratio $z/z'$ moves by that
factor out of its species' band, into the band of a neighboring
oxidation state or into none, so the bands of Fig.~\ref{fig:charges} are
a check on the labels that a bond-length fit cannot make. An audit of the
3680 hosts against a geometry-based assignment~\cite{jablonka2021} has
not been done.

\emph{The flux construction and the charge convention.}
Equation~\eqref{eq:flux} is Brown's identification of the valence with
the electrostatic flux linking cation and anion~\cite{preiser1999},
tested there with point ions and a capacitance construction and carried
out here on the electron density itself, through the faces the density
defines, without point charges or a chosen partition of the Coulomb
field. The one convention in the construction, counting the basin charge
as the formal charge, has a physical reading. In an insulator $z$ is the
integer number of polarization quanta, or Wannier centers, carried along
when the sublattice is translated by a lattice vector~\cite{jiang2012},
while $q$ is the static flux through the basin surface, so that $z$ and
$q$ are different theorems about the same crystal and $q/z$ is their
ratio.

\emph{The bond-point and bond-order routes.} The bond point is where the
valence crosses, the circumstance under which $\rho_b$ tracks it, so that
whether a bond path ``is'' a chemical bond~\cite{bader2009} does not
arise. The flux decides, and it decides in favor of the cation--anion
faces over the anion--anion faces, genuine bond points that carry little
flux. The most developed form of the bond-point route, the bond-order
reading of Levi, Aurbach and Gatti~\cite{levi2020}, ties the valence to a
quantum-chemical bond order at $\rho_b$ and carries the exponential form
to metal--metal and cluster compounds beyond the cation--anion bonds
measured here, but imposes the sum rule on a point value rather than
deriving it. Theirs widens the range of bonds the exponential form
covers, and this one supplies the surface on which the sum rule is a
theorem.

\emph{The loop rule.} The path, or loop, rule is not addressed by the
data here, and on the present construction it acquires a candidate in the
potential differences between basins.

\emph{The equal-split record.} Across some five thousand oxides the equal
split is met by about a fifth of the oxygen atoms, by about three fifths
of those bonded to Al or Si, and nearly exactly on the few hundred
structures whose polyhedra are undistorted~\cite{george2020}.

\subsection{Supporting figures and tables}

\ifiucrj\else Figure~\ref{fig:flow} shows how the corpus was defined.
\fi Tables~\ref{tab:families} and~\ref{tab:paulingcorpus} list every
family on which the law is measured. The other tables and the figures
cover the families of the transferability test of
Sec.~\ref{sec:transfer}, all grouped by anion.

\newcommand{\captionfamilies}{Bond-valence parameters per cation--anion family (hydrogen-free hosts), for every family with six or more crystallographically independent distorted sites, the families of the transferability test marked with a dagger, grouped by anion and ordered within an anion by the atomic number of the cation. The columns give the hosts, the bonded faces and their bond-length range, the within-site $B$ with its 68\% bootstrap interval, $R_0$ at that $B$ with the face-to-face spread, the rms deviation of the faces from the law in $s$, $\Delta R_0$, the shift of $R_0$ when Pauling's equal split $z/n$ replaces the flux valence on every face, and the rms deviation of the faces from that equal split. The oxidation state is written where an element occurs in more than one.}
\newcommand{\captiontransfer}{Transferability across hosts for every cation--anion family that meets the data requirement (hydrogen-free hosts), grouped by anion and ordered within an anion by $D$. The columns give the hosts, the cation sites and the distorted sites (crystallographically independent ones in parentheses), the within-site $B$ with its 68\% bootstrap interval, the coordination-branch $B^{*}$ with its bootstrap interval over hosts, $D=B-B^{*}$, the host-to-host spread $\sigma_n$ of the regular bond length in units of $\ln s$, and the within-site scatter of $\ln s$ about the family law. A family passes when $|D|\leq\max(0.03~\text{\AA},\text{half-width})$. The oxidation state is written where an element occurs in more than one.}
\newcommand{\captionbcn}{The per-site softness by coordination number, one row for each family of Table~\ref{tab:transfer} and each coordination number populated by at least five independent distorted sites, giving the number of such sites, the median of their per-site $B$ with the interquartile range in brackets, the flux-pooled $B$ at that coordination number with its 68\% bootstrap interval and, in the family's first row, $B^{*}$. A dash marks a coordination number with too few sites for the pooled bootstrap.}
\newcommand{\captionpaulingcorpus}{Pauling's rule with a distance exponent, Eq.~\eqref{eq:pauling}, for every family of Table~\ref{tab:families} (the families of the transferability test marked with a dagger, grouped by anion). The columns give the values of $z/n$ the family populates, the faces and sites with $z\neq n$, the rms of $\ln s$ of the measured flux valence about the form with the softness-weighted mean $d_B$, about the form with the arithmetic mean $\langle d\rangle$, about the exponential law at the family $R_0$ and $B$ and about the equal split $z/n$, and the closure of the arithmetic form, $|\sum_j s_j/z-1|$, mean and maximum over the family's sites (\%). $R_0$ and $B$ are those of Table~\ref{tab:families}, and the faces of sites with $z=n$ are excluded.}
\newcommand{\captionhsens}{Hydrogen sensitivity of the family softness test, for each family of Table~\ref{tab:transfer}, giving the hosts without ($-$H, the hydrogen-free set) and with ($+$H) the 244 hydrogen-bearing hosts, each hydrogen merged into its bonded anion, then $B$, $B^{*}$ and $D=B-B^{*}$ on each set and whether the family passes on each.}
\newcommand{\captiontransfercomp}{Transferability across hosts, sites and coordination numbers for every family of the test (hydrogen-free hosts, bonded faces), grouped by anion and ordered within an anion by the host spread of $R_0$. The columns give the hosts and sites, the within-site $B$, the two-parameter fit $B$ and $B^{*}$, the rms residual of the two-parameter fit in $\ln s$ and its host, site-within-host and face-within-site components, the host and site components times $B$ (the host-to-host and site-to-site spreads of $R_0$, \AA{}), and the held-out ratio of the two-parameter fit.}

\ifiucrj\else
\begin{figure*}[t]
\includegraphics[width=\textwidth]{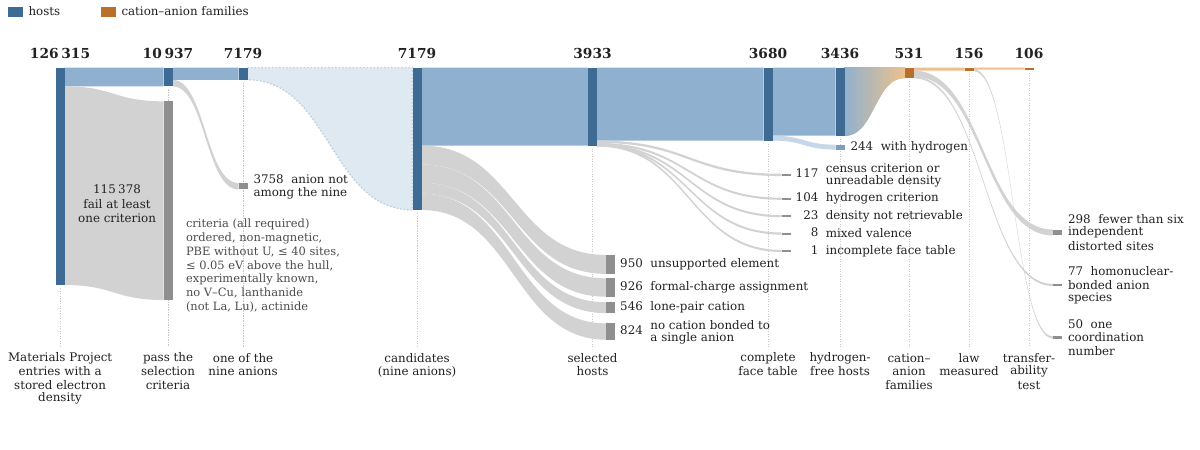}
\caption{\captionflow}
\label{fig:flow}
\end{figure*}
\fi

\clearpage
\onecolumngrid
\addtolength{\textheight}{-18pt}
\makeatletter\@maxdepth=0pt\maxdepth=0pt\makeatother
{\footnotesize\setlength{\tabcolsep}{1.9pt}%
\begin{longtable}{lrrcccccc}
\caption{\captionfamilies}\label{tab:families}\\
\toprule
pair & hosts & faces & $d$ range (\AA) & $B$ ($R_0$-free, \AA) & $R_0$ (flux, \AA) & scatter in $s$ (\%) & $\Delta R_0$ (\AA) & scatter, equal split (\%) \\
\midrule
\endfirsthead
\multicolumn{9}{c}{\tablename~\thetable{} (continued)}\\
\toprule
pair & hosts & faces & $d$ range (\AA) & $B$ ($R_0$-free, \AA) & $R_0$ (flux, \AA) & scatter in $s$ (\%) & $\Delta R_0$ (\AA) & scatter, equal split (\%) \\
\midrule
\endhead
\bottomrule
\endlastfoot
\multicolumn{9}{l}{\emph{F$^{-}$} (12 families)} \\
Li--F$^{\dagger}$ & 46 & 734 & 1.85--2.44 & 0.44 $\pm$ 0.03 & 1.295 $\pm$ 0.032 & 7.4 & +0.014 & 19 \\
Na--F$^{\dagger}$ & 56 & 1070 & 2.22--3.11 & 0.46 $\pm$ 0.02 & 1.522 $\pm$ 0.047 & 10 & +0.019 & 24 \\
Mg--F & 21 & 330 & 1.94--2.31 & 0.44 $\pm$ 0.02 & 1.543 $\pm$ 0.019 & 4.3 & +0.002 & 9.6 \\
Al--F & 45 & 834 & 1.76--1.93 & 0.38 $\pm$ 0.01 & 1.571 $\pm$ 0.008 & 2.2 & +0.001 & 7.9 \\
K--F$^{\dagger}$ & 86 & 2343 & 2.52--3.72 & 0.51 $\pm$ 0.01 & 1.724 $\pm$ 0.078 & 15 & +0.035 & 33 \\
Ca--F$^{\dagger}$ & 22 & 388 & 2.23--2.84 & 0.51 $\pm$ 0.03 & 1.699 $\pm$ 0.030 & 5.8 & +0.009 & 18 \\
Zn--F & 23 & 380 & 1.95--3.07 & 0.469 $\pm$ 0.002 & 1.556 $\pm$ 0.019 & 4.0 & +0.022 & 36 \\
Rb--F$^{\dagger}$ & 22 & 646 & 2.73--3.65 & 0.49 $\pm$ 0.02 & 1.890 $\pm$ 0.101 & 21 & +0.027 & 25 \\
Zr--F$^{\dagger}$ & 11 & 216 & 1.98--2.43 & 0.48 $\pm$ 0.01 & 1.823 $\pm$ 0.013 & 2.8 & +0.006 & 15 \\
Cs--F$^{\dagger}$ & 22 & 747 & 2.97--3.69 & 0.46 $\pm$ 0.04 & 2.120 $\pm$ 0.072 & 16 & +0.026 & 30 \\
Ba--F$^{\dagger}$ & 17 & 676 & 2.60--3.56 & 0.55 $\pm$ 0.02 & 1.911 $\pm$ 0.051 & 9.3 & +0.038 & 37 \\
Au--F & 6 & 128 & 1.94--3.28 & 0.61 $\pm$ 0.02 & 1.707 $\pm$ 0.099 & 16 & +0.221 & 98 \\
\midrule
\multicolumn{9}{l}{\emph{Cl$^{-}$} (16 families)} \\
Li--Cl$^{\dagger}$ & 15 & 278 & 2.33--2.89 & 0.51 $\pm$ 0.02 & 1.712 $\pm$ 0.042 & 8.3 & +0.006 & 16 \\
Na--Cl$^{\dagger}$ & 17 & 408 & 2.73--3.54 & 0.47 $\pm$ 0.01 & 2.034 $\pm$ 0.047 & 9.9 & +0.022 & 31 \\
Al--Cl & 15 & 212 & 2.10--2.34 & 0.44 $\pm$ 0.02 & 2.037 $\pm$ 0.009 & 2.1 & +0.001 & 7.3 \\
K--Cl$^{\dagger}$ & 27 & 1160 & 3.04--4.02 & 0.64 $\pm$ 0.02 & 2.008 $\pm$ 0.080 & 13 & +0.033 & 29 \\
Sc--Cl & 9 & 162 & 2.41--2.63 & 0.57 $\pm$ 0.03 & 2.114 $\pm$ 0.009 & 1.6 & +0.003 & 10 \\
Se$^{4+}$--Cl & 7 & 144 & 2.17--3.12 & 0.88 $\pm$ 0.02 & 2.110 $\pm$ 0.035 & 3.9 & +0.055 & 36 \\
Rb--Cl$^{\dagger}$ & 22 & 742 & 3.25--4.45 & 0.72 $\pm$ 0.03 & 1.914 $\pm$ 0.078 & 11 & +0.021 & 21 \\
Nb--Cl & 9 & 168 & 2.26--2.67 & 0.71 $\pm$ 0.01 & 2.261 $\pm$ 0.010 & 1.4 & +0.009 & 16 \\
Mo--Cl & 3 & 180 & 2.39--2.61 & 1.83 $\pm$ 1.00 & 0.762 $\pm$ 0.401 & 22 & +0.047 & 22 \\
Pd--Cl$^{\dagger}$ & 15 & 172 & 2.30--3.76 & 0.61 $\pm$ 0.07 & 1.908 $\pm$ 0.108 & 18 & +0.067 & 58 \\
Ag--Cl & 10 & 116 & 2.40--3.34 & 0.54 $\pm$ 0.03 & 1.851 $\pm$ 0.061 & 11 & +0.013 & 32 \\
Te$^{4+}$--Cl & 8 & 104 & 2.33--3.46 & 1.02 $\pm$ 0.05 & 2.193 $\pm$ 0.058 & 5.6 & +0.060 & 34 \\
Cs--Cl$^{\dagger}$ & 49 & 1754 & 3.43--4.60 & 0.67 $\pm$ 0.04 & 2.189 $\pm$ 0.099 & 15 & +0.025 & 25 \\
Ba--Cl$^{\dagger}$ & 9 & 339 & 3.02--3.94 & 0.66 $\pm$ 0.04 & 2.305 $\pm$ 0.054 & 8.3 & +0.025 & 27 \\
Au--Cl$^{\dagger}$ & 13 & 206 & 2.28--4.09 & 0.75 $\pm$ 0.04 & 2.095 $\pm$ 0.223 & 30 & +0.213 & 94 \\
Hg$^{2+}$--Cl & 8 & 148 & 2.32--3.80 & 0.89 $\pm$ 0.04 & 1.846 $\pm$ 0.078 & 8.8 & +0.101 & 49 \\
\midrule
\multicolumn{9}{l}{\emph{O$^{2-}$} (44 families)} \\
Li--O$^{\dagger}$ & 214 & 4618 & 1.89--2.62 & 0.47 $\pm$ 0.01 & 1.349 $\pm$ 0.044 & 9.4 & +0.012 & 17 \\
Be--O$^{\dagger}$ & 34 & 498 & 1.50--1.75 & 0.49 $\pm$ 0.02 & 1.311 $\pm$ 0.022 & 4.4 & +0.001 & 7.4 \\
B--O$^{\dagger}$ & 183 & 2782 & 1.32--1.60 & 0.38 $\pm$ 0.01 & 1.383 $\pm$ 0.010 & 2.6 & +0.001 & 6.9 \\
N--O$^{\dagger}$ & 16 & 267 & 1.14--3.15 & 0.75 $\pm$ 0.05 & 1.627 $\pm$ 0.050 & 6.7 & +0.064 & 48 \\
Na--O$^{\dagger}$ & 276 & 6834 & 2.20--3.50 & 0.49 $\pm$ 0.02 & 1.601 $\pm$ 0.078 & 16 & +0.038 & 30 \\
Mg--O$^{\dagger}$ & 95 & 1734 & 1.92--2.51 & 0.52 $\pm$ 0.02 & 1.563 $\pm$ 0.033 & 6.5 & +0.005 & 13 \\
Al--O$^{\dagger}$ & 123 & 2350 & 1.68--2.27 & 0.42 $\pm$ 0.01 & 1.645 $\pm$ 0.019 & 4.5 & +0.003 & 12 \\
Si--O$^{\dagger}$ & 260 & 4866 & 1.58--1.85 & 0.328 $\pm$ 0.004 & 1.643 $\pm$ 0.013 & 4.1 & +0.001 & 7.0 \\
P$^{5+}$--O & 251 & 4552 & 1.45--1.71 & 0.354 $\pm$ 0.002 & 1.631 $\pm$ 0.009 & 2.7 & +0.002 & 11 \\
S--O & 52 & 764 & 1.41--1.87 & 0.34 $\pm$ 0.02 & 1.629 $\pm$ 0.010 & 2.9 & +0.004 & 16 \\
K--O$^{\dagger}$ & 209 & 6189 & 2.53--3.93 & 0.57 $\pm$ 0.01 & 1.748 $\pm$ 0.113 & 20 & +0.060 & 32 \\
Ca--O$^{\dagger}$ & 157 & 3816 & 2.20--3.08 & 0.54 $\pm$ 0.01 & 1.776 $\pm$ 0.054 & 10 & +0.025 & 26 \\
Sc--O & 39 & 654 & 2.01--2.60 & 0.52 $\pm$ 0.05 & 1.762 $\pm$ 0.026 & 4.9 & +0.005 & 14 \\
Ti--O$^{\dagger}$ & 117 & 2360 & 1.68--3.14 & 0.59 $\pm$ 0.01 & 1.751 $\pm$ 0.036 & 6.1 & +0.012 & 22 \\
Zn--O$^{\dagger}$ & 110 & 1898 & 1.87--3.37 & 0.52 $\pm$ 0.02 & 1.614 $\pm$ 0.052 & 10 & +0.012 & 25 \\
Ga--O$^{\dagger}$ & 63 & 1007 & 1.78--2.95 & 0.52 $\pm$ 0.03 & 1.686 $\pm$ 0.036 & 6.9 & +0.009 & 18 \\
Ge--O$^{\dagger}$ & 116 & 1744 & 1.71--3.13 & 0.51 $\pm$ 0.03 & 1.760 $\pm$ 0.034 & 6.6 & +0.008 & 17 \\
As$^{5+}$--O$^{\dagger}$ & 53 & 816 & 1.66--2.98 & 0.52 $\pm$ 0.02 & 1.826 $\pm$ 0.027 & 5.1 & +0.012 & 20 \\
As$^{3+}$--O & 6 & 122 & 1.72--3.53 & 0.63 $\pm$ 0.03 & 1.792 $\pm$ 0.061 & 9.8 & +0.140 & 83 \\
Se$^{6+}$--O & 16 & 214 & 1.64--1.87 & 0.48 $\pm$ 0.01 & 1.869 $\pm$ 0.012 & 2.5 & +0.006 & 8.2 \\
Se$^{4+}$--O$^{\dagger}$ & 25 & 714 & 1.65--3.65 & 0.63 $\pm$ 0.01 & 1.851 $\pm$ 0.074 & 12 & +0.262 & 109 \\
Rb--O$^{\dagger}$ & 65 & 2221 & 2.69--3.79 & 0.53 $\pm$ 0.02 & 1.989 $\pm$ 0.112 & 21 & +0.067 & 38 \\
Sr--O$^{\dagger}$ & 175 & 4948 & 2.37--3.97 & 0.53 $\pm$ 0.01 & 1.918 $\pm$ 0.058 & 11 & +0.035 & 31 \\
Y--O$^{\dagger}$ & 60 & 1420 & 2.19--3.10 & 0.54 $\pm$ 0.02 & 1.896 $\pm$ 0.040 & 7.5 & +0.010 & 18 \\
Zr--O$^{\dagger}$ & 52 & 982 & 1.93--2.50 & 0.54 $\pm$ 0.01 & 1.893 $\pm$ 0.031 & 5.8 & +0.005 & 13 \\
Nb--O$^{\dagger}$ & 112 & 2345 & 1.78--2.81 & 0.62 $\pm$ 0.01 & 1.908 $\pm$ 0.031 & 5.0 & +0.015 & 23 \\
Pd--O$^{\dagger}$ & 22 & 264 & 2.02--3.77 & 0.55 $\pm$ 0.02 & 1.700 $\pm$ 0.090 & 16 & +0.051 & 75 \\
Ag--O$^{\dagger}$ & 32 & 788 & 2.07--3.49 & 0.50 $\pm$ 0.02 & 1.714 $\pm$ 0.102 & 20 & +0.048 & 50 \\
Cd--O$^{\dagger}$ & 40 & 805 & 2.17--3.37 & 0.53 $\pm$ 0.03 & 1.790 $\pm$ 0.068 & 13 & +0.021 & 29 \\
In--O$^{\dagger}$ & 47 & 746 & 2.05--3.05 & 0.46 $\pm$ 0.07 & 1.891 $\pm$ 0.047 & 10 & +0.005 & 17 \\
Sn--O & 37 & 502 & 1.96--2.17 & 0.52 $\pm$ 0.07 & 1.877 $\pm$ 0.034 & 6.5 & 0.000 & 5.7 \\
Sb--O & 49 & 774 & 1.92--2.17 & 0.62 $\pm$ 0.04 & 1.911 $\pm$ 0.023 & 3.6 & 0.000 & 7.5 \\
Te$^{6+}$--O & 29 & 368 & 1.85--2.11 & 0.50 $\pm$ 0.01 & 1.970 $\pm$ 0.020 & 4.1 & +0.002 & 11 \\
Te$^{4+}$--O$^{\dagger}$ & 24 & 720 & 1.84--4.03 & 0.72 $\pm$ 0.01 & 1.968 $\pm$ 0.084 & 12 & +0.219 & 90 \\
Cs--O$^{\dagger}$ & 65 & 2158 & 2.80--4.14 & 0.54 $\pm$ 0.02 & 2.125 $\pm$ 0.123 & 23 & +0.075 & 42 \\
Ba--O$^{\dagger}$ & 205 & 5951 & 2.54--4.07 & 0.58 $\pm$ 0.02 & 1.984 $\pm$ 0.084 & 14 & +0.041 & 32 \\
La--O$^{\dagger}$ & 74 & 1970 & 2.25--3.59 & 0.55 $\pm$ 0.02 & 1.990 $\pm$ 0.058 & 10 & +0.030 & 32 \\
Lu--O$^{\dagger}$ & 17 & 378 & 2.14--2.74 & 0.60 $\pm$ 0.05 & 1.798 $\pm$ 0.043 & 7.1 & +0.008 & 15 \\
Hf--O$^{\dagger}$ & 11 & 228 & 1.97--2.43 & 0.59 $\pm$ 0.02 & 1.831 $\pm$ 0.035 & 5.9 & +0.006 & 14 \\
Ta$^{5+}$--O$^{\dagger}$ & 79 & 1404 & 1.79--2.52 & 0.60 $\pm$ 0.02 & 1.894 $\pm$ 0.027 & 4.6 & +0.006 & 15 \\
Hg$^{+}$--O & 4 & 140 & 2.15--3.61 & 0.63 $\pm$ 0.04 & 1.702 $\pm$ 0.086 & 14 & +0.136 & 61 \\
Hg$^{2+}$--O$^{\dagger}$ & 15 & 208 & 2.00--3.43 & 0.66 $\pm$ 0.02 & 1.722 $\pm$ 0.111 & 17 & +0.103 & 53 \\
Tl--O$^{\dagger}$ & 11 & 236 & 2.11--2.78 & 0.76 $\pm$ 0.04 & 1.852 $\pm$ 0.106 & 14 & 0.000 & 15 \\
Pb--O$^{\dagger}$ & 25 & 324 & 2.06--3.84 & 0.57 $\pm$ 0.03 & 2.032 $\pm$ 0.056 & 9.9 & +0.007 & 26 \\
\midrule
\multicolumn{9}{l}{\emph{S$^{2-}$} (28 families)} \\
Li--S$^{\dagger}$ & 37 & 746 & 2.40--3.10 & 0.42 $\pm$ 0.01 & 1.916 $\pm$ 0.054 & 13 & +0.009 & 20 \\
B--S$^{\dagger}$ & 18 & 260 & 1.78--1.97 & 0.26 $\pm$ 0.01 & 1.830 $\pm$ 0.019 & 7.4 & 0.000 & 7.3 \\
Na--S$^{\dagger}$ & 50 & 1502 & 2.71--3.73 & 0.52 $\pm$ 0.02 & 2.071 $\pm$ 0.074 & 14 & +0.032 & 28 \\
Mg--S$^{\dagger}$ & 11 & 228 & 2.46--2.75 & 0.67 $\pm$ 0.09 & 1.902 $\pm$ 0.050 & 7.5 & +0.001 & 5.5 \\
Al--S$^{\dagger}$ & 20 & 372 & 2.17--2.53 & 0.41 $\pm$ 0.04 & 2.156 $\pm$ 0.017 & 4.2 & +0.002 & 9.6 \\
P$^{5+}$--S & 24 & 280 & 1.91--2.16 & 0.151 $\pm$ 0.004 & 2.092 $\pm$ 0.016 & 10 & +0.009 & 34 \\
P$^{3+}$--S & 2 & 82 & 1.93--4.09 & 0.65 $\pm$ 0.04 & 2.116 $\pm$ 0.297 & 46 & +0.287 & 118 \\
K--S$^{\dagger}$ & 57 & 2004 & 3.09--4.23 & 0.71 $\pm$ 0.02 & 1.997 $\pm$ 0.096 & 14 & +0.054 & 28 \\
Ca--S$^{\dagger}$ & 10 & 318 & 2.78--3.42 & 0.51 $\pm$ 0.02 & 2.296 $\pm$ 0.030 & 5.8 & +0.011 & 23 \\
Zn--S & 62 & 2396 & 2.30--2.59 & 1.10 $\pm$ 0.62 & 1.601 $\pm$ 0.032 & 2.9 & 0.000 & 1.9 \\
Ga--S & 27 & 492 & 2.22--3.61 & 0.47 $\pm$ 0.04 & 2.170 $\pm$ 0.020 & 4.3 & +0.008 & 25 \\
Ge--S & 46 & 680 & 2.14--3.22 & 0.50 $\pm$ 0.01 & 2.239 $\pm$ 0.017 & 3.4 & +0.006 & 16 \\
As$^{3+}$--S$^{\dagger}$ & 15 & 248 & 2.16--4.27 & 0.63 $\pm$ 0.02 & 2.232 $\pm$ 0.107 & 17 & +0.183 & 92 \\
As$^{2+}$--S$^{\dagger}$ & 5 & 216 & 2.21--4.19 & 0.76 $\pm$ 0.01 & 2.158 $\pm$ 0.200 & 26 & +0.335 & 110 \\
Rb--S$^{\dagger}$ & 40 & 1480 & 3.19--4.72 & 0.73 $\pm$ 0.03 & 2.088 $\pm$ 0.120 & 17 & +0.058 & 28 \\
Sr--S & 19 & 566 & 2.89--3.59 & 0.59 $\pm$ 0.03 & 2.320 $\pm$ 0.055 & 9.3 & +0.020 & 23 \\
Y--S$^{\dagger}$ & 23 & 816 & 2.65--3.35 & 0.68 $\pm$ 0.06 & 2.248 $\pm$ 0.048 & 7.0 & +0.004 & 11 \\
Zr--S & 11 & 218 & 2.47--2.76 & 0.59 $\pm$ 0.04 & 2.317 $\pm$ 0.028 & 4.8 & +0.002 & 8.4 \\
Ag--S$^{\dagger}$ & 76 & 1422 & 2.39--3.88 & 0.60 $\pm$ 0.03 & 1.856 $\pm$ 0.136 & 23 & +0.028 & 49 \\
Cd--S$^{\dagger}$ & 24 & 304 & 2.48--4.10 & 0.65 $\pm$ 0.07 & 2.104 $\pm$ 0.056 & 8.6 & +0.037 & 39 \\
In--S$^{\dagger}$ & 37 & 846 & 2.42--3.37 & 0.69 $\pm$ 0.01 & 2.231 $\pm$ 0.061 & 8.9 & +0.012 & 16 \\
Sn--S$^{\dagger}$ & 33 & 490 & 2.37--3.65 & 0.55 $\pm$ 0.01 & 2.403 $\pm$ 0.034 & 6.2 & +0.028 & 37 \\
Cs--S$^{\dagger}$ & 28 & 886 & 3.41--4.86 & 0.74 $\pm$ 0.03 & 2.190 $\pm$ 0.126 & 17 & +0.051 & 27 \\
Ba--S$^{\dagger}$ & 49 & 1926 & 3.03--4.47 & 0.69 $\pm$ 0.02 & 2.337 $\pm$ 0.081 & 12 & +0.034 & 27 \\
La--S$^{\dagger}$ & 26 & 1048 & 2.82--3.78 & 0.66 $\pm$ 0.02 & 2.369 $\pm$ 0.047 & 7.2 & +0.017 & 22 \\
Lu--S$^{\dagger}$ & 16 & 544 & 2.56--3.33 & 0.55 $\pm$ 0.03 & 2.308 $\pm$ 0.030 & 5.5 & +0.005 & 14 \\
Ta$^{5+}$--S$^{\dagger}$ & 15 & 218 & 2.27--2.66 & 0.69 $\pm$ 0.04 & 2.377 $\pm$ 0.083 & 12 & +0.004 & 10 \\
Hg$^{2+}$--S$^{\dagger}$ & 24 & 436 & 2.35--4.27 & 0.85 $\pm$ 0.06 & 2.069 $\pm$ 0.196 & 23 & +0.097 & 62 \\
\midrule
\multicolumn{9}{l}{\emph{Se$^{2-}$} (20 families)} \\
Li--Se$^{\dagger}$ & 16 & 284 & 2.52--3.01 & 0.45 $\pm$ 0.05 & 2.019 $\pm$ 0.051 & 11 & +0.005 & 14 \\
Na--Se$^{\dagger}$ & 38 & 1372 & 2.86--3.76 & 0.53 $\pm$ 0.02 & 2.172 $\pm$ 0.075 & 14 & +0.036 & 29 \\
Mg--Se$^{\dagger}$ & 13 & 230 & 2.58--2.86 & 0.83 $\pm$ 0.08 & 1.877 $\pm$ 0.070 & 8.4 & +0.001 & 4.7 \\
Al--Se$^{\dagger}$ & 17 & 272 & 2.31--2.60 & 0.52 $\pm$ 0.03 & 2.261 $\pm$ 0.022 & 4.3 & +0.001 & 6.3 \\
K--Se$^{\dagger}$ & 40 & 1384 & 3.25--4.26 & 0.68 $\pm$ 0.03 & 2.181 $\pm$ 0.104 & 15 & +0.059 & 28 \\
Ga--Se & 17 & 252 & 2.37--2.65 & 0.83 $\pm$ 0.12 & 2.215 $\pm$ 0.044 & 5.3 & $-$0.007 & 5.8 \\
Ge--Se & 22 & 256 & 2.29--2.49 & 0.50 $\pm$ 0.07 & 2.389 $\pm$ 0.024 & 4.9 & +0.002 & 8.4 \\
As$^{3+}$--Se$^{\dagger}$ & 14 & 342 & 2.33--4.27 & 0.65 $\pm$ 0.03 & 2.400 $\pm$ 0.170 & 26 & +0.149 & 90 \\
Rb--Se$^{\dagger}$ & 20 & 682 & 3.42--4.35 & 0.67 $\pm$ 0.05 & 2.301 $\pm$ 0.110 & 16 & +0.066 & 30 \\
Sr--Se & 8 & 196 & 3.02--3.62 & 0.65 $\pm$ 0.02 & 2.376 $\pm$ 0.044 & 6.7 & +0.013 & 17 \\
Y--Se & 16 & 356 & 2.77--3.11 & 0.83 $\pm$ 0.11 & 2.309 $\pm$ 0.041 & 5.0 & +0.002 & 5.3 \\
Ag--Se$^{\dagger}$ & 36 & 580 & 2.54--3.69 & 0.59 $\pm$ 0.02 & 1.990 $\pm$ 0.127 & 21 & $-$0.015 & 37 \\
Cd--Se & 21 & 292 & 2.62--3.59 & 0.71 $\pm$ 0.03 & 2.202 $\pm$ 0.047 & 6.6 & +0.012 & 19 \\
In--Se$^{\dagger}$ & 22 & 336 & 2.55--3.87 & 0.77 $\pm$ 0.03 & 2.386 $\pm$ 0.057 & 7.5 & +0.013 & 13 \\
Sn--Se$^{\dagger}$ & 26 & 330 & 2.50--3.72 & 0.61 $\pm$ 0.18 & 2.563 $\pm$ 0.032 & 5.3 & +0.016 & 22 \\
Cs--Se$^{\dagger}$ & 20 & 784 & 3.43--5.16 & 0.68 $\pm$ 0.04 & 2.460 $\pm$ 0.133 & 20 & +0.062 & 36 \\
Ba--Se$^{\dagger}$ & 23 & 744 & 3.24--4.29 & 0.63 $\pm$ 0.04 & 2.563 $\pm$ 0.062 & 9.9 & +0.018 & 21 \\
La--Se$^{\dagger}$ & 11 & 404 & 2.96--3.42 & 0.70 $\pm$ 0.06 & 2.460 $\pm$ 0.050 & 7.2 & +0.014 & 12 \\
Ta$^{4+}$--Se & 4 & 102 & 2.48--2.77 & 0.77 $\pm$ 0.01 & 2.313 $\pm$ 0.029 & 3.8 & +0.004 & 10 \\
Hg$^{2+}$--Se & 19 & 242 & 2.51--4.46 & 0.95 $\pm$ 0.29 & 2.105 $\pm$ 0.262 & 28 & +0.010 & 44 \\
\midrule
\multicolumn{9}{l}{\emph{Te$^{2-}$} (7 families)} \\
Na--Te$^{\dagger}$ & 9 & 248 & 3.12--3.80 & 0.58 $\pm$ 0.05 & 2.330 $\pm$ 0.066 & 11 & +0.034 & 27 \\
K--Te$^{\dagger}$ & 16 & 558 & 3.45--4.48 & 0.77 $\pm$ 0.04 & 2.247 $\pm$ 0.110 & 14 & +0.043 & 28 \\
Sc--Te & 4 & 174 & 2.87--3.06 & 0.85 $\pm$ 0.21 & 2.354 $\pm$ 0.030 & 3.5 & +0.001 & 5.5 \\
Rb--Te & 8 & 144 & 3.56--4.51 & 0.87 $\pm$ 0.21 & 2.101 $\pm$ 0.140 & 16 & +0.037 & 21 \\
Ag--Te$^{\dagger}$ & 9 & 186 & 2.77--3.48 & 0.66 $\pm$ 0.10 & 2.123 $\pm$ 0.145 & 22 & $-$0.048 & 27 \\
Cs--Te & 9 & 196 & 3.78--5.72 & 0.63 $\pm$ 0.05 & 2.779 $\pm$ 0.153 & 24 & +0.044 & 41 \\
Ba--Te & 11 & 300 & 3.47--4.04 & 0.63 $\pm$ 0.07 & 2.759 $\pm$ 0.057 & 9.0 & +0.014 & 18 \\
\midrule
\multicolumn{9}{l}{\emph{N$^{3-}$} (11 families)} \\
Li--N$^{\dagger}$ & 38 & 744 & 1.92--2.59 & 0.65 $\pm$ 0.04 & 1.269 $\pm$ 0.077 & 12 & +0.017 & 16 \\
Na--N$^{\dagger}$ & 12 & 178 & 2.35--2.89 & 0.59 $\pm$ 0.07 & 1.601 $\pm$ 0.103 & 18 & +0.043 & 20 \\
Mg--N$^{\dagger}$ & 16 & 280 & 2.06--2.97 & 0.43 $\pm$ 0.05 & 1.831 $\pm$ 0.049 & 11 & +0.010 & 21 \\
Al--N & 8 & 192 & 1.83--2.05 & 0.44 $\pm$ 0.05 & 1.788 $\pm$ 0.033 & 7.5 & +0.001 & 7.5 \\
Si--N & 30 & 692 & 1.67--1.98 & 0.52 $\pm$ 0.04 & 1.756 $\pm$ 0.023 & 4.4 & +0.001 & 7.7 \\
P$^{5+}$--N & 9 & 152 & 1.56--1.75 & 0.43 $\pm$ 0.01 & 1.741 $\pm$ 0.013 & 3.0 & +0.002 & 10 \\
Ca--N$^{\dagger}$ & 32 & 1266 & 2.26--3.51 & 0.66 $\pm$ 0.02 & 1.876 $\pm$ 0.087 & 13 & +0.031 & 23 \\
Ga--N$^{\dagger}$ & 10 & 154 & 1.88--2.09 & 0.96 $\pm$ 0.27 & 1.798 $\pm$ 0.096 & 10 & $-$0.045 & 7.5 \\
Sr--N$^{\dagger}$ & 32 & 1124 & 2.47--3.77 & 0.61 $\pm$ 0.03 & 2.059 $\pm$ 0.081 & 13 & +0.053 & 32 \\
Ba--N$^{\dagger}$ & 17 & 624 & 2.65--3.62 & 0.58 $\pm$ 0.04 & 2.246 $\pm$ 0.094 & 16 & +0.062 & 40 \\
La--N & 8 & 236 & 2.40--3.31 & 0.62 $\pm$ 0.03 & 2.074 $\pm$ 0.077 & 12 & +0.038 & 31 \\
\midrule
\multicolumn{9}{l}{\emph{P$^{3-}$} (7 families)} \\
Na--P$^{\dagger}$ & 15 & 428 & 2.86--3.67 & 0.65 $\pm$ 0.02 & 2.059 $\pm$ 0.076 & 12 & +0.035 & 21 \\
K--P$^{\dagger}$ & 17 & 290 & 3.24--4.87 & 0.85 $\pm$ 0.07 & 2.063 $\pm$ 0.196 & 23 & +0.137 & 40 \\
Ca--P$^{\dagger}$ & 12 & 235 & 2.83--4.11 & 0.74 $\pm$ 0.06 & 2.155 $\pm$ 0.145 & 20 & +0.068 & 28 \\
Zn--P & 16 & 316 & 2.23--2.97 & 0.95 $\pm$ 0.02 & 1.846 $\pm$ 0.159 & 17 & $-$0.013 & 18 \\
Sr--P & 13 & 232 & 3.01--4.52 & 0.71 $\pm$ 0.11 & 2.345 $\pm$ 0.148 & 21 & +0.069 & 32 \\
Cd--P & 11 & 172 & 2.42--3.23 & 0.97 $\pm$ 0.04 & 2.110 $\pm$ 0.394 & 40 & $-$0.066 & 40 \\
Ba--P$^{\dagger}$ & 16 & 444 & 3.17--4.71 & 0.70 $\pm$ 0.05 & 2.540 $\pm$ 0.109 & 16 & +0.064 & 30 \\
\midrule
\multicolumn{9}{l}{\emph{As$^{3-}$} (11 families)} \\
Li--As$^{\dagger}$ & 14 & 225 & 2.53--3.08 & 0.63 $\pm$ 0.01 & 1.855 $\pm$ 0.068 & 11 & +0.008 & 12 \\
Na--As$^{\dagger}$ & 23 & 806 & 2.88--3.68 & 0.65 $\pm$ 0.03 & 2.142 $\pm$ 0.097 & 15 & +0.043 & 22 \\
K--As$^{\dagger}$ & 31 & 816 & 3.19--4.96 & 0.94 $\pm$ 0.06 & 1.926 $\pm$ 0.205 & 22 & +0.150 & 33 \\
Ca--As$^{\dagger}$ & 15 & 276 & 2.92--4.07 & 0.75 $\pm$ 0.05 & 2.275 $\pm$ 0.135 & 18 & +0.066 & 27 \\
Zn--As & 24 & 485 & 2.44--3.02 & 1.21 $\pm$ 0.07 & 1.804 $\pm$ 0.146 & 12 & $-$0.054 & 14 \\
Rb--As$^{\dagger}$ & 10 & 236 & 3.34--5.23 & 0.84 $\pm$ 0.09 & 2.204 $\pm$ 0.273 & 33 & +0.157 & 42 \\
Sr--As & 14 & 314 & 3.09--4.62 & 0.66 $\pm$ 0.04 & 2.503 $\pm$ 0.118 & 18 & +0.058 & 33 \\
Cd--As & 20 & 330 & 2.64--4.25 & 0.85 $\pm$ 0.19 & 2.315 $\pm$ 0.197 & 23 & $-$0.103 & 32 \\
Te$^{2+}$--As & 3 & 130 & 2.67--3.26 & 0.85 $\pm$ 0.34 & 2.308 $\pm$ 0.224 & 26 & +0.008 & 17 \\
Cs--As$^{\dagger}$ & 9 & 280 & 3.56--4.99 & 0.88 $\pm$ 0.07 & 2.397 $\pm$ 0.280 & 32 & +0.178 & 38 \\
Ba--As$^{\dagger}$ & 18 & 456 & 3.23--5.01 & 0.78 $\pm$ 0.05 & 2.479 $\pm$ 0.178 & 23 & +0.094 & 31 \\
\end{longtable}
}
\clearpage
{\footnotesize\setlength{\tabcolsep}{3.5pt}%
\begin{longtable}{lrrcccccccc}
\caption{\captionhsens}\label{tab:hsens}\\
\toprule
 & \multicolumn{2}{c}{hosts} & \multicolumn{3}{c}{without H} & \multicolumn{3}{c}{with H} & \multicolumn{2}{c}{passes} \\
\cmidrule(lr){2-3}\cmidrule(lr){4-6}\cmidrule(lr){7-9}\cmidrule(lr){10-11}
family & $-$H & $+$H & $B$ & $B^{*}$ & $D$ & $B$ & $B^{*}$ & $D$ & $-$H & $+$H \\
\midrule
\endfirsthead
\multicolumn{11}{c}{\tablename~\thetable{} (continued)}\\
\toprule
 & \multicolumn{2}{c}{hosts} & \multicolumn{3}{c}{without H} & \multicolumn{3}{c}{with H} & \multicolumn{2}{c}{passes} \\
\cmidrule(lr){2-3}\cmidrule(lr){4-6}\cmidrule(lr){7-9}\cmidrule(lr){10-11}
family & $-$H & $+$H & $B$ & $B^{*}$ & $D$ & $B$ & $B^{*}$ & $D$ & $-$H & $+$H \\
\midrule
\endhead
\bottomrule
\endlastfoot
\multicolumn{11}{l}{\emph{F$^{-}$} (8 families)} \\
Ba--F & 17 & 17 & 0.55 & 0.41 & +0.137 & 0.55 & 0.41 & +0.137 &  &  \\
Ca--F & 22 & 22 & 0.51 & 0.36 & +0.147 & 0.51 & 0.36 & +0.147 &  &  \\
Cs--F & 22 & 22 & 0.46 & 0.32 & +0.142 & 0.46 & 0.32 & +0.142 &  &  \\
K--F & 86 & 88 & 0.51 & 0.39 & +0.119 & 0.51 & 0.39 & +0.122 &  &  \\
Li--F & 46 & 46 & 0.44 & 0.36 & +0.077 & 0.44 & 0.36 & +0.077 &  &  \\
Na--F & 56 & 56 & 0.46 & 0.46 & +0.004 & 0.46 & 0.46 & +0.004 & \checkmark & \checkmark \\
Rb--F & 22 & 22 & 0.49 & 0.25 & +0.239 & 0.49 & 0.25 & +0.239 &  &  \\
Zr--F & 11 & 11 & 0.48 & 0.36 & +0.122 & 0.48 & 0.36 & +0.122 &  &  \\
\midrule
\multicolumn{11}{l}{\emph{Cl$^{-}$} (8 families)} \\
Au--Cl & 13 & 13 & 0.75 & 0.09 & +0.654 & 0.75 & 0.09 & +0.654 &  &  \\
Ba--Cl & 9 & 9 & 0.66 & 0.57 & +0.085 & 0.66 & 0.57 & +0.085 & \checkmark & \checkmark \\
Cs--Cl & 49 & 49 & 0.67 & 0.44 & +0.224 & 0.67 & 0.44 & +0.224 &  &  \\
K--Cl & 27 & 27 & 0.64 & 0.74 & $-$0.106 & 0.64 & 0.74 & $-$0.106 &  &  \\
Li--Cl & 15 & 15 & 0.51 & 0.59 & $-$0.080 & 0.51 & 0.59 & $-$0.080 &  &  \\
Na--Cl & 17 & 17 & 0.47 & 0.50 & $-$0.035 & 0.47 & 0.50 & $-$0.035 & \checkmark & \checkmark \\
Pd--Cl & 15 & 15 & 0.61 & $-$0.18 & +0.792 & 0.61 & $-$0.18 & +0.792 &  &  \\
Rb--Cl & 22 & 22 & 0.72 & 0.53 & +0.192 & 0.72 & 0.53 & +0.192 &  &  \\
\midrule
\multicolumn{11}{l}{\emph{O$^{2-}$} (35 families)} \\
Ag--O & 32 & 32 & 0.50 & 0.34 & +0.165 & 0.50 & 0.34 & +0.165 &  &  \\
Al--O & 123 & 145 & 0.42 & 0.40 & +0.019 & 0.42 & 0.39 & +0.027 & \checkmark & \checkmark \\
As$^{5+}$--O & 53 & 66 & 0.52 & 0.30 & +0.220 & 0.52 & 0.31 & +0.211 &  &  \\
B--O & 183 & 199 & 0.38 & 0.34 & +0.039 & 0.38 & 0.34 & +0.039 &  &  \\
Ba--O & 205 & 210 & 0.58 & 0.32 & +0.261 & 0.58 & 0.32 & +0.262 &  &  \\
Be--O & 34 & 39 & 0.49 & 0.33 & +0.159 & 0.52 & 0.33 & +0.194 &  &  \\
Ca--O & 157 & 186 & 0.54 & 0.38 & +0.153 & 0.53 & 0.38 & +0.154 &  &  \\
Cd--O & 40 & 42 & 0.53 & 0.33 & +0.193 & 0.53 & 0.33 & +0.195 &  &  \\
Cs--O & 65 & 73 & 0.54 & 0.32 & +0.213 & 0.53 & 0.33 & +0.200 &  &  \\
Ga--O & 63 & 69 & 0.52 & 0.35 & +0.174 & 0.52 & 0.35 & +0.172 &  &  \\
Ge--O & 116 & 118 & 0.51 & 0.32 & +0.185 & 0.51 & 0.32 & +0.185 &  &  \\
Hf--O & 11 & 12 & 0.59 & 0.40 & +0.190 & 0.59 & 0.40 & +0.182 &  &  \\
Hg$^{2+}$--O & 15 & 17 & 0.66 & 0.44 & +0.225 & 0.67 & 0.44 & +0.233 &  &  \\
In--O & 47 & 50 & 0.46 & 0.25 & +0.213 & 0.46 & 0.25 & +0.215 &  &  \\
K--O & 209 & 238 & 0.57 & 0.25 & +0.323 & 0.57 & 0.25 & +0.318 &  &  \\
La--O & 74 & 74 & 0.55 & 0.39 & +0.165 & 0.55 & 0.39 & +0.165 &  &  \\
Li--O & 214 & 239 & 0.47 & 0.43 & +0.042 & 0.47 & 0.43 & +0.040 &  &  \\
Lu--O & 17 & 17 & 0.60 & 0.37 & +0.225 & 0.60 & 0.37 & +0.225 &  &  \\
Mg--O & 95 & 117 & 0.52 & 0.37 & +0.151 & 0.52 & 0.36 & +0.163 &  &  \\
N--O & 16 & 20 & 0.75 & $-$0.70 & +1.447 & 0.77 & $-$0.85 & +1.619 &  &  \\
Na--O & 276 & 314 & 0.49 & 0.31 & +0.185 & 0.50 & 0.31 & +0.190 &  &  \\
Nb--O & 112 & 112 & 0.62 & 0.28 & +0.338 & 0.62 & 0.28 & +0.338 &  &  \\
Pb--O & 25 & 26 & 0.57 & 0.25 & +0.320 & 0.57 & 0.25 & +0.320 &  &  \\
Pd--O & 22 & 22 & 0.55 & $-$0.25 & +0.805 & 0.55 & $-$0.25 & +0.805 &  &  \\
Rb--O & 65 & 71 & 0.53 & 0.32 & +0.209 & 0.53 & 0.32 & +0.215 &  &  \\
Se$^{4+}$--O & 25 & 34 & 0.63 & 0.18 & +0.446 & 0.63 & 0.19 & +0.448 &  &  \\
Si--O & 260 & 282 & 0.33 & 0.35 & $-$0.023 & 0.33 & 0.35 & $-$0.021 & \checkmark & \checkmark \\
Sr--O & 175 & 185 & 0.53 & 0.41 & +0.117 & 0.53 & 0.41 & +0.118 &  &  \\
Ta$^{5+}$--O & 79 & 79 & 0.60 & 0.27 & +0.328 & 0.60 & 0.27 & +0.328 &  &  \\
Te$^{4+}$--O & 24 & 29 & 0.72 & 0.15 & +0.569 & 0.71 & 0.13 & +0.582 &  &  \\
Ti--O & 117 & 121 & 0.59 & 0.31 & +0.275 & 0.59 & 0.31 & +0.279 &  &  \\
Tl--O & 11 & 12 & 0.76 & 0.44 & +0.321 & 0.76 & 0.44 & +0.326 &  &  \\
Y--O & 60 & 61 & 0.54 & 0.38 & +0.160 & 0.54 & 0.38 & +0.161 &  &  \\
Zn--O & 110 & 143 & 0.52 & 0.32 & +0.202 & 0.53 & 0.33 & +0.199 &  &  \\
Zr--O & 52 & 53 & 0.54 & 0.34 & +0.199 & 0.54 & 0.34 & +0.199 &  &  \\
\midrule
\multicolumn{11}{l}{\emph{S$^{2-}$} (21 families)} \\
Ag--S & 76 & 76 & 0.60 & 0.25 & +0.345 & 0.60 & 0.25 & +0.345 &  &  \\
Al--S & 20 & 20 & 0.41 & 0.42 & $-$0.010 & 0.41 & 0.42 & $-$0.010 & \checkmark & \checkmark \\
As$^{2+}$--S & 5 & 5 & 0.76 & 0.19 & +0.570 & 0.76 & 0.19 & +0.570 &  &  \\
As$^{3+}$--S & 15 & 15 & 0.63 & 0.06 & +0.569 & 0.63 & 0.06 & +0.569 &  &  \\
B--S & 18 & 18 & 0.26 & 0.39 & $-$0.132 & 0.26 & 0.39 & $-$0.132 &  &  \\
Ba--S & 49 & 49 & 0.69 & 0.43 & +0.266 & 0.69 & 0.43 & +0.266 &  &  \\
Ca--S & 10 & 10 & 0.51 & 0.43 & +0.081 & 0.51 & 0.43 & +0.081 &  &  \\
Cd--S & 24 & 24 & 0.65 & 0.27 & +0.378 & 0.65 & 0.27 & +0.378 &  &  \\
Cs--S & 28 & 28 & 0.74 & 0.32 & +0.417 & 0.74 & 0.32 & +0.417 &  &  \\
Hg$^{2+}$--S & 24 & 24 & 0.85 & 0.16 & +0.686 & 0.85 & 0.16 & +0.686 &  &  \\
In--S & 37 & 37 & 0.69 & 0.36 & +0.327 & 0.69 & 0.36 & +0.327 &  &  \\
K--S & 57 & 57 & 0.71 & 0.36 & +0.344 & 0.71 & 0.36 & +0.344 &  &  \\
La--S & 26 & 26 & 0.66 & 0.33 & +0.327 & 0.66 & 0.33 & +0.327 &  &  \\
Li--S & 37 & 37 & 0.42 & 0.43 & $-$0.010 & 0.42 & 0.43 & $-$0.010 & \checkmark & \checkmark \\
Lu--S & 16 & 16 & 0.55 & 0.40 & +0.145 & 0.55 & 0.40 & +0.145 &  &  \\
Mg--S & 11 & 11 & 0.67 & 0.35 & +0.322 & 0.67 & 0.35 & +0.322 &  &  \\
Na--S & 50 & 50 & 0.52 & 0.35 & +0.169 & 0.52 & 0.35 & +0.169 &  &  \\
Rb--S & 40 & 40 & 0.73 & 0.33 & +0.395 & 0.73 & 0.33 & +0.395 &  &  \\
Sn--S & 33 & 33 & 0.55 & 0.32 & +0.227 & 0.55 & 0.32 & +0.227 &  &  \\
Ta$^{5+}$--S & 15 & 15 & 0.69 & 0.40 & +0.298 & 0.69 & 0.40 & +0.298 &  &  \\
Y--S & 23 & 23 & 0.68 & 0.41 & +0.273 & 0.68 & 0.41 & +0.273 &  &  \\
\midrule
\multicolumn{11}{l}{\emph{Se$^{2-}$} (13 families)} \\
Ag--Se & 36 & 36 & 0.59 & 0.24 & +0.353 & 0.59 & 0.24 & +0.353 &  &  \\
Al--Se & 17 & 17 & 0.52 & 0.42 & +0.097 & 0.52 & 0.42 & +0.097 &  &  \\
As$^{3+}$--Se & 14 & 14 & 0.65 & 0.34 & +0.308 & 0.65 & 0.34 & +0.308 &  &  \\
Ba--Se & 23 & 23 & 0.63 & 0.56 & +0.074 & 0.63 & 0.56 & +0.074 & \checkmark & \checkmark \\
Cs--Se & 20 & 20 & 0.68 & 0.29 & +0.383 & 0.68 & 0.29 & +0.383 &  &  \\
In--Se & 22 & 22 & 0.77 & 0.36 & +0.410 & 0.77 & 0.36 & +0.410 &  &  \\
K--Se & 40 & 40 & 0.68 & 0.36 & +0.327 & 0.68 & 0.36 & +0.327 &  &  \\
La--Se & 11 & 11 & 0.70 & 0.40 & +0.296 & 0.70 & 0.40 & +0.296 &  &  \\
Li--Se & 16 & 16 & 0.45 & 0.54 & $-$0.096 & 0.45 & 0.54 & $-$0.096 &  &  \\
Mg--Se & 13 & 13 & 0.83 & 0.40 & +0.434 & 0.83 & 0.40 & +0.434 &  &  \\
Na--Se & 38 & 38 & 0.53 & 0.25 & +0.284 & 0.53 & 0.25 & +0.284 &  &  \\
Rb--Se & 20 & 20 & 0.67 & 0.34 & +0.335 & 0.67 & 0.34 & +0.335 &  &  \\
Sn--Se & 26 & 26 & 0.61 & 0.29 & +0.321 & 0.61 & 0.29 & +0.321 &  &  \\
\midrule
\multicolumn{11}{l}{\emph{Te$^{2-}$} (3 families)} \\
Ag--Te & 9 & 9 & 0.66 & 0.23 & +0.427 & 0.66 & 0.23 & +0.427 &  &  \\
K--Te & 16 & 16 & 0.77 & 0.32 & +0.453 & 0.77 & 0.32 & +0.453 &  &  \\
Na--Te & 9 & 9 & 0.58 & 0.53 & +0.049 & 0.58 & 0.53 & +0.049 & \checkmark & \checkmark \\
\midrule
\multicolumn{11}{l}{\emph{N$^{3-}$} (7 families)} \\
Ba--N & 17 & 17 & 0.58 & 0.35 & +0.222 & 0.58 & 0.35 & +0.222 &  &  \\
Ca--N & 32 & 32 & 0.66 & 0.28 & +0.388 & 0.66 & 0.28 & +0.388 &  &  \\
Ga--N & 10 & 11 & 0.96 & 0.34 & +0.615 & 0.96 & 0.33 & +0.628 &  &  \\
Li--N & 38 & 46 & 0.65 & 0.36 & +0.283 & 0.65 & 0.36 & +0.292 &  &  \\
Mg--N & 16 & 17 & 0.43 & 0.18 & +0.250 & 0.43 & 0.19 & +0.236 &  &  \\
Na--N & 12 & 15 & 0.59 & 0.03 & +0.556 & 0.61 & 0.08 & +0.523 &  &  \\
Sr--N & 32 & 34 & 0.61 & 0.34 & +0.278 & 0.61 & 0.33 & +0.280 &  &  \\
\midrule
\multicolumn{11}{l}{\emph{P$^{3-}$} (4 families)} \\
Ba--P & 16 & 16 & 0.70 & 0.19 & +0.503 & 0.70 & 0.19 & +0.503 &  &  \\
Ca--P & 12 & 12 & 0.74 & 0.34 & +0.402 & 0.74 & 0.34 & +0.402 &  &  \\
K--P & 17 & 17 & 0.85 & 0.04 & +0.812 & 0.85 & 0.04 & +0.812 &  &  \\
Na--P & 15 & 15 & 0.65 & 0.45 & +0.198 & 0.65 & 0.45 & +0.198 &  &  \\
\midrule
\multicolumn{11}{l}{\emph{As$^{3-}$} (7 families)} \\
Ba--As & 18 & 18 & 0.78 & 0.29 & +0.490 & 0.78 & 0.29 & +0.490 &  &  \\
Ca--As & 15 & 15 & 0.75 & 0.42 & +0.327 & 0.75 & 0.42 & +0.327 &  &  \\
Cs--As & 9 & 9 & 0.88 & 0.12 & +0.766 & 0.88 & 0.12 & +0.766 &  &  \\
K--As & 31 & 31 & 0.94 & 0.14 & +0.795 & 0.94 & 0.14 & +0.795 &  &  \\
Li--As & 14 & 14 & 0.63 & 0.57 & +0.061 & 0.63 & 0.57 & +0.061 & \checkmark & \checkmark \\
Na--As & 23 & 23 & 0.65 & 0.45 & +0.203 & 0.65 & 0.45 & +0.203 &  &  \\
Rb--As & 10 & 10 & 0.84 & 0.08 & +0.758 & 0.84 & 0.08 & +0.758 &  &  \\
\end{longtable}
}
\clearpage

\begin{figure*}[t]
\includegraphics[width=\textwidth,height=0.8\textheight,keepaspectratio]{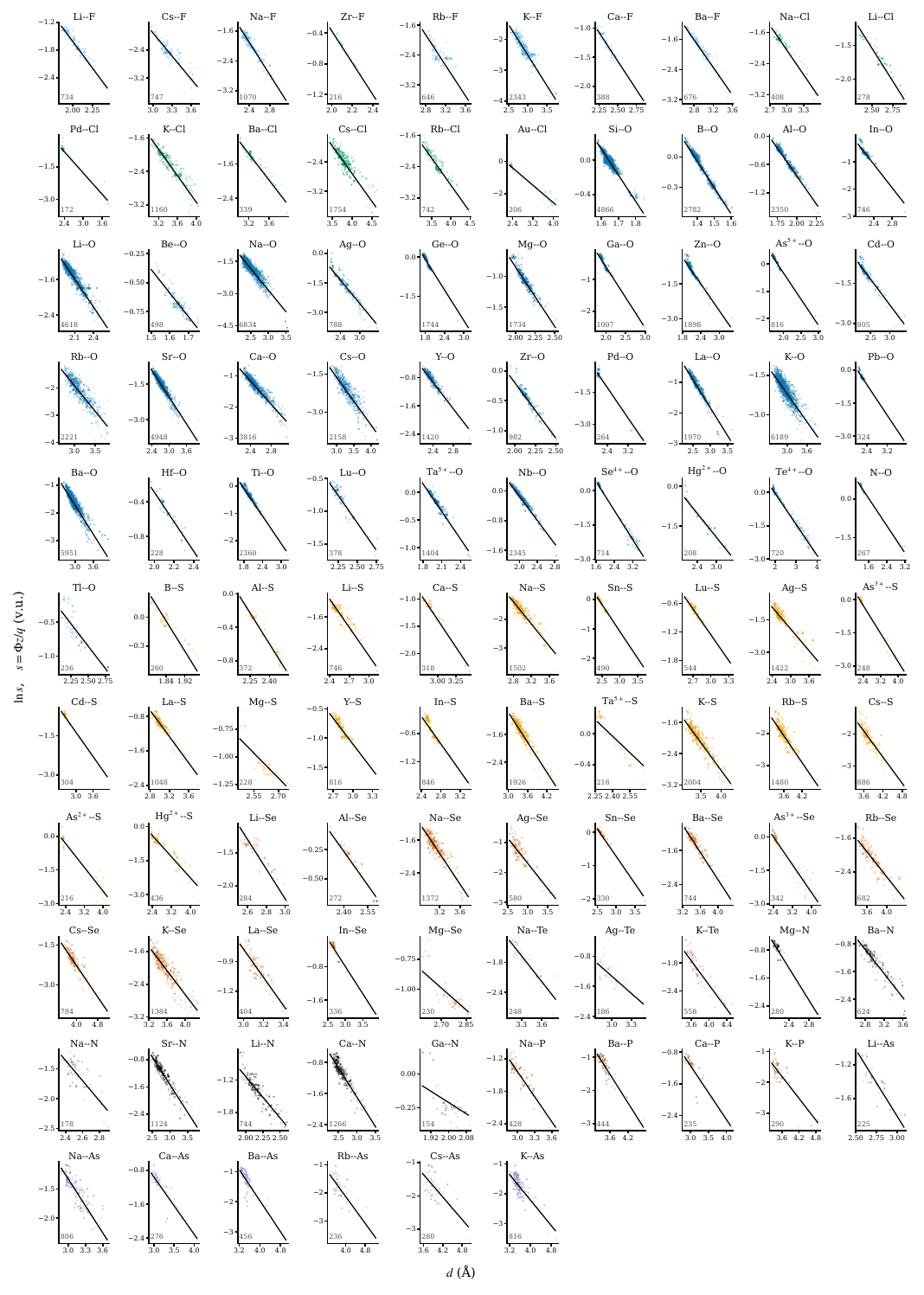}
\caption{The bond-valence law family by family, as the flux valence $\ln s$ of every bonded cation--anion face (hydrogen-free hosts) against bond length, one panel per family of the transferability test, grouped by anion and ordered within an anion by $B$, with the number of faces. The line is Eq.~\eqref{eq:law} at the within-site $B$ of Table~\ref{tab:transfer} and $R_0=\langle d+B\ln s\rangle$ over the family's faces.}
\label{fig:law}
\end{figure*}

\clearpage
{\footnotesize\setlength{\tabcolsep}{3.5pt}%
\begin{longtable}{lrrrcccccc}
\caption{\captiontransfer}\label{tab:transfer}\\
\toprule
family & hosts & sites & distorted & $B$ (\AA) & $B^{*}$ (\AA) & $D$ (\AA) & $\sigma_n$ & rms & transfers \\
\midrule
\endfirsthead
\multicolumn{10}{c}{\tablename~\thetable{} (continued)}\\
\toprule
family & hosts & sites & distorted & $B$ (\AA) & $B^{*}$ (\AA) & $D$ (\AA) & $\sigma_n$ & rms & transfers \\
\midrule
\endhead
\bottomrule
\endlastfoot
\multicolumn{10}{l}{\emph{F$^{-}$} (8 families)} \\
Na--F & 56 & 160 & 134 (45) & 0.46 $\pm$ 0.02 & 0.46 $\pm$ 0.05 & 0.00 & 0.10 & 0.10 & \checkmark \\
Li--F & 46 & 141 & 89 (25) & 0.44 $\pm$ 0.03 & 0.36 $\pm$ 0.03 & +0.08 & 0.07 & 0.07 &  \\
K--F & 86 & 249 & 190 (70) & 0.51 $\pm$ 0.01 & 0.39 $\pm$ 0.05 & +0.12 & 0.17 & 0.15 &  \\
Zr--F & 11 & 29 & 28 (10) & 0.48 $\pm$ 0.01 & 0.36 $\pm$ 0.02 & +0.12 & 0.01 & 0.03 &  \\
Ba--F & 17 & 57 & 56 (21) & 0.55 $\pm$ 0.02 & 0.41 $\pm$ 0.09 & +0.14 & 0.05 & 0.09 &  \\
Cs--F & 22 & 67 & 53 (19) & 0.46 $\pm$ 0.04 & 0.32 $\pm$ 0.05 & +0.14 & 0.10 & 0.16 &  \\
Ca--F & 22 & 55 & 34 (13) & 0.51 $\pm$ 0.03 & 0.36 $\pm$ 0.02 & +0.15 & 0.03 & 0.06 &  \\
Rb--F & 22 & 60 & 33 (13) & 0.49 $\pm$ 0.02 & 0.25 $\pm$ 0.10 & +0.24 & 0.21 & 0.21 &  \\
\midrule
\multicolumn{10}{l}{\emph{Cl$^{-}$} (8 families)} \\
K--Cl & 27 & 132 & 118 (33) & 0.64 $\pm$ 0.02 & 0.74 $\pm$ 0.05 & $-$0.11 & 0.09 & 0.13 &  \\
Li--Cl & 15 & 52 & 34 (9) & 0.51 $\pm$ 0.02 & 0.59 $\pm$ 0.06 & $-$0.08 & 0.08 & 0.08 &  \\
Na--Cl & 17 & 65 & 46 (15) & 0.47 $\pm$ 0.01 & 0.50 $\pm$ 0.12 & $-$0.04 & 0.09 & 0.10 & \checkmark \\
Ba--Cl & 9 & 38 & 27 (10) & 0.66 $\pm$ 0.04 & 0.57 $\pm$ 0.08 & +0.09 & 0.02 & 0.08 & \checkmark \\
Rb--Cl & 22 & 76 & 54 (16) & 0.72 $\pm$ 0.03 & 0.53 $\pm$ 0.05 & +0.19 & 0.05 & 0.11 &  \\
Cs--Cl & 49 & 165 & 132 (42) & 0.67 $\pm$ 0.04 & 0.44 $\pm$ 0.04 & +0.22 & 0.10 & 0.15 &  \\
Au--Cl & 13 & 41 & 33 (13) & 0.75 $\pm$ 0.04 & 0.09 $\pm$ 0.26 & +0.65 & 0.15 & 0.30 &  \\
Pd--Cl & 15 & 41 & 22 (6) & 0.61 $\pm$ 0.07 & $-$0.18 $\pm$ 0.11 & +0.79 & 0.01 & 0.18 &  \\
\midrule
\multicolumn{10}{l}{\emph{O$^{2-}$} (35 families)} \\
Si--O & 260 & 1205 & 584 (177) & 0.328 $\pm$ 0.004 & 0.35 $\pm$ 0.01 & $-$0.02 & 0.03 & 0.04 & \checkmark \\
Al--O & 123 & 502 & 235 (95) & 0.42 $\pm$ 0.01 & 0.40 $\pm$ 0.01 & +0.02 & 0.04 & 0.05 & \checkmark \\
B--O & 183 & 858 & 356 (104) & 0.38 $\pm$ 0.01 & 0.343 $\pm$ 0.003 & +0.04 & 0.02 & 0.03 &  \\
Li--O & 214 & 1052 & 783 (233) & 0.47 $\pm$ 0.01 & 0.43 $\pm$ 0.02 & +0.04 & 0.08 & 0.09 &  \\
Sr--O & 175 & 590 & 535 (183) & 0.53 $\pm$ 0.01 & 0.41 $\pm$ 0.01 & +0.12 & 0.09 & 0.11 &  \\
Mg--O & 95 & 304 & 209 (67) & 0.52 $\pm$ 0.02 & 0.37 $\pm$ 0.02 & +0.15 & 0.05 & 0.06 &  \\
Ca--O & 157 & 529 & 468 (180) & 0.54 $\pm$ 0.01 & 0.38 $\pm$ 0.02 & +0.15 & 0.07 & 0.10 &  \\
Be--O & 34 & 128 & 70 (19) & 0.49 $\pm$ 0.02 & 0.33 $\pm$ 0.03 & +0.16 & 0.03 & 0.04 &  \\
Y--O & 60 & 205 & 175 (56) & 0.54 $\pm$ 0.02 & 0.38 $\pm$ 0.02 & +0.16 & 0.04 & 0.07 &  \\
La--O & 74 & 227 & 209 (81) & 0.55 $\pm$ 0.02 & 0.39 $\pm$ 0.01 & +0.16 & 0.06 & 0.10 &  \\
Ag--O & 32 & 195 & 145 (43) & 0.50 $\pm$ 0.02 & 0.34 $\pm$ 0.06 & +0.17 & 0.15 & 0.20 &  \\
Ga--O & 63 & 212 & 106 (37) & 0.52 $\pm$ 0.03 & 0.35 $\pm$ 0.01 & +0.17 & 0.05 & 0.07 &  \\
Na--O & 276 & 1223 & 1080 (382) & 0.49 $\pm$ 0.02 & 0.31 $\pm$ 0.01 & +0.18 & 0.12 & 0.16 &  \\
Ge--O & 116 & 400 & 207 (74) & 0.51 $\pm$ 0.03 & 0.32 $\pm$ 0.01 & +0.18 & 0.04 & 0.07 &  \\
Hf--O & 11 & 34 & 23 (9) & 0.59 $\pm$ 0.02 & 0.40 $\pm$ 0.03 & +0.19 & 0.03 & 0.06 &  \\
Cd--O & 40 & 137 & 131 (47) & 0.53 $\pm$ 0.03 & 0.33 $\pm$ 0.05 & +0.19 & 0.06 & 0.13 &  \\
Zr--O & 52 & 157 & 105 (41) & 0.54 $\pm$ 0.01 & 0.34 $\pm$ 0.03 & +0.20 & 0.03 & 0.06 &  \\
Zn--O & 110 & 424 & 236 (70) & 0.52 $\pm$ 0.02 & 0.32 $\pm$ 0.01 & +0.20 & 0.06 & 0.10 &  \\
Rb--O & 65 & 319 & 303 (102) & 0.53 $\pm$ 0.02 & 0.32 $\pm$ 0.03 & +0.21 & 0.13 & 0.21 &  \\
Cs--O & 65 & 294 & 278 (95) & 0.54 $\pm$ 0.02 & 0.32 $\pm$ 0.01 & +0.21 & 0.13 & 0.23 &  \\
In--O & 47 & 130 & 80 (28) & 0.46 $\pm$ 0.07 & 0.25 $\pm$ 0.03 & +0.21 & 0.05 & 0.10 &  \\
As$^{5+}$--O & 53 & 192 & 112 (42) & 0.52 $\pm$ 0.02 & 0.30 $\pm$ 0.03 & +0.22 & 0.05 & 0.05 &  \\
Hg$^{2+}$--O & 15 & 37 & 32 (11) & 0.66 $\pm$ 0.02 & 0.44 $\pm$ 0.06 & +0.23 & 0.14 & 0.17 &  \\
Lu--O & 17 & 58 & 48 (14) & 0.60 $\pm$ 0.05 & 0.37 $\pm$ 0.05 & +0.23 & 0.04 & 0.07 &  \\
Ba--O & 205 & 622 & 534 (206) & 0.58 $\pm$ 0.02 & 0.32 $\pm$ 0.01 & +0.26 & 0.11 & 0.14 &  \\
Ti--O & 117 & 409 & 320 (128) & 0.59 $\pm$ 0.01 & 0.31 $\pm$ 0.01 & +0.28 & 0.04 & 0.06 &  \\
Pb--O & 25 & 60 & 25 (13) & 0.57 $\pm$ 0.03 & 0.25 $\pm$ 0.03 & +0.32 & 0.09 & 0.10 &  \\
Tl--O & 11 & 42 & 40 (11) & 0.76 $\pm$ 0.04 & 0.44 $\pm$ 0.15 & +0.32 & 0.03 & 0.14 &  \\
K--O & 209 & 893 & 824 (285) & 0.57 $\pm$ 0.01 & 0.25 $\pm$ 0.02 & +0.32 & 0.14 & 0.20 &  \\
Ta$^{5+}$--O & 79 & 231 & 166 (52) & 0.60 $\pm$ 0.02 & 0.27 $\pm$ 0.01 & +0.33 & 0.03 & 0.05 &  \\
Nb--O & 112 & 387 & 328 (111) & 0.62 $\pm$ 0.01 & 0.28 $\pm$ 0.02 & +0.34 & 0.05 & 0.05 &  \\
Se$^{4+}$--O & 25 & 150 & 122 (35) & 0.63 $\pm$ 0.01 & 0.18 $\pm$ 0.06 & +0.45 & 0.08 & 0.12 &  \\
Te$^{4+}$--O & 24 & 132 & 124 (44) & 0.72 $\pm$ 0.01 & 0.15 $\pm$ 0.05 & +0.57 & 0.14 & 0.12 &  \\
Pd--O & 22 & 59 & 17 (10) & 0.55 $\pm$ 0.02 & $-$0.25 $\pm$ 0.13 & +0.80 & 0.09 & 0.16 &  \\
N--O & 16 & 84 & 24 (6) & 0.75 $\pm$ 0.05 & $-$0.70 $\pm$ 0.39 & +1.45 & 0.08 & 0.07 &  \\
\midrule
\multicolumn{10}{l}{\emph{S$^{2-}$} (21 families)} \\
B--S & 18 & 83 & 28 (7) & 0.26 $\pm$ 0.01 & 0.39 $\pm$ 0.01 & $-$0.13 & 0.04 & 0.07 &  \\
Li--S & 37 & 165 & 114 (32) & 0.42 $\pm$ 0.01 & 0.43 $\pm$ 0.05 & $-$0.01 & 0.12 & 0.13 & \checkmark \\
Al--S & 20 & 75 & 30 (8) & 0.41 $\pm$ 0.04 & 0.42 $\pm$ 0.02 & $-$0.01 & 0.03 & 0.04 & \checkmark \\
Ca--S & 10 & 47 & 46 (12) & 0.51 $\pm$ 0.02 & 0.43 $\pm$ 0.05 & +0.08 & 0.03 & 0.06 &  \\
Lu--S & 16 & 85 & 64 (18) & 0.55 $\pm$ 0.03 & 0.40 $\pm$ 0.03 & +0.14 & 0.03 & 0.06 &  \\
Na--S & 50 & 286 & 255 (71) & 0.52 $\pm$ 0.02 & 0.35 $\pm$ 0.03 & +0.17 & 0.13 & 0.14 &  \\
Sn--S & 33 & 105 & 60 (17) & 0.55 $\pm$ 0.01 & 0.32 $\pm$ 0.06 & +0.23 & 0.10 & 0.06 &  \\
Ba--S & 49 & 226 & 214 (67) & 0.69 $\pm$ 0.02 & 0.43 $\pm$ 0.04 & +0.27 & 0.08 & 0.12 &  \\
Y--S & 23 & 121 & 98 (31) & 0.68 $\pm$ 0.06 & 0.41 $\pm$ 0.02 & +0.27 & 0.03 & 0.07 &  \\
Ta$^{5+}$--S & 15 & 44 & 15 (8) & 0.69 $\pm$ 0.04 & 0.40 $\pm$ 0.01 & +0.30 & 0.02 & 0.12 &  \\
Mg--S & 11 & 40 & 24 (6) & 0.67 $\pm$ 0.09 & 0.35 $\pm$ 0.03 & +0.32 & 0.03 & 0.07 &  \\
In--S & 37 & 170 & 77 (31) & 0.69 $\pm$ 0.01 & 0.36 $\pm$ 0.01 & +0.33 & 0.04 & 0.09 &  \\
La--S & 26 & 130 & 126 (32) & 0.66 $\pm$ 0.02 & 0.33 $\pm$ 0.05 & +0.33 & 0.05 & 0.07 &  \\
K--S & 57 & 295 & 276 (76) & 0.71 $\pm$ 0.02 & 0.36 $\pm$ 0.03 & +0.34 & 0.09 & 0.14 &  \\
Ag--S & 76 & 360 & 296 (103) & 0.60 $\pm$ 0.03 & 0.25 $\pm$ 0.03 & +0.34 & 0.21 & 0.23 &  \\
Cd--S & 24 & 68 & 30 (11) & 0.65 $\pm$ 0.07 & 0.27 $\pm$ 0.11 & +0.38 & 0.05 & 0.09 &  \\
Rb--S & 40 & 211 & 206 (64) & 0.73 $\pm$ 0.03 & 0.33 $\pm$ 0.03 & +0.40 & 0.10 & 0.17 &  \\
Cs--S & 28 & 116 & 112 (38) & 0.74 $\pm$ 0.03 & 0.32 $\pm$ 0.06 & +0.42 & 0.10 & 0.17 &  \\
As$^{3+}$--S & 15 & 60 & 42 (13) & 0.63 $\pm$ 0.02 & 0.06 $\pm$ 0.07 & +0.57 & 0.16 & 0.17 &  \\
As$^{2+}$--S & 5 & 64 & 54 (14) & 0.76 $\pm$ 0.01 & 0.19 $\pm$ 0.22 & +0.57 & 0.33 & 0.26 &  \\
Hg$^{2+}$--S & 24 & 92 & 79 (28) & 0.85 $\pm$ 0.06 & 0.16 $\pm$ 0.13 & +0.69 & 0.15 & 0.23 &  \\
\midrule
\multicolumn{10}{l}{\emph{Se$^{2-}$} (13 families)} \\
Li--Se & 16 & 64 & 38 (11) & 0.45 $\pm$ 0.05 & 0.54 $\pm$ 0.05 & $-$0.10 & 0.11 & 0.11 &  \\
Ba--Se & 23 & 89 & 82 (28) & 0.63 $\pm$ 0.04 & 0.56 $\pm$ 0.06 & +0.07 & 0.07 & 0.10 & \checkmark \\
Al--Se & 17 & 62 & 24 (10) & 0.52 $\pm$ 0.03 & 0.42 $\pm$ 0.02 & +0.10 & 0.02 & 0.04 &  \\
Na--Se & 38 & 253 & 224 (71) & 0.53 $\pm$ 0.02 & 0.25 $\pm$ 0.04 & +0.28 & 0.12 & 0.14 &  \\
La--Se & 11 & 52 & 48 (10) & 0.70 $\pm$ 0.06 & 0.40 $\pm$ 0.07 & +0.30 & 0.04 & 0.07 &  \\
As$^{3+}$--Se & 14 & 76 & 66 (22) & 0.65 $\pm$ 0.03 & 0.34 $\pm$ 0.09 & +0.31 & 0.12 & 0.26 &  \\
Sn--Se & 26 & 79 & 42 (13) & 0.61 $\pm$ 0.18 & 0.29 $\pm$ 0.16 & +0.32 & 0.04 & 0.05 &  \\
K--Se & 40 & 210 & 198 (61) & 0.68 $\pm$ 0.03 & 0.36 $\pm$ 0.03 & +0.33 & 0.11 & 0.15 &  \\
Rb--Se & 20 & 96 & 88 (25) & 0.67 $\pm$ 0.05 & 0.34 $\pm$ 0.04 & +0.34 & 0.08 & 0.16 &  \\
Ag--Se & 36 & 149 & 98 (35) & 0.59 $\pm$ 0.02 & 0.24 $\pm$ 0.04 & +0.35 & 0.14 & 0.21 &  \\
Cs--Se & 20 & 106 & 106 (34) & 0.68 $\pm$ 0.04 & 0.29 $\pm$ 0.09 & +0.38 & 0.11 & 0.20 &  \\
In--Se & 22 & 79 & 12 (6) & 0.77 $\pm$ 0.03 & 0.36 $\pm$ 0.05 & +0.41 & 0.02 & 0.07 &  \\
Mg--Se & 13 & 41 & 25 (8) & 0.83 $\pm$ 0.08 & 0.40 $\pm$ 0.03 & +0.43 & 0.03 & 0.08 &  \\
\midrule
\multicolumn{10}{l}{\emph{Te$^{2-}$} (3 families)} \\
Na--Te & 9 & 46 & 32 (8) & 0.58 $\pm$ 0.05 & 0.53 $\pm$ 0.04 & +0.05 & 0.08 & 0.11 & \checkmark \\
Ag--Te & 9 & 50 & 38 (10) & 0.66 $\pm$ 0.10 & 0.23 $\pm$ 0.09 & +0.43 & 0.10 & 0.22 &  \\
K--Te & 16 & 88 & 80 (20) & 0.77 $\pm$ 0.04 & 0.32 $\pm$ 0.04 & +0.45 & 0.08 & 0.14 &  \\
\midrule
\multicolumn{10}{l}{\emph{N$^{3-}$} (7 families)} \\
Ba--N & 17 & 94 & 88 (27) & 0.58 $\pm$ 0.04 & 0.35 $\pm$ 0.05 & +0.22 & 0.11 & 0.16 &  \\
Mg--N & 16 & 67 & 51 (10) & 0.43 $\pm$ 0.05 & 0.18 $\pm$ 0.07 & +0.25 & 0.09 & 0.11 &  \\
Sr--N & 32 & 191 & 172 (45) & 0.61 $\pm$ 0.03 & 0.34 $\pm$ 0.04 & +0.28 & 0.08 & 0.13 &  \\
Li--N & 38 & 194 & 146 (36) & 0.65 $\pm$ 0.04 & 0.36 $\pm$ 0.06 & +0.28 & 0.07 & 0.12 &  \\
Ca--N & 32 & 243 & 203 (59) & 0.66 $\pm$ 0.02 & 0.28 $\pm$ 0.03 & +0.39 & 0.08 & 0.13 &  \\
Na--N & 12 & 40 & 32 (13) & 0.59 $\pm$ 0.07 & 0.03 $\pm$ 0.08 & +0.56 & 0.10 & 0.17 &  \\
Ga--N & 10 & 40 & 24 (8) & 0.96 $\pm$ 0.27 & 0.34 $\pm$ 0.05 & +0.62 & 0.02 & 0.10 &  \\
\midrule
\multicolumn{10}{l}{\emph{P$^{3-}$} (4 families)} \\
Na--P & 15 & 100 & 78 (21) & 0.65 $\pm$ 0.02 & 0.45 $\pm$ 0.10 & +0.20 & 0.11 & 0.12 &  \\
Ca--P & 12 & 40 & 31 (10) & 0.74 $\pm$ 0.06 & 0.34 $\pm$ 0.10 & +0.40 & 0.09 & 0.20 &  \\
Ba--P & 16 & 71 & 61 (14) & 0.70 $\pm$ 0.05 & 0.19 $\pm$ 0.10 & +0.50 & 0.09 & 0.16 &  \\
K--P & 17 & 66 & 48 (15) & 0.85 $\pm$ 0.07 & 0.04 $\pm$ 0.05 & +0.81 & 0.17 & 0.23 &  \\
\midrule
\multicolumn{10}{l}{\emph{As$^{3-}$} (7 families)} \\
Li--As & 14 & 57 & 32 (7) & 0.63 $\pm$ 0.01 & 0.57 $\pm$ 0.10 & +0.06 & 0.08 & 0.11 & \checkmark \\
Na--As & 23 & 186 & 145 (33) & 0.65 $\pm$ 0.03 & 0.45 $\pm$ 0.07 & +0.20 & 0.14 & 0.15 &  \\
Ca--As & 15 & 48 & 39 (19) & 0.75 $\pm$ 0.05 & 0.42 $\pm$ 0.07 & +0.33 & 0.09 & 0.18 &  \\
Ba--As & 18 & 72 & 62 (16) & 0.78 $\pm$ 0.05 & 0.29 $\pm$ 0.05 & +0.49 & 0.08 & 0.23 &  \\
Rb--As & 10 & 46 & 42 (13) & 0.84 $\pm$ 0.09 & 0.08 $\pm$ 0.11 & +0.76 & 0.16 & 0.33 &  \\
Cs--As & 9 & 64 & 52 (19) & 0.88 $\pm$ 0.07 & 0.12 $\pm$ 0.08 & +0.77 & 0.14 & 0.32 &  \\
K--As & 31 & 173 & 154 (46) & 0.94 $\pm$ 0.06 & 0.14 $\pm$ 0.05 & +0.79 & 0.11 & 0.22 &  \\
\end{longtable}
}
\clearpage

\begin{figure*}[p]
\includegraphics[width=\textwidth,height=0.80\textheight,keepaspectratio]{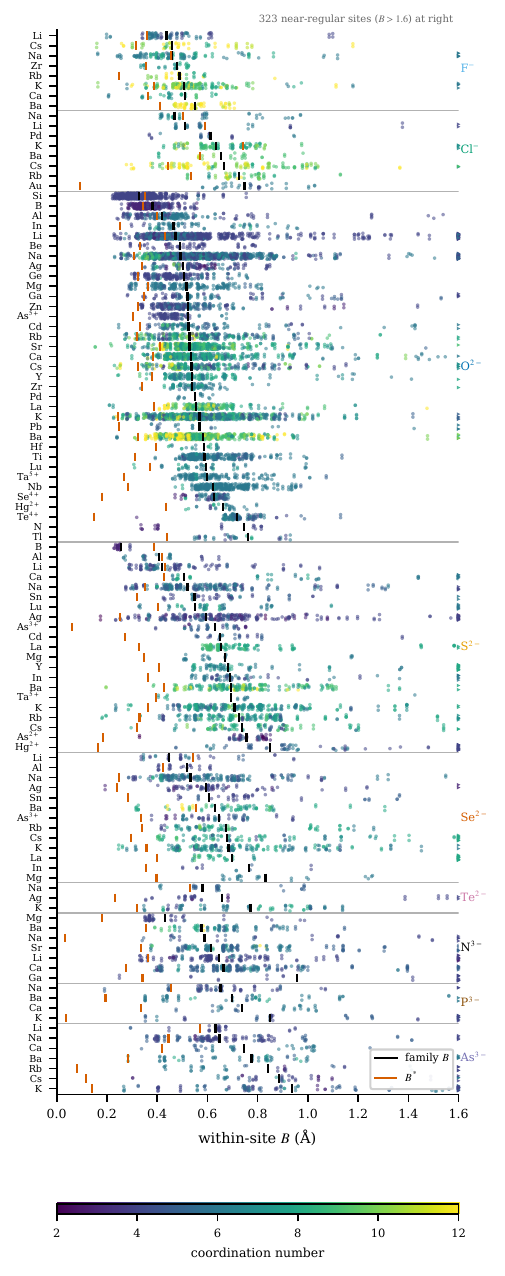}
\caption{The per-site softness $B$ of every crystallographically independent distorted cation site, one row per family of Table~\ref{tab:transfer} (cation at the left, anion at the right, ordered within an anion by the family $B$), colored by the site's coordination number. The black mark is the family $B$ and the orange mark $B^{*}$. Near-regular sites, whose slope is undetermined, are collected at the right edge.}
\label{fig:bsites}
\end{figure*}

\begin{figure*}[p]
\includegraphics[width=\textwidth,height=0.70\textheight,keepaspectratio]{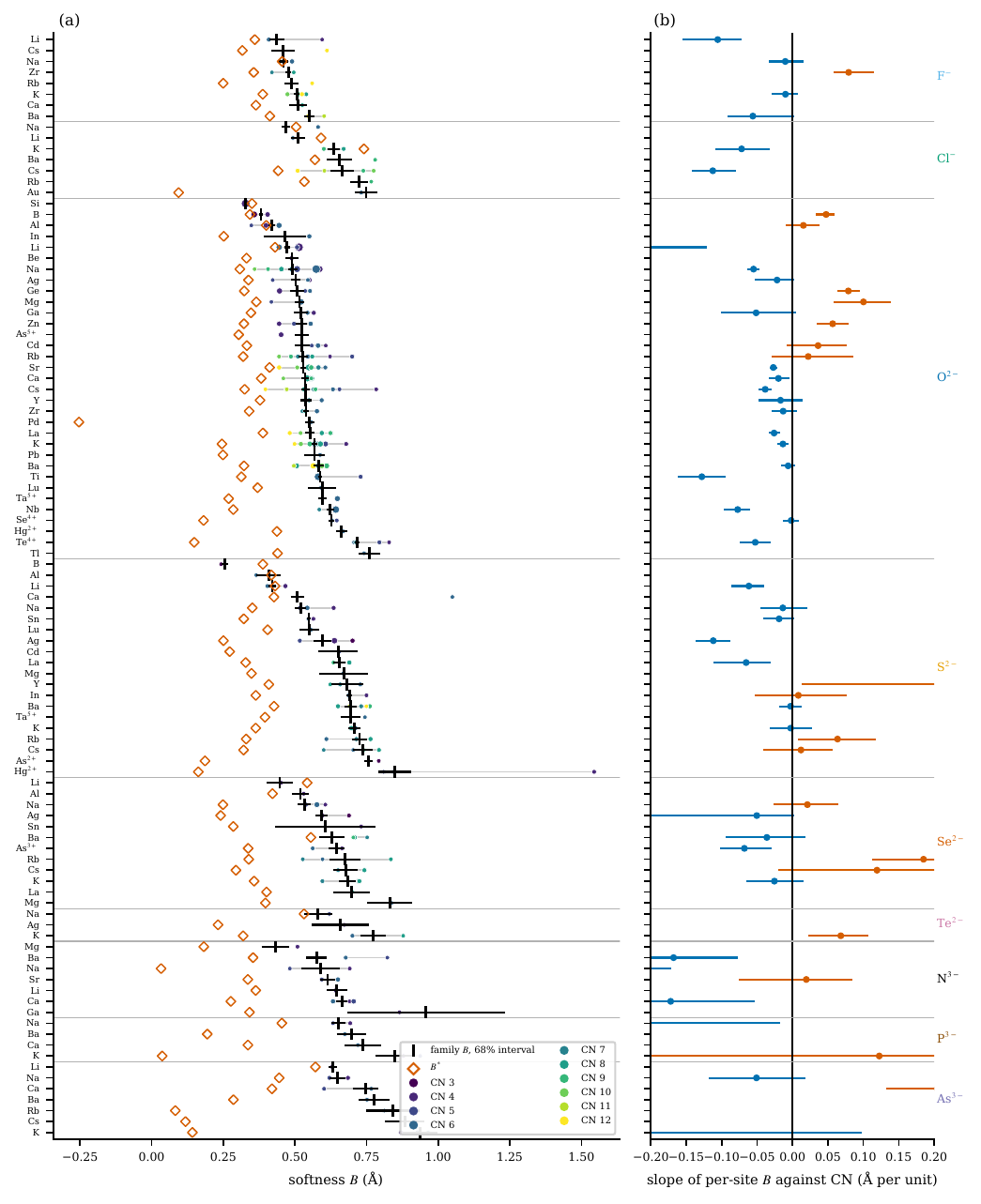}
\caption{The softness by coordination number, one row per family of Table~\ref{tab:transfer} with a coordination number populated by at least five independent distorted sites (cation at the left, anion at the right, ordered within an anion by $B$). (a)~The family $B$ with its 68\% bootstrap interval (black bar), the flux-pooled $B$ at each populated coordination number (dots, colored by coordination number, sized by the number of sites), and the coordination-branch softness $B^{*}$ (orange diamond). (b)~The slope of the per-site $B$ against coordination number with its 68\% bootstrap interval, positive (orange) or negative (blue).}
\label{fig:bcn}
\end{figure*}

\clearpage
\begin{longtable}{llrccc}
\caption{\captionbcn}\label{tab:bcn}\\
\toprule
family & CN & sites & median $B$ [IQR] (\AA) & pooled $B$ (\AA) & $B^{*}$ (\AA) \\
\midrule
\endfirsthead
\multicolumn{6}{c}{\tablename~\thetable{} (continued)}\\
\toprule
family & CN & sites & median $B$ [IQR] (\AA) & pooled $B$ (\AA) & $B^{*}$ (\AA) \\
\midrule
\endhead
\bottomrule
\endlastfoot
Ba--F & 11 & 6 & 0.56 [0.52, 0.64] & 0.60 [0.56, 0.63] & 0.41 \\
 & 12 & 12 & 0.56 [0.53, 0.60] & 0.55 [0.52, 0.57] &  \\
\midrule
Ca--F & 8 & 7 & 0.50 [0.46, 0.57] & 0.53 [0.50, 0.55] & 0.36 \\
\midrule
Cs--F & 12 & 7 & 0.53 [0.50, 0.66] & 0.61 [0.54, 0.71] & 0.32 \\
\midrule
K--F & 8 & 11 & 0.62 [0.52, 0.66] & 0.54 [0.50, 0.64] & 0.39 \\
 & 9 & 16 & 0.55 [0.50, 0.62] & 0.51 [0.49, 0.54] &  \\
 & 10 & 17 & 0.48 [0.46, 0.52] & 0.47 [0.46, 0.48] &  \\
 & 12 & 17 & 0.49 [0.42, 0.56] & 0.52 [0.49, 0.55] &  \\
\midrule
Li--F & 4 & 5 & 0.50 [0.42, 0.61] & 0.60 [0.52, 0.63] & 0.36 \\
 & 6 & 16 & 0.39 [0.37, 0.46] & 0.41 [0.39, 0.43] &  \\
\midrule
Na--F & 6 & 24 & 0.41 [0.32, 0.60] & 0.49 [0.43, 0.51] & 0.46 \\
 & 7 & 7 & 0.50 [0.41, 0.52] & 0.45 [0.42, 0.49] &  \\
 & 8 & 13 & 0.46 [0.42, 0.55] & 0.46 [0.44, 0.49] &  \\
\midrule
Rb--F & 12 & 6 & 0.49 [0.47, 0.56] & 0.56 [0.47, 0.67] & 0.25 \\
\midrule
Zr--F & 7 & 5 & 0.41 [0.35, 0.43] & 0.42 [0.39, 0.44] & 0.36 \\
 & 8 & 5 & 0.52 [0.49, 0.52] & 0.50 [0.49, 0.51] &  \\
\midrule
Au--Cl & 6 & 11 & 0.72 [0.67, 0.85] & 0.73 [0.70, 0.77] & 0.09 \\
\midrule
Ba--Cl & 9 & 6 & 0.78 [0.74, 0.89] & 0.78 [0.76, 0.81] & 0.57 \\
\midrule
Cs--Cl & 9 & 8 & 0.93 [0.78, 0.98] & 0.74 [0.63, 0.85] & 0.44 \\
 & 10 & 9 & 0.75 [0.63, 0.85] & 0.77 [0.72, 0.81] &  \\
 & 11 & 5 & 0.61 [0.52, 0.73] & 0.60 [0.55, 0.68] &  \\
 & 12 & 16 & 0.52 [0.41, 0.68] & 0.51 [0.48, 0.53] &  \\
\midrule
K--Cl & 8 & 12 & 0.71 [0.61, 0.75] & 0.67 [0.64, 0.70] & 0.74 \\
 & 9 & 12 & 0.60 [0.58, 0.63] & 0.60 [0.59, 0.62] &  \\
\midrule
Li--Cl & 6 & 5 & 0.52 [0.47, 0.57] & 0.49 [0.48, 0.52] & 0.59 \\
\midrule
Na--Cl & 6 & 9 & 0.60 [0.58, 0.81] & 0.58 [0.52, 0.68] & 0.50 \\
\midrule
\midrule
Rb--Cl & 9 & 5 & 0.79 [0.75, 0.82] & 0.77 [0.70, 0.84] & 0.53 \\
\midrule
Ag--O & 4 & 19 & 0.57 [0.53, 0.62] & 0.55 [0.53, 0.58] & 0.34 \\
 & 5 & 7 & 0.42 [0.39, 0.54] & 0.42 [0.39, 0.45] &  \\
 & 6 & 7 & 0.60 [0.52, 0.63] & 0.55 [0.50, 0.59] &  \\
\midrule
Al--O & 4 & 40 & 0.37 [0.29, 0.65] & 0.40 [0.38, 0.42] & 0.40 \\
 & 5 & 5 & 0.35 [0.32, 0.38] & 0.35 [0.34, 0.36] &  \\
 & 6 & 50 & 0.46 [0.43, 0.49] & 0.44 [0.44, 0.46] &  \\
\midrule
As$^{5+}$--O & 4 & 36 & 0.44 [0.42, 0.47] & 0.45 [0.45, 0.46] & 0.30 \\
\midrule
B--O & 3 & 71 & 0.36 [0.34, 0.38] & 0.36 [0.35, 0.36] & 0.34 \\
 & 4 & 33 & 0.40 [0.35, 0.43] & 0.40 [0.40, 0.41] &  \\
\midrule
Ba--O & 7 & 12 & 0.68 [0.58, 0.77] & 0.57 [0.54, 0.63] & 0.32 \\
 & 8 & 40 & 0.54 [0.47, 0.59] & 0.51 [0.49, 0.53] &  \\
 & 9 & 43 & 0.61 [0.56, 0.69] & 0.61 [0.58, 0.64] &  \\
 & 10 & 38 & 0.57 [0.52, 0.68] & 0.59 [0.58, 0.61] &  \\
 & 11 & 10 & 0.52 [0.45, 0.53] & 0.50 [0.47, 0.52] &  \\
 & 12 & 57 & 0.51 [0.47, 0.60] & 0.56 [0.53, 0.60] &  \\
\midrule
Be--O & 4 & 18 & 0.52 [0.45, 0.58] & 0.49 [0.47, 0.51] & 0.33 \\
\midrule
Ca--O & 6 & 49 & 0.58 [0.50, 0.68] & 0.56 [0.53, 0.59] & 0.38 \\
 & 7 & 32 & 0.56 [0.48, 0.68] & 0.56 [0.53, 0.60] &  \\
 & 8 & 81 & 0.54 [0.50, 0.63] & 0.54 [0.53, 0.56] &  \\
 & 9 & 6 & 0.54 [0.46, 0.64] & 0.56 [0.51, 0.62] &  \\
 & 10 & 5 & 0.51 [0.44, 0.55] & 0.46 [0.44, 0.49] &  \\
\midrule
Cd--O & 4 & 5 & 0.56 [0.45, 0.63] & 0.61 [0.57, 0.65] & 0.33 \\
 & 5 & 8 & 0.54 [0.43, 0.61] & 0.56 [0.51, 0.59] &  \\
 & 6 & 29 & 0.54 [0.47, 0.67] & 0.58 [0.56, 0.61] &  \\
\midrule
Cs--O & 4 & 8 & 0.72 [0.60, 0.95] & 0.78 [0.65, 0.91] & 0.32 \\
 & 5 & 9 & 0.67 [0.60, 0.78] & 0.66 [0.63, 0.70] &  \\
 & 6 & 13 & 0.57 [0.54, 0.75] & 0.63 [0.59, 0.67] &  \\
 & 7 & 11 & 0.54 [0.51, 0.64] & 0.57 [0.54, 0.61] &  \\
 & 8 & 23 & 0.53 [0.48, 0.65] & 0.53 [0.51, 0.55] &  \\
 & 9 & 10 & 0.59 [0.51, 0.62] & 0.57 [0.55, 0.60] &  \\
 & 10 & 8 & 0.42 [0.33, 0.60] & 0.47 [0.42, 0.53] &  \\
 & 11 & 6 & 0.47 [0.45, 0.49] & 0.47 [0.45, 0.49] &  \\
 & 12 & 5 & 0.39 [0.38, 0.46] & 0.40 [0.38, 0.42] &  \\
\midrule
Ga--O & 4 & 18 & 0.52 [0.43, 0.81] & 0.57 [0.50, 0.63] & 0.35 \\
 & 6 & 15 & 0.61 [0.53, 0.64] & 0.54 [0.54, 0.60] &  \\
\midrule
Ge--O & 4 & 53 & 0.44 [0.40, 0.49] & 0.45 [0.44, 0.46] & 0.32 \\
 & 5 & 6 & 0.50 [0.48, 0.54] & 0.54 [0.52, 0.55] &  \\
 & 6 & 14 & 0.60 [0.57, 0.69] & 0.55 [0.54, 0.57] &  \\
\midrule
\midrule
Hg$^{2+}$--O & 6 & 7 & 0.69 [0.66, 0.73] & 0.67 [0.63, 0.71] & 0.44 \\
\midrule
In--O & 6 & 23 & 0.48 [0.44, 0.65] & 0.55 [0.51, 0.59] & 0.25 \\
\midrule
K--O & 4 & 14 & 0.54 [0.42, 0.77] & 0.68 [0.62, 0.73] & 0.25 \\
 & 5 & 41 & 0.61 [0.55, 0.75] & 0.61 [0.58, 0.64] &  \\
 & 6 & 70 & 0.60 [0.51, 0.74] & 0.57 [0.55, 0.59] &  \\
 & 7 & 38 & 0.54 [0.47, 0.66] & 0.58 [0.56, 0.60] &  \\
 & 8 & 50 & 0.55 [0.48, 0.68] & 0.59 [0.57, 0.61] &  \\
 & 9 & 39 & 0.51 [0.45, 0.72] & 0.55 [0.52, 0.59] &  \\
 & 10 & 21 & 0.54 [0.46, 0.63] & 0.52 [0.49, 0.56] &  \\
 & 12 & 8 & 0.49 [0.48, 0.50] & 0.50 [0.49, 0.51] &  \\
\midrule
La--O & 7 & 10 & 0.60 [0.54, 0.74] & 0.63 [0.61, 0.65] & 0.39 \\
 & 8 & 25 & 0.58 [0.54, 0.64] & 0.59 [0.58, 0.61] &  \\
 & 9 & 16 & 0.65 [0.55, 0.68] & 0.62 [0.60, 0.66] &  \\
 & 10 & 10 & 0.54 [0.52, 0.57] & 0.52 [0.51, 0.54] &  \\
 & 12 & 16 & 0.60 [0.44, 0.62] & 0.48 [0.46, 0.51] &  \\
\midrule
Li--O & 4 & 154 & 0.55 [0.44, 0.71] & 0.51 [0.50, 0.53] & 0.43 \\
 & 5 & 21 & 0.47 [0.39, 0.54] & 0.51 [0.48, 0.54] &  \\
 & 6 & 56 & 0.46 [0.40, 0.51] & 0.45 [0.44, 0.45] &  \\
\midrule
Lu--O & 6 & 8 & 0.58 [0.51, 0.65] & 0.59 [0.56, 0.62] & 0.37 \\
\midrule
Mg--O & 5 & 5 & 0.42 [0.40, 0.43] & 0.42 [0.39, 0.44] & 0.37 \\
 & 6 & 62 & 0.52 [0.43, 0.64] & 0.52 [0.51, 0.54] &  \\
\midrule
\midrule
Na--O & 4 & 74 & 0.58 [0.49, 0.74] & 0.59 [0.57, 0.61] & 0.31 \\
 & 5 & 83 & 0.52 [0.44, 0.64] & 0.51 [0.47, 0.55] &  \\
 & 6 & 146 & 0.61 [0.53, 0.70] & 0.57 [0.56, 0.59] &  \\
 & 7 & 33 & 0.48 [0.40, 0.58] & 0.49 [0.47, 0.51] &  \\
 & 8 & 32 & 0.45 [0.39, 0.54] & 0.45 [0.44, 0.47] &  \\
 & 9 & 8 & 0.40 [0.37, 0.43] & 0.41 [0.39, 0.42] &  \\
 & 10 & 5 & 0.36 [0.36, 0.37] & 0.36 [0.36, 0.37] &  \\
\midrule
Nb--O & 6 & 101 & 0.67 [0.61, 0.74] & 0.64 [0.63, 0.66] & 0.28 \\
 & 7 & 6 & 0.58 [0.56, 0.61] & 0.59 [0.57, 0.60] &  \\
\midrule
Pb--O & 6 & 10 & 0.59 [0.50, 0.88] & 0.59 [0.54, 0.63] & 0.25 \\
\midrule
Pd--O & 6 & 8 & 0.57 [0.50, 0.59] & 0.56 [0.54, 0.57] & $-$0.25 \\
\midrule
Rb--O & 4 & 6 & 0.58 [0.47, 0.63] & 0.62 [0.58, 0.69] & 0.32 \\
 & 5 & 16 & 0.69 [0.59, 0.79] & 0.70 [0.67, 0.74] &  \\
 & 6 & 22 & 0.58 [0.45, 0.64] & 0.55 [0.52, 0.58] &  \\
 & 7 & 14 & 0.48 [0.37, 0.59] & 0.51 [0.47, 0.55] &  \\
 & 8 & 18 & 0.62 [0.50, 0.74] & 0.56 [0.51, 0.61] &  \\
 & 9 & 13 & 0.47 [0.38, 0.55] & 0.49 [0.44, 0.54] &  \\
 & 10 & 9 & 0.52 [0.35, 0.56] & 0.44 [0.41, 0.49] &  \\
\midrule
Se$^{4+}$--O & 4 & 8 & 0.63 [0.62, 0.65] & 0.63 [0.61, 0.64] & 0.18 \\
 & 5 & 10 & 0.64 [0.63, 0.66] & 0.65 [0.64, 0.65] &  \\
 & 6 & 11 & 0.64 [0.61, 0.66] & 0.63 [0.62, 0.64] &  \\
\midrule
Si--O & 4 & 176 & 0.33 [0.30, 0.37] & 0.33 [0.32, 0.33] & 0.35 \\
\midrule
Sr--O & 6 & 14 & 0.65 [0.38, 0.96] & 0.61 [0.55, 0.67] & 0.41 \\
 & 7 & 25 & 0.55 [0.53, 0.65] & 0.58 [0.53, 0.62] &  \\
 & 8 & 74 & 0.55 [0.50, 0.61] & 0.55 [0.54, 0.56] &  \\
 & 9 & 30 & 0.58 [0.51, 0.66] & 0.56 [0.54, 0.57] &  \\
 & 10 & 14 & 0.49 [0.44, 0.53] & 0.51 [0.48, 0.54] &  \\
 & 12 & 22 & 0.43 [0.42, 0.48] & 0.45 [0.44, 0.45] &  \\
\midrule
Ta$^{5+}$--O & 6 & 47 & 0.62 [0.55, 0.75] & 0.65 [0.63, 0.67] & 0.27 \\
\midrule
Te$^{4+}$--O & 4 & 7 & 0.80 [0.78, 0.84] & 0.83 [0.80, 0.86] & 0.15 \\
 & 5 & 14 & 0.75 [0.71, 0.79] & 0.79 [0.76, 0.83] &  \\
 & 6 & 17 & 0.70 [0.69, 0.73] & 0.71 [0.70, 0.71] &  \\
 & 7 & 5 & 0.70 [0.69, 0.71] & 0.71 [0.69, 0.73] &  \\
\midrule
Ti--O & 4 & 5 & 0.51 [0.51, 0.63] & 0.58 [0.52, 0.63] & 0.31 \\
 & 5 & 16 & 0.72 [0.64, 0.76] & 0.73 [0.71, 0.76] &  \\
 & 6 & 106 & 0.59 [0.54, 0.68] & 0.58 [0.57, 0.59] &  \\
\midrule
Tl--O & 6 & 7 & 0.68 [0.59, 0.70] & 0.74 [0.63, 0.81] & 0.44 \\
\midrule
Y--O & 6 & 18 & 0.58 [0.49, 0.67] & 0.59 [0.56, 0.63] & 0.38 \\
 & 7 & 18 & 0.53 [0.48, 0.64] & 0.55 [0.53, 0.57] &  \\
 & 8 & 18 & 0.54 [0.45, 0.61] & 0.54 [0.51, 0.57] &  \\
\midrule
Zn--O & 4 & 28 & 0.45 [0.42, 0.48] & 0.45 [0.43, 0.46] & 0.32 \\
 & 5 & 15 & 0.49 [0.46, 0.56] & 0.50 [0.46, 0.53] &  \\
 & 6 & 25 & 0.53 [0.50, 0.61] & 0.56 [0.54, 0.57] &  \\
\midrule
Zr--O & 6 & 25 & 0.58 [0.54, 0.61] & 0.58 [0.55, 0.60] & 0.34 \\
 & 7 & 8 & 0.54 [0.49, 0.56] & 0.53 [0.51, 0.54] &  \\
 & 8 & 6 & 0.54 [0.49, 0.57] & 0.54 [0.52, 0.55] &  \\
\midrule
Ag--S & 3 & 23 & 0.65 [0.49, 0.88] & 0.70 [0.66, 0.72] & 0.25 \\
 & 4 & 61 & 0.67 [0.51, 0.92] & 0.64 [0.61, 0.66] &  \\
 & 5 & 14 & 0.54 [0.48, 0.59] & 0.52 [0.50, 0.54] &  \\
\midrule
Al--S & 6 & 5 & 0.35 [0.34, 0.40] & 0.36 [0.31, 0.41] & 0.42 \\
\midrule
As$^{2+}$--S & 3 & 6 & 0.81 [0.75, 0.83] & 0.79 [0.77, 0.81] & 0.19 \\
\midrule
\midrule
B--S & 3 & 6 & 0.24 [0.24, 0.25] & 0.24 [0.24, 0.25] & 0.39 \\
\midrule
Ba--S & 7 & 15 & 0.77 [0.71, 0.91] & 0.73 [0.70, 0.77] & 0.43 \\
 & 8 & 20 & 0.65 [0.58, 0.76] & 0.65 [0.62, 0.68] &  \\
 & 9 & 18 & 0.82 [0.73, 0.99] & 0.76 [0.72, 0.81] &  \\
 & 11 & 5 & 0.71 [0.61, 0.74] & 0.69 [0.64, 0.72] &  \\
 & 12 & 5 & 0.74 [0.71, 0.80] & 0.75 [0.71, 0.77] &  \\
\midrule
Ca--S & 6 & 6 & 1.24 [0.97, 1.56] & 1.05 [0.90, 1.27] & 0.43 \\
\midrule
Cd--S & 5 & 6 & 0.68 [0.62, 0.72] & 0.65 [0.61, 0.72] & 0.27 \\
\midrule
Cs--S & 6 & 5 & 0.67 [0.63, 0.80] & 0.70 [0.61, 0.87] & 0.32 \\
 & 7 & 5 & 0.61 [0.48, 0.69] & 0.60 [0.53, 0.68] &  \\
 & 8 & 13 & 0.80 [0.70, 1.17] & 0.79 [0.74, 0.85] &  \\
 & 9 & 8 & 0.86 [0.69, 0.92] & 0.74 [0.69, 0.80] &  \\
\midrule
Hg$^{2+}$--S & 4 & 12 & 2.22 [1.33, 3.55] & 1.54 [1.38, 1.95] & 0.16 \\
 & 5 & 8 & 0.80 [0.74, 0.90] & 0.81 [0.75, 0.86] &  \\
 & 6 & 8 & 0.96 [0.86, 1.10] & 0.84 [0.76, 0.98] &  \\
\midrule
In--S & 4 & 9 & 0.70 [0.64, 0.84] & 0.75 [0.71, 0.80] & 0.36 \\
 & 6 & 22 & 0.68 [0.64, 0.78] & 0.69 [0.68, 0.70] &  \\
\midrule
K--S & 6 & 23 & 0.69 [0.57, 0.93] & 0.71 [0.67, 0.76] & 0.36 \\
 & 7 & 17 & 0.74 [0.67, 0.81] & 0.70 [0.67, 0.73] &  \\
 & 8 & 27 & 0.78 [0.57, 0.95] & 0.70 [0.67, 0.73] &  \\
\midrule
La--S & 8 & 22 & 0.73 [0.66, 0.79] & 0.69 [0.67, 0.71] & 0.33 \\
 & 9 & 6 & 0.63 [0.60, 0.64] & 0.63 [0.61, 0.66] &  \\
\midrule
Li--S & 4 & 20 & 0.47 [0.36, 0.64] & 0.47 [0.43, 0.51] & 0.43 \\
 & 6 & 10 & 0.39 [0.36, 0.40] & 0.40 [0.39, 0.41] &  \\
\midrule
Lu--S & 6 & 12 & 0.63 [0.56, 0.82] & 0.56 [0.53, 0.59] & 0.40 \\
\midrule
Mg--S & 6 & 6 & 0.59 [0.53, 0.66] & 0.67 [0.59, 0.76] & 0.35 \\
\midrule
Na--S & 4 & 17 & 0.63 [0.45, 0.74] & 0.64 [0.59, 0.68] & 0.35 \\
 & 5 & 19 & 0.52 [0.44, 0.64] & 0.52 [0.49, 0.55] &  \\
 & 6 & 31 & 0.60 [0.49, 0.67] & 0.54 [0.51, 0.58] &  \\
\midrule
Rb--S & 6 & 14 & 0.55 [0.52, 0.65] & 0.61 [0.57, 0.66] & 0.33 \\
 & 7 & 19 & 0.78 [0.63, 0.85] & 0.71 [0.67, 0.76] &  \\
 & 8 & 20 & 0.70 [0.64, 0.87] & 0.76 [0.72, 0.81] &  \\
\midrule
Sn--S & 4 & 9 & 0.63 [0.55, 0.68] & 0.56 [0.53, 0.61] & 0.32 \\
 & 6 & 6 & 0.62 [0.55, 0.69] & 0.55 [0.55, 0.55] &  \\
\midrule
Ta$^{5+}$--S & 6 & 5 & 0.75 [0.75, 0.76] & 0.74 [0.74, 0.76] & 0.40 \\
\midrule
Y--S & 6 & 13 & 0.67 [0.62, 0.79] & 0.73 [0.69, 0.78] & 0.41 \\
 & 7 & 12 & 0.66 [0.60, 0.69] & 0.66 [0.64, 0.68] &  \\
 & 8 & 6 & 0.66 [0.57, 3.32] & 0.62 [0.58, 0.71] &  \\
\midrule
Ag--Se & 3 & 13 & 0.56 [0.47, 0.68] & 0.69 [0.65, 1.01] & 0.24 \\
 & 4 & 18 & 0.60 [0.58, 0.72] & 0.60 [0.57, 0.62] &  \\
\midrule
Al--Se & 4 & 10 & 0.52 [0.45, 0.56] & 0.53 [0.50, 0.56] & 0.42 \\
\midrule
As$^{3+}$--Se & 4 & 8 & 0.63 [0.56, 0.78] & 0.66 [0.61, 0.71] & 0.34 \\
 & 6 & 9 & 0.52 [0.50, 0.63] & 0.56 [0.52, 0.60] &  \\
\midrule
Ba--Se & 7 & 5 & 0.75 [0.67, 0.76] & 0.75 [0.69, 0.81] & 0.56 \\
 & 8 & 14 & 0.73 [0.64, 0.83] & 0.71 [0.64, 0.81] &  \\
 & 9 & 5 & 0.68 [0.60, 0.85] & 0.70 [0.65, 0.77] &  \\
\midrule
Cs--Se & 7 & 10 & 0.63 [0.58, 0.90] & 0.65 [0.63, 0.69] & 0.29 \\
 & 8 & 16 & 0.79 [0.66, 1.45] & 0.74 [0.68, 0.82] &  \\
\midrule
\midrule
K--Se & 5 & 5 & 0.66 [0.59, 0.80] & --- & 0.36 \\
 & 6 & 23 & 0.76 [0.58, 1.04] & 0.72 [0.66, 0.79] &  \\
 & 7 & 12 & 0.61 [0.58, 0.71] & 0.60 [0.57, 0.63] &  \\
 & 8 & 18 & 0.69 [0.55, 0.89] & 0.72 [0.68, 0.77] &  \\
\midrule
La--Se & 8 & 6 & 0.68 [0.62, 1.77] & 0.70 [0.66, 0.77] & 0.40 \\
\midrule
Li--Se & 4 & 7 & 0.49 [0.41, 0.59] & 0.45 [0.41, 0.51] & 0.54 \\
\midrule
Mg--Se & 6 & 8 & 0.77 [0.71, 0.95] & 0.84 [0.78, 0.91] & 0.40 \\
\midrule
Na--Se & 4 & 8 & 0.58 [0.36, 0.79] & 0.61 [0.52, 0.68] & 0.25 \\
 & 5 & 18 & 0.53 [0.45, 0.72] & 0.54 [0.51, 0.57] &  \\
 & 6 & 43 & 0.58 [0.46, 0.69] & 0.58 [0.55, 0.60] &  \\
\midrule
Rb--Se & 6 & 5 & 0.59 [0.56, 0.70] & 0.60 [0.57, 0.64] & 0.34 \\
 & 7 & 8 & 0.53 [0.49, 0.59] & 0.53 [0.51, 0.54] &  \\
 & 8 & 9 & 0.71 [0.62, 1.01] & 0.83 [0.75, 0.92] &  \\
\midrule
Sn--Se & 4 & 10 & 0.73 [0.70, 0.78] & 0.73 [0.70, 0.76] & 0.29 \\
\midrule
Ag--Te & 4 & 8 & 0.56 [0.51, 1.53] & 0.67 [0.58, 0.82] & 0.23 \\
\midrule
K--Te & 6 & 12 & 0.67 [0.50, 0.94] & 0.70 [0.64, 0.76] & 0.32 \\
 & 8 & 5 & 0.85 [0.80, 0.93] & 0.88 [0.83, 0.91] &  \\
\midrule
Na--Te & 5 & 6 & 0.64 [0.57, 0.83] & 0.62 [0.59, 0.69] & 0.53 \\
\midrule
Ba--N & 5 & 5 & 0.86 [0.71, 0.95] & 0.82 [0.76, 0.91] & 0.35 \\
 & 6 & 11 & 0.71 [0.62, 0.75] & 0.68 [0.65, 0.70] &  \\
\midrule
Ca--N & 4 & 5 & 0.83 [0.41, 0.90] & 0.69 [0.50, 0.82] & 0.28 \\
 & 5 & 28 & 0.72 [0.63, 0.93] & 0.71 [0.68, 0.73] &  \\
 & 6 & 24 & 0.68 [0.63, 0.81] & 0.63 [0.60, 0.67] &  \\
\midrule
Ga--N & 4 & 8 & 1.16 [0.62, 1.83] & 0.86 [0.65, 1.24] & 0.34 \\
\midrule
Li--N & 4 & 29 & 0.63 [0.58, 0.78] & 0.64 [0.60, 0.69] & 0.36 \\
\midrule
Mg--N & 4 & 8 & 0.46 [0.37, 0.54] & 0.51 [0.43, 0.56] & 0.18 \\
\midrule
Na--N & 4 & 8 & 0.90 [0.60, 1.20] & 0.69 [0.61, 0.83] & 0.03 \\
 & 5 & 5 & 0.47 [0.46, 0.67] & 0.48 [0.46, 0.52] &  \\
\midrule
Sr--N & 5 & 13 & 0.70 [0.51, 1.14] & 0.59 [0.56, 0.65] & 0.34 \\
 & 6 & 23 & 0.70 [0.63, 0.89] & 0.65 [0.62, 0.69] &  \\
\midrule
Ba--P & 6 & 10 & 0.74 [0.45, 1.09] & 0.67 [0.61, 0.72] & 0.19 \\
\midrule
Ca--P & 6 & 7 & 0.95 [0.78, 0.98] & 0.72 [0.69, 0.83] & 0.34 \\
\midrule
K--P & 4 & 7 & 1.44 [0.86, 2.02] & 0.88 [0.86, 0.99] & 0.04 \\
 & 5 & 7 & 1.00 [0.81, 1.23] & 0.94 [0.89, 1.32] &  \\
\midrule
Na--P & 4 & 12 & 0.72 [0.65, 0.86] & 0.69 [0.67, 0.73] & 0.45 \\
 & 5 & 9 & 0.57 [0.45, 0.63] & 0.63 [0.60, 0.67] &  \\
\midrule
Ba--As & 6 & 10 & 0.75 [0.53, 0.85] & 0.75 [0.70, 0.82] & 0.29 \\
\midrule
Ca--As & 5 & 9 & 0.58 [0.56, 0.71] & 0.60 [0.58, 0.63] & 0.42 \\
 & 6 & 10 & 0.93 [0.75, 0.99] & 0.77 [0.73, 0.83] &  \\
\midrule
Cs--As & 4 & 7 & 0.95 [0.92, 1.25] & 0.92 [0.87, 1.02] & 0.12 \\
\midrule
K--As & 4 & 14 & 0.78 [0.72, 2.47] & 0.87 [0.78, 1.04] & 0.14 \\
 & 5 & 23 & 1.00 [0.75, 1.23] & 0.93 [0.85, 1.03] &  \\
 & 6 & 7 & 0.77 [0.62, 1.23] & 0.97 [0.78, 1.18] &  \\
\midrule
Li--As & 4 & 6 & 0.64 [0.61, 0.67] & 0.63 [0.61, 0.65] & 0.57 \\
\midrule
Na--As & 4 & 18 & 0.66 [0.61, 0.80] & 0.69 [0.65, 0.72] & 0.45 \\
 & 5 & 14 & 0.58 [0.48, 0.80] & 0.62 [0.59, 0.65] &  \\
\midrule
Rb--As & 6 & 7 & 0.58 [0.55, 0.65] & 0.81 [0.60, 1.00] & 0.08 \\
\end{longtable}
 \clearpage
{\footnotesize\setlength{\tabcolsep}{2.3pt}%
\begin{longtable}{lcrccccc}
\caption{\captionpaulingcorpus}\label{tab:paulingcorpus}\\
\toprule
pair & $z/n$ & faces (sites) & Pauling, $d_B$ & Pauling, $\langle d\rangle$ & exponential, $B$ & equal split & closure, $\langle d\rangle$ mean/max (\%) \\
\midrule
\endfirsthead
\multicolumn{8}{c}{\tablename~\thetable{} (continued)}\\
\toprule
pair & $z/n$ & faces (sites) & Pauling, $d_B$ & Pauling, $\langle d\rangle$ & exponential, $B$ & equal split & closure, $\langle d\rangle$ mean/max (\%) \\
\midrule
\endhead
\bottomrule
\endlastfoot
\multicolumn{8}{l}{\emph{F$^{-}$} (12 families)} \\
Li--F$^{\dagger}$ & 0.17--0.25 & 734 (141) & 6.2 & 7.4 & 7.4 & 19 & 1.4/7.1 \\
Na--F$^{\dagger}$ & 0.08--0.25 & 1070 (160) & 7.5 & 9.2 & 10 & 24 & 2.0/15 \\
Mg--F & 0.33 & 330 (55) & 2.4 & 2.6 & 4.3 & 9.6 & 0.39/2.9 \\
Al--F & 0.5 & 834 (139) & 1.5 & 1.5 & 2.2 & 7.9 & 0.30/1.1 \\
K--F$^{\dagger}$ & 0.06--0.17 & 2343 (249) & 11 & 14 & 15 & 33 & 3.4/34 \\
Ca--F$^{\dagger}$ & 0.22--0.33 & 388 (55) & 4.5 & 5.0 & 5.8 & 18 & 1.2/6.5 \\
Zn--F & 0.25--0.33 & 380 (62) & 2.6 & 18 & 4.0 & 36 & 2.4/35 \\
Rb--F$^{\dagger}$ & 0.08--0.12 & 646 (60) & 8.5 & 11 & 21 & 25 & 2.2/11 \\
Zr--F$^{\dagger}$ & 0.5--0.67 & 216 (29) & 2.2 & 2.6 & 2.8 & 15 & 0.94/2.9 \\
Cs--F$^{\dagger}$ & 0.07--0.17 & 747 (67) & 12 & 14 & 16 & 30 & 2.9/13 \\
Ba--F$^{\dagger}$ & 0.15--0.18 & 676 (57) & 9.0 & 12 & 9.3 & 37 & 5.3/10 \\
Au--F & 0.38--0.5 & 128 (18) & 15 & 48 & 16 & 98 & 19/26 \\
\midrule
\multicolumn{8}{l}{\emph{Cl$^{-}$} (16 families)} \\
Li--Cl$^{\dagger}$ & 0.17--0.25 & 278 (52) & 3.3 & 4.0 & 8.3 & 16 & 0.99/4.0 \\
Na--Cl$^{\dagger}$ & 0.12--0.2 & 408 (65) & 6.2 & 9.3 & 9.9 & 31 & 3.0/11 \\
Al--Cl & 0.5--0.75 & 212 (52) & 1.8 & 2.0 & 2.1 & 7.3 & 0.23/1.1 \\
K--Cl$^{\dagger}$ & 0.08--0.17 & 1160 (132) & 9.0 & 11 & 13 & 29 & 3.2/10 \\
Sc--Cl & 0.5 & 162 (27) & 1.4 & 1.7 & 1.6 & 10 & 0.48/1.6 \\
Se$^{4+}$--Cl & 0.67 & 144 (24) & 2.5 & 9.1 & 3.9 & 36 & 4.5/6.3 \\
Rb--Cl$^{\dagger}$ & 0.08--0.17 & 742 (76) & 5.7 & 6.8 & 11 & 21 & 1.7/6.8 \\
Nb--Cl & 0.83 & 168 (28) & 1.4 & 2.8 & 1.4 & 16 & 1.0/2.1 \\
Mo--Cl & 0.4 & 180 (36) & 22 & 22 & 22 & 22 & 0.013/0.037 \\
Pd--Cl$^{\dagger}$ & 0.33--0.5 & 172 (41) & 16 & 26 & 16 & 59 & 5.9/27 \\
Ag--Cl & 0.14--0.25 & 116 (21) & 7.6 & 9.9 & 11 & 32 & 4.0/13 \\
Te$^{4+}$--Cl & 0.57--0.67 & 104 (17) & 3.3 & 9.3 & 5.6 & 34 & 3.9/7.6 \\
Cs--Cl$^{\dagger}$ & 0.08--0.17 & 1754 (165) & 9.2 & 10 & 15 & 25 & 2.1/8.6 \\
Ba--Cl$^{\dagger}$ & 0.18--0.25 & 339 (38) & 8.1 & 9.6 & 8.3 & 27 & 2.8/7.0 \\
Au--Cl$^{\dagger}$ & 0.5--3 & 206 (41) & 23 & 47 & 30 & 94 & 14/21 \\
Hg$^{2+}$--Cl & 0.25--0.5 & 148 (25) & 7.3 & 19 & 8.8 & 49 & 7.9/20 \\
\midrule
\multicolumn{8}{l}{\emph{O$^{2-}$} (44 families)} \\
Li--O$^{\dagger}$ & 0.17--0.33 & 4618 (1052) & 6.3 & 6.8 & 9.4 & 17 & 1.0/8.1 \\
Be--O$^{\dagger}$ & 0.5--0.67 & 498 (128) & 2.5 & 2.5 & 4.4 & 7.4 & 0.24/1.1 \\
B--O$^{\dagger}$ & 0.75 & 2782 (858) & 2.2 & 2.1 & 3.1 & 8.5 & 0.40/1.9 \\
N--O$^{\dagger}$ & 1.25--2.5 & 267 (84) & 7.8 & 122 & 6.7 & 48 & 72/750 \\
Na--O$^{\dagger}$ & 0.08--0.5 & 6834 (1223) & 13 & 15 & 16 & 30 & 2.6/25 \\
Mg--O$^{\dagger}$ & 0.33--0.5 & 1734 (304) & 3.7 & 3.8 & 6.5 & 13 & 0.71/4.7 \\
Al--O$^{\dagger}$ & 0.5--0.75 & 2350 (502) & 3.1 & 3.6 & 4.5 & 12 & 0.46/5.2 \\
Si--O$^{\dagger}$ & 0.67 & 4866 (1205) & 2.1 & 2.1 & 6.2 & 2.6 & 0.077/0.21 \\
P$^{5+}$--O & 1.25 & 4552 (1138) & 1.9 & 2.6 & 2.7 & 11 & 0.79/4.1 \\
S--O & 1.5 & 764 (191) & 2.1 & 6.4 & 2.9 & 16 & 1.9/33 \\
K--O$^{\dagger}$ & 0.08--0.25 & 6189 (893) & 15 & 17 & 20 & 32 & 2.8/20 \\
Ca--O$^{\dagger}$ & 0.17--0.5 & 3816 (529) & 8.2 & 9.8 & 10 & 26 & 2.3/12 \\
Sc--O & 0.38--0.5 & 654 (107) & 3.4 & 4.1 & 4.9 & 14 & 0.72/6.4 \\
Ti--O$^{\dagger}$ & 0.57--0.8 & 2360 (409) & 3.8 & 6.3 & 5.1 & 22 & 1.7/9.5 \\
Zn--O$^{\dagger}$ & 0.33--0.67 & 1898 (424) & 5.6 & 8.2 & 10 & 25 & 1.6/20 \\
Ga--O$^{\dagger}$ & 0.5--0.75 & 1007 (212) & 3.7 & 7.9 & 6.9 & 18 & 0.67/16 \\
Ge--O$^{\dagger}$ & 0.5--0.8 & 1744 (400) & 5.0 & 17 & 11 & 31 & 1.6/21 \\
As$^{3+}$--O & 0.6--0.75 & 122 (34) & 17 & 79 & 13 & 122 & 14/17 \\
As$^{5+}$--O$^{\dagger}$ & 0.83--1.25 & 816 (192) & 4.6 & 10 & 5.1 & 15 & 0.53/5.1 \\
Se$^{4+}$--O$^{\dagger}$ & 0.5--1.33 & 714 (150) & 14 & 76 & 12 & 111 & 7.4/22 \\
Se$^{6+}$--O & 1.2--1.5 & 214 (52) & 2.5 & 2.8 & 2.5 & 8.2 & 0.33/2.5 \\
Rb--O$^{\dagger}$ & 0.08--0.25 & 2221 (319) & 17 & 20 & 21 & 38 & 3.9/16 \\
Sr--O$^{\dagger}$ & 0.17--0.4 & 4948 (590) & 9.2 & 12 & 11 & 31 & 3.2/16 \\
Y--O$^{\dagger}$ & 0.33--0.5 & 1420 (205) & 5.9 & 6.5 & 7.5 & 18 & 1.2/7.9 \\
Zr--O$^{\dagger}$ & 0.5--0.8 & 982 (157) & 3.1 & 3.5 & 5.1 & 13 & 0.71/7.2 \\
Nb--O$^{\dagger}$ & 0.62--0.83 & 2345 (387) & 5.3 & 8.8 & 4.9 & 23 & 1.6/12 \\
Pd--O$^{\dagger}$ & 0.33--0.5 & 264 (59) & 16 & 37 & 16 & 75 & 8.0/41 \\
Ag--O$^{\dagger}$ & 0.1--0.5 & 788 (195) & 15 & 22 & 20 & 50 & 6.9/32 \\
Cd--O$^{\dagger}$ & 0.22--0.5 & 805 (137) & 6.5 & 10 & 13 & 29 & 2.4/15 \\
In--O$^{\dagger}$ & 0.38--0.75 & 746 (130) & 4.9 & 7.1 & 10 & 17 & 0.86/18 \\
Sn--O & 0.67--0.8 & 502 (85) & 2.8 & 2.8 & 5.9 & 5.6 & 0.11/0.98 \\
Sb--O & 0.83 & 774 (129) & 2.7 & 2.7 & 3.6 & 7.5 & 0.20/0.80 \\
Te$^{4+}$--O$^{\dagger}$ & 0.57--1.33 & 720 (132) & 15 & 55 & 12 & 93 & 9.2/17 \\
Te$^{6+}$--O & 1.2 & 368 (62) & 5.5 & 5.6 & 12 & 5.7 & 0.62/0.62 \\
Cs--O$^{\dagger}$ & 0.08--0.33 & 2158 (294) & 17 & 21 & 23 & 42 & 4.8/28 \\
Ba--O$^{\dagger}$ & 0.15--0.33 & 5951 (622) & 11 & 13 & 14 & 32 & 3.1/21 \\
La--O$^{\dagger}$ & 0.25--0.5 & 1970 (227) & 8.3 & 11 & 10 & 32 & 3.3/14 \\
Lu--O$^{\dagger}$ & 0.38--0.5 & 378 (58) & 4.8 & 5.2 & 7.1 & 15 & 0.79/3.7 \\
Hf--O$^{\dagger}$ & 0.5--0.67 & 228 (34) & 3.8 & 4.1 & 5.9 & 14 & 0.84/2.5 \\
Ta$^{5+}$--O$^{\dagger}$ & 0.62--0.83 & 1404 (231) & 2.9 & 3.5 & 4.3 & 15 & 0.83/4.1 \\
Hg$^{+}$--O & 0.2--0.33 & 140 (40) & 15 & 29 & 14 & 61 & 8.6/25 \\
Hg$^{2+}$--O$^{\dagger}$ & 0.25--0.5 & 208 (37) & 12 & 24 & 13 & 54 & 8.8/16 \\
Tl--O$^{\dagger}$ & 0.38--0.75 & 236 (42) & 4.9 & 5.0 & 14 & 15 & 0.87/3.0 \\
Pb--O$^{\dagger}$ & 0.67--0.8 & 324 (60) & 5.3 & 14 & 8.2 & 28 & 1.1/12 \\
\midrule
\multicolumn{8}{l}{\emph{S$^{2-}$} (28 families)} \\
Li--S$^{\dagger}$ & 0.17--0.33 & 746 (165) & 5.3 & 6.9 & 13 & 20 & 1.2/18 \\
B--S$^{\dagger}$ & 0.75 & 260 (83) & 1.8 & 1.9 & 12 & 8.0 & 0.20/0.46 \\
Na--S$^{\dagger}$ & 0.12--0.25 & 1502 (286) & 11 & 12 & 14 & 28 & 2.3/13 \\
Mg--S$^{\dagger}$ & 0.33--0.5 & 228 (40) & 3.3 & 3.3 & 7.5 & 5.5 & 0.091/0.35 \\
Al--S$^{\dagger}$ & 0.5--0.75 & 372 (75) & 2.7 & 2.8 & 4.2 & 9.6 & 0.38/1.8 \\
P$^{3+}$--S & 0.75 & 82 (24) & 66 & 91 & 48 & 101 & 9.9/10 \\
P$^{5+}$--S & 1.25 & 280 (70) & 15 & 25 & 10 & 34 & 21/192 \\
K--S$^{\dagger}$ & 0.1--0.25 & 2004 (295) & 13 & 14 & 14 & 28 & 2.5/10 \\
Ca--S$^{\dagger}$ & 0.25--0.33 & 318 (47) & 4.6 & 6.6 & 5.8 & 23 & 1.8/8.6 \\
Zn--S & 0.33--0.5 & 2396 (597) & 1.7 & 1.6 & 2.9 & 1.9 & 0.0037/0.16 \\
Ga--S & 0.6--0.75 & 492 (122) & 3.8 & 12 & 4.3 & 25 & 0.82/17 \\
Ge--S & 0.8 & 680 (169) & 11 & 38 & 3.4 & 79 & 7.7/7.8 \\
As$^{2+}$--S$^{\dagger}$ & 0.4--0.67 & 216 (64) & 36 & 69 & 26 & 116 & 19/42 \\
As$^{3+}$--S$^{\dagger}$ & 0.5--0.75 & 248 (60) & 19 & 66 & 19 & 111 & 14/22 \\
Rb--S$^{\dagger}$ & 0.1--0.25 & 1480 (211) & 14 & 15 & 17 & 28 & 2.2/10 \\
Sr--S & 0.22--0.33 & 566 (72) & 7.0 & 7.8 & 9.3 & 23 & 2.0/6.0 \\
Y--S$^{\dagger}$ & 0.38--0.5 & 816 (121) & 4.7 & 4.8 & 7.0 & 11 & 0.43/2.5 \\
Zr--S & 0.57--0.67 & 218 (35) & 3.9 & 3.9 & 4.8 & 8.4 & 0.27/0.79 \\
Ag--S$^{\dagger}$ & 0.17--0.5 & 1422 (360) & 16 & 21 & 23 & 49 & 5.9/31 \\
Cd--S$^{\dagger}$ & 0.33--0.5 & 304 (68) & 7.5 & 14 & 8.6 & 39 & 3.4/19 \\
In--S$^{\dagger}$ & 0.5--0.75 & 846 (170) & 3.6 & 4.8 & 8.9 & 16 & 0.86/12 \\
Sn--S$^{\dagger}$ & 0.67--0.8 & 490 (105) & 7.3 & 30 & 8.2 & 55 & 3.8/13 \\
Cs--S$^{\dagger}$ & 0.1--0.25 & 886 (116) & 13 & 14 & 17 & 27 & 2.2/10 \\
Ba--S$^{\dagger}$ & 0.17--0.33 & 1926 (226) & 9.8 & 11 & 12 & 27 & 2.3/15 \\
La--S$^{\dagger}$ & 0.33--0.5 & 1048 (130) & 5.8 & 6.7 & 7.2 & 22 & 1.7/7.2 \\
Lu--S$^{\dagger}$ & 0.38--0.5 & 544 (85) & 4.1 & 4.4 & 5.5 & 14 & 0.73/4.2 \\
Ta$^{5+}$--S$^{\dagger}$ & 0.62--1.25 & 218 (44) & 5.2 & 4.0 & 12 & 10 & 0.61/3.8 \\
Hg$^{2+}$--S$^{\dagger}$ & 0.33--0.5 & 436 (92) & 19 & 28 & 23 & 62 & 9.8/26 \\
\midrule
\multicolumn{8}{l}{\emph{Se$^{2-}$} (20 families)} \\
Li--Se$^{\dagger}$ & 0.17--0.25 & 284 (64) & 5.0 & 5.3 & 11 & 14 & 0.70/2.4 \\
Na--Se$^{\dagger}$ & 0.12--0.25 & 1372 (253) & 11 & 12 & 14 & 29 & 2.6/18 \\
Mg--Se$^{\dagger}$ & 0.33--0.5 & 230 (41) & 3.0 & 3.0 & 8.4 & 4.7 & 0.067/0.23 \\
Al--Se$^{\dagger}$ & 0.5--0.75 & 272 (62) & 1.4 & 1.5 & 4.3 & 6.3 & 0.16/0.81 \\
K--Se$^{\dagger}$ & 0.11--0.25 & 1384 (210) & 14 & 15 & 15 & 28 & 2.3/11 \\
Ga--Se & 0.5--0.75 & 252 (62) & 3.8 & 3.8 & 5.3 & 5.8 & 0.099/0.33 \\
Ge--Se & 1 & 256 (64) & --- & --- & --- & --- & --- \\
As$^{3+}$--Se$^{\dagger}$ & 0.5--0.75 & 342 (76) & 25 & 52 & 28 & 97 & 13/23 \\
Rb--Se$^{\dagger}$ & 0.1--0.25 & 682 (96) & 16 & 17 & 16 & 30 & 2.5/8.2 \\
Sr--Se & 0.25--0.33 & 196 (25) & 5.3 & 5.7 & 6.7 & 17 & 1.2/4.3 \\
Y--Se & 0.38--0.5 & 356 (58) & 3.4 & 3.4 & 5.0 & 5.3 & 0.078/0.69 \\
Ag--Se$^{\dagger}$ & 0.17--0.33 & 580 (149) & 19 & 20 & 21 & 37 & 3.5/17 \\
Cd--Se & 0.33--0.5 & 292 (69) & 4.8 & 6.4 & 6.6 & 19 & 1.0/8.0 \\
In--Se$^{\dagger}$ & 0.5--0.75 & 336 (79) & 4.1 & 5.3 & 7.5 & 13 & 0.35/10 \\
Sn--Se$^{\dagger}$ & 0.67--0.8 & 330 (79) & 10 & 26 & 9.5 & 47 & 3.5/8.0 \\
Cs--Se$^{\dagger}$ & 0.11--0.25 & 784 (106) & 16 & 18 & 20 & 36 & 3.7/27 \\
Ba--Se$^{\dagger}$ & 0.17--0.33 & 744 (89) & 7.9 & 8.9 & 9.9 & 21 & 1.6/6.8 \\
La--Se$^{\dagger}$ & 0.33--0.5 & 404 (52) & 5.7 & 5.9 & 7.2 & 12 & 0.54/1.3 \\
Ta$^{4+}$--Se & 0.67 & 102 (17) & 3.0 & 3.0 & 3.8 & 10 & 0.50/1.1 \\
Hg$^{2+}$--Se & 0.29--0.5 & 242 (54) & 23 & 26 & 28 & 44 & 4.3/23 \\
\midrule
\multicolumn{8}{l}{\emph{Te$^{2-}$} (7 families)} \\
Na--Te$^{\dagger}$ & 0.12--0.25 & 248 (46) & 9.0 & 11 & 11 & 27 & 2.4/6.5 \\
K--Te$^{\dagger}$ & 0.12--0.25 & 558 (88) & 10 & 11 & 14 & 28 & 2.6/7.0 \\
Sc--Te & 0.5 & 174 (29) & 3.2 & 3.2 & 3.5 & 5.5 & 0.090/0.39 \\
Rb--Te & 0.1--0.25 & 144 (21) & 9.3 & 10 & 16 & 21 & 1.4/3.8 \\
Ag--Te$^{\dagger}$ & 0.25--0.33 & 186 (50) & 21 & 21 & 22 & 27 & 1.4/7.0 \\
Cs--Te & 0.12--0.25 & 196 (28) & 12 & 18 & 24 & 41 & 3.8/18 \\
Ba--Te & 0.25--0.33 & 300 (39) & 6.2 & 6.8 & 9.0 & 18 & 1.2/3.5 \\
\midrule
\multicolumn{8}{l}{\emph{N$^{3-}$} (11 families)} \\
Li--N$^{\dagger}$ & 0.17--0.5 & 744 (194) & 7.0 & 7.3 & 12 & 16 & 0.94/4.1 \\
Na--N$^{\dagger}$ & 0.17--0.25 & 178 (40) & 11 & 12 & 18 & 20 & 1.1/4.4 \\
Mg--N$^{\dagger}$ & 0.33--0.5 & 280 (67) & 3.9 & 8.1 & 11 & 21 & 0.96/17 \\
Al--N & 0.5--0.75 & 192 (46) & 3.4 & 3.4 & 7.5 & 7.5 & 0.23/0.89 \\
Si--N & 1 & 692 (173) & --- & --- & --- & --- & --- \\
P$^{5+}$--N & 1.25 & 152 (38) & 1.8 & 2.0 & 3.0 & 10 & 0.56/1.5 \\
Ca--N$^{\dagger}$ & 0.25--0.5 & 1266 (243) & 8.6 & 9.8 & 13 & 23 & 1.7/8.9 \\
Ga--N$^{\dagger}$ & 0.75 & 154 (40) & 5.4 & 5.4 & 5.7 & 6.2 & 0.10/0.34 \\
Sr--N$^{\dagger}$ & 0.17--0.5 & 1124 (191) & 11 & 14 & 13 & 32 & 3.0/13 \\
Ba--N$^{\dagger}$ & 0.17--0.4 & 624 (94) & 14 & 18 & 16 & 40 & 4.5/14 \\
La--N & 0.33--0.5 & 236 (30) & 9.4 & 11 & 12 & 31 & 3.2/7.3 \\
\midrule
\multicolumn{8}{l}{\emph{P$^{3-}$} (7 families)} \\
Na--P$^{\dagger}$ & 0.2--0.33 & 428 (100) & 9.2 & 10 & 12 & 21 & 1.3/8.2 \\
K--P$^{\dagger}$ & 0.17--0.5 & 290 (66) & 28 & 32 & 23 & 40 & 1.8/13 \\
Ca--P$^{\dagger}$ & 0.25--0.4 & 235 (40) & 19 & 20 & 20 & 28 & 1.4/8.3 \\
Zn--P & 0.5--0.67 & 316 (80) & 5.1 & 5.1 & 7.7 & 14 & 0.85/3.1 \\
Sr--P & 0.25--0.33 & 232 (38) & 19 & 21 & 21 & 32 & 1.9/9.1 \\
Cd--P & 0.5 & 172 (44) & 11 & 11 & 12 & 20 & 1.5/3.8 \\
Ba--P$^{\dagger}$ & 0.25--0.4 & 444 (71) & 16 & 17 & 16 & 30 & 1.9/11 \\
\midrule
\multicolumn{8}{l}{\emph{As$^{3-}$} (11 families)} \\
Li--As$^{\dagger}$ & 0.17--0.33 & 225 (57) & 3.7 & 3.9 & 11 & 12 & 0.66/2.0 \\
Na--As$^{\dagger}$ & 0.17--0.33 & 806 (186) & 12 & 13 & 15 & 22 & 1.3/7.0 \\
K--As$^{\dagger}$ & 0.14--0.5 & 816 (173) & 24 & 25 & 22 & 33 & 1.4/13 \\
Ca--As$^{\dagger}$ & 0.25--0.5 & 276 (48) & 17 & 17 & 18 & 27 & 1.5/7.6 \\
Zn--As & 0.5--0.67 & 485 (122) & 11 & 11 & 12 & 14 & 0.36/1.8 \\
Rb--As$^{\dagger}$ & 0.12--0.33 & 236 (46) & 28 & 30 & 33 & 42 & 2.8/16 \\
Sr--As & 0.25--0.4 & 314 (52) & 17 & 19 & 18 & 33 & 2.3/14 \\
Cd--As & 0.4--0.5 & 330 (82) & 27 & 27 & 23 & 32 & 1.8/20 \\
Te$^{2+}$--As & 0.33--0.67 & 130 (33) & 9.5 & 9.1 & 26 & 17 & 0.70/4.1 \\
Cs--As$^{\dagger}$ & 0.12--0.5 & 280 (64) & 31 & 32 & 32 & 38 & 1.4/9.2 \\
Ba--As$^{\dagger}$ & 0.25--0.4 & 456 (72) & 21 & 22 & 23 & 31 & 1.6/13 \\
\end{longtable}
}
\clearpage

\begin{figure*}[t]
\includegraphics[width=0.6\textwidth]{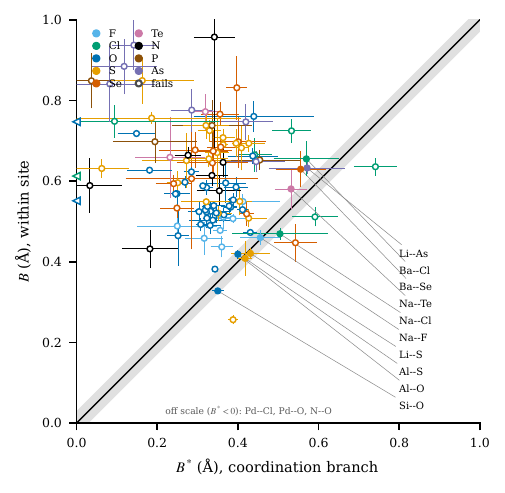}
\caption{The within-site softness $B$ against the coordination-branch softness $B^{*}$ for the 106 families of Table~\ref{tab:transfer}, colored by anion, with bootstrap intervals. Filled marks pass and are labeled, open marks fail, the line is $B=B^{*}$ and the band $\pm0.03$~\AA{}. Three families with $B^{*}<0$ (N$^{5+}$--O and the square-planar Pd$^{2+}$) are drawn at the left edge.}
\label{fig:transfer}
\end{figure*}

\begin{table*}[t]
\caption{The within-site softness $B$ on seven partitions of the same electron density against the coordination-branch softness $B^{*}$, which depends on bond lengths and coordination numbers only and is the same on every partition (hydrogen-free hosts, 68\% bootstrap intervals). The partitions are the density's own zero-flux surfaces, the power diagram at the four weightings $t$ of the notes above, and the density basins eroded and dilated by one voxel. The families that pass are listed first, then failing examples of each anion group. Dashes mark the anions for which the power diagrams are not reported.}
\label{tab:partitions}
\centering\footnotesize\setlength{\tabcolsep}{4pt}
\fitwidth{%
\begin{tabular}{lcccccccc}
\toprule
family & $B^{*}$ & density & $t=0$ & $t=0.5$ & $t=1$ & $t=1.5$ & erode & dilate \\
\midrule
Na--F & 0.46 $\pm$ 0.05 & 0.46 $\pm$ 0.02 & 0.57 $\pm$ 0.03 & 0.53 $\pm$ 0.02 & 0.50 $\pm$ 0.02 & 0.46 $\pm$ 0.02 & 0.45 $\pm$ 0.02 & 0.49 $\pm$ 0.02 \\
Ba--Cl & 0.57 $\pm$ 0.08 & 0.66 $\pm$ 0.04 & --- & --- & --- & --- & 0.64 $\pm$ 0.04 & 0.70 $\pm$ 0.05 \\
Na--Cl & 0.50 $\pm$ 0.12 & 0.47 $\pm$ 0.01 & --- & --- & --- & --- & 0.46 $\pm$ 0.02 & 0.51 $\pm$ 0.02 \\
Al--O & 0.40 $\pm$ 0.01 & 0.42 $\pm$ 0.01 & 0.59 $\pm$ 0.02 & 0.53 $\pm$ 0.02 & 0.49 $\pm$ 0.01 & 0.44 $\pm$ 0.01 & 0.42 $\pm$ 0.01 & 0.45 $\pm$ 0.01 \\
Si--O & 0.35 $\pm$ 0.01 & 0.328 $\pm$ 0.004 & 0.48 $\pm$ 0.02 & 0.432 $\pm$ 0.003 & 0.40 $\pm$ 0.01 & 0.396 $\pm$ 0.004 & 0.326 $\pm$ 0.004 & 0.322 $\pm$ 0.004 \\
Al--S & 0.42 $\pm$ 0.02 & 0.41 $\pm$ 0.04 & --- & --- & --- & --- & 0.40 $\pm$ 0.04 & 0.47 $\pm$ 0.02 \\
Li--S & 0.43 $\pm$ 0.05 & 0.42 $\pm$ 0.01 & --- & --- & --- & --- & 0.41 $\pm$ 0.01 & 0.47 $\pm$ 0.01 \\
Ba--Se & 0.56 $\pm$ 0.06 & 0.63 $\pm$ 0.04 & --- & --- & --- & --- & 0.62 $\pm$ 0.04 & 0.67 $\pm$ 0.05 \\
Na--Te & 0.53 $\pm$ 0.04 & 0.58 $\pm$ 0.05 & --- & --- & --- & --- & 0.58 $\pm$ 0.05 & 0.62 $\pm$ 0.05 \\
Li--As & 0.57 $\pm$ 0.10 & 0.63 $\pm$ 0.01 & --- & --- & --- & --- & 0.63 $\pm$ 0.01 & 0.65 $\pm$ 0.02 \\
Li--F & 0.36 $\pm$ 0.03 & 0.44 $\pm$ 0.03 & 0.69 $\pm$ 0.10 & 0.56 $\pm$ 0.05 & 0.50 $\pm$ 0.03 & 0.44 $\pm$ 0.02 & 0.43 $\pm$ 0.03 & 0.46 $\pm$ 0.03 \\
Li--O & 0.43 $\pm$ 0.02 & 0.47 $\pm$ 0.01 & 0.73 $\pm$ 0.05 & 0.57 $\pm$ 0.02 & 0.52 $\pm$ 0.01 & 0.48 $\pm$ 0.01 & 0.47 $\pm$ 0.01 & 0.50 $\pm$ 0.01 \\
Mg--O & 0.37 $\pm$ 0.02 & 0.52 $\pm$ 0.02 & 0.63 $\pm$ 0.04 & 0.58 $\pm$ 0.03 & 0.53 $\pm$ 0.03 & 0.49 $\pm$ 0.02 & 0.51 $\pm$ 0.02 & 0.54 $\pm$ 0.02 \\
Zr--O & 0.34 $\pm$ 0.03 & 0.54 $\pm$ 0.01 & 0.51 $\pm$ 0.01 & 0.53 $\pm$ 0.01 & 0.55 $\pm$ 0.01 & 0.56 $\pm$ 0.01 & 0.53 $\pm$ 0.01 & 0.55 $\pm$ 0.01 \\
Ca--O & 0.38 $\pm$ 0.02 & 0.54 $\pm$ 0.01 & 0.57 $\pm$ 0.01 & 0.57 $\pm$ 0.01 & 0.57 $\pm$ 0.01 & 0.56 $\pm$ 0.01 & 0.52 $\pm$ 0.01 & 0.57 $\pm$ 0.01 \\
K--O & 0.25 $\pm$ 0.02 & 0.57 $\pm$ 0.01 & 0.54 $\pm$ 0.01 & 0.56 $\pm$ 0.01 & 0.57 $\pm$ 0.01 & 0.58 $\pm$ 0.01 & 0.56 $\pm$ 0.01 & 0.59 $\pm$ 0.01 \\
Na--S & 0.35 $\pm$ 0.03 & 0.52 $\pm$ 0.02 & --- & --- & --- & --- & 0.52 $\pm$ 0.02 & 0.56 $\pm$ 0.02 \\
Ca--S & 0.43 $\pm$ 0.05 & 0.51 $\pm$ 0.02 & --- & --- & --- & --- & 0.50 $\pm$ 0.02 & 0.55 $\pm$ 0.03 \\
K--Cl & 0.74 $\pm$ 0.05 & 0.64 $\pm$ 0.02 & --- & --- & --- & --- & 0.62 $\pm$ 0.02 & 0.69 $\pm$ 0.02 \\
Ba--S & 0.43 $\pm$ 0.04 & 0.69 $\pm$ 0.02 & --- & --- & --- & --- & 0.68 $\pm$ 0.02 & 0.74 $\pm$ 0.02 \\
Mg--N & 0.18 $\pm$ 0.07 & 0.43 $\pm$ 0.05 & --- & --- & --- & --- & 0.43 $\pm$ 0.04 & 0.46 $\pm$ 0.05 \\
K--As & 0.14 $\pm$ 0.05 & 0.94 $\pm$ 0.06 & --- & --- & --- & --- & 0.92 $\pm$ 0.06 & 0.98 $\pm$ 0.07 \\
\bottomrule
\end{tabular}
}
\end{table*}

\begin{figure*}[t]
\includegraphics[width=\textwidth]{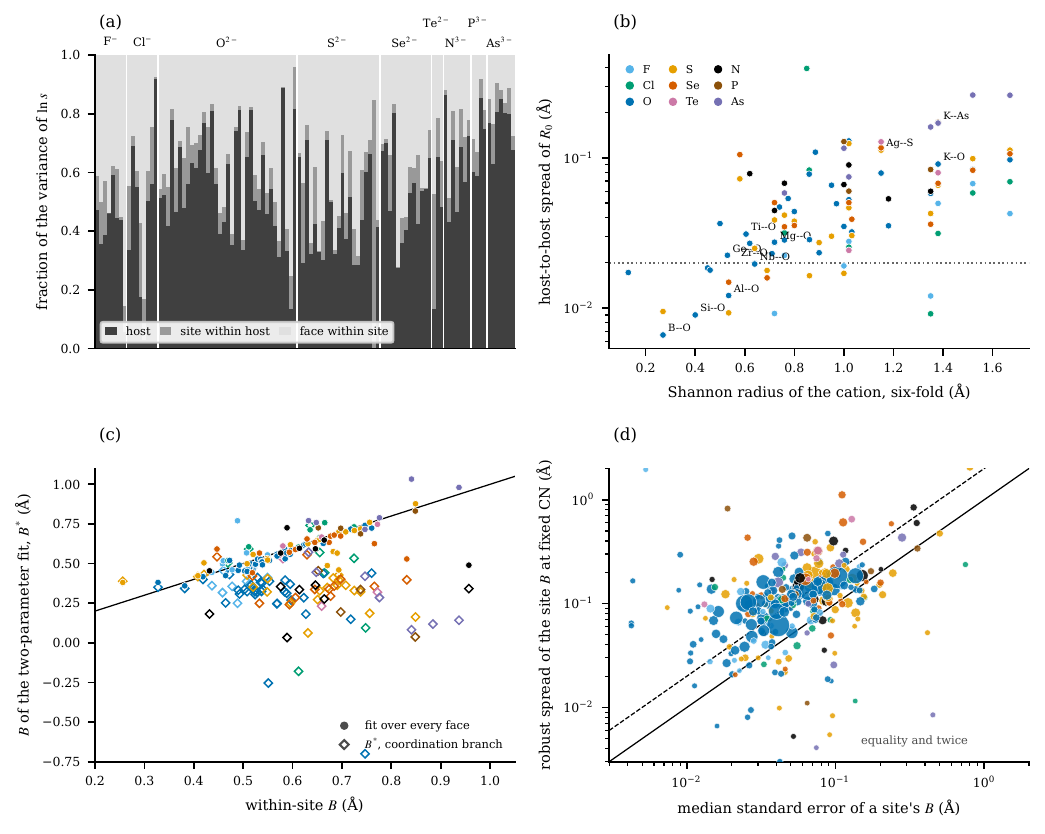}
\caption{What transfers between hosts, for the 106 families of Table~\ref{tab:transfercomp} (hydrogen-free hosts). (a)~The scatter of $\ln s$ about the two-parameter family law, split into its host, site-within-host and face-within-site parts (variance fractions), one bar per family, grouped by anion and ordered within an anion by the within-site $B$. (b)~The host-to-host spread of $R_0$ at the within-site $B$ against the cation's Shannon six-fold radius, colored by anion, with a dotted line at $0.02$~\AA{} and the compact oxide families labeled. (c)~The softness of the two-parameter fit (filled circles) and $B^{*}$ (open diamonds) against the within-site $B$, with the line of equality. (d)~The robust spread of the per-site $B$ at fixed coordination number against the median standard error of a single site's slope, marker area proportional to the number of sites, with the lines of equality and of twice the error.}
\label{fig:transfercomp}
\end{figure*}

\begin{table*}[t]
\caption{Transferability of $(R_0,B)$ for nineteen families, from Table~\ref{tab:transfercomp}, giving the hosts and sites, the within-site $B$, the two-parameter fit $B$ and $B^{*}$, the rms residual of the two-parameter law and its host, site-within-host and face-within-site components in $\ln s$, and the host-to-host and site-to-site spreads of $R_0$.}
\label{tab:transferselect}
\centering\footnotesize\setlength{\tabcolsep}{4pt}
\fitwidth{%
\begin{tabular}{lrrcccccccccc}
\toprule
 & & & \multicolumn{3}{c}{$B$ (\AA)} & \multicolumn{4}{c}{scatter of $\ln s$} & \multicolumn{2}{c}{spread of $R_0$ (\AA)} \\
\cmidrule(lr){4-6}\cmidrule(lr){7-10}\cmidrule(lr){11-12}
family & hosts & sites & within & fit & $B^{*}$ & total & host & site & face & host & site \\
\midrule
Si--O & 260 & 1205 & 0.33 & 0.38 & 0.35 & 0.039 & 0.027 & 0.007 & 0.026 & 0.009 & 0.002 \\
B--O & 183 & 858 & 0.38 & 0.36 & 0.34 & 0.024 & 0.017 & 0.007 & 0.015 & 0.007 & 0.002 \\
Al--O & 123 & 502 & 0.42 & 0.42 & 0.40 & 0.045 & 0.029 & 0.016 & 0.030 & 0.012 & 0.007 \\
Mg--O & 95 & 304 & 0.52 & 0.50 & 0.37 & 0.064 & 0.053 & 0.020 & 0.037 & 0.027 & 0.010 \\
Ti--O & 117 & 409 & 0.59 & 0.57 & 0.31 & 0.061 & 0.053 & 0.013 & 0.035 & 0.031 & 0.008 \\
Ge--O & 116 & 400 & 0.51 & 0.45 & 0.32 & 0.060 & 0.044 & 0.014 & 0.031 & 0.022 & 0.007 \\
Zr--O & 52 & 157 & 0.54 & 0.49 & 0.34 & 0.056 & 0.051 & 0.007 & 0.029 & 0.027 & 0.004 \\
Nb--O & 112 & 387 & 0.62 & 0.62 & 0.28 & 0.050 & 0.032 & 0.014 & 0.035 & 0.020 & 0.009 \\
Zn--O & 110 & 424 & 0.52 & 0.48 & 0.32 & 0.097 & 0.090 & 0.020 & 0.045 & 0.047 & 0.011 \\
Li--O & 214 & 1052 & 0.47 & 0.52 & 0.43 & 0.091 & 0.060 & 0.034 & 0.055 & 0.028 & 0.016 \\
Na--O & 276 & 1223 & 0.49 & 0.49 & 0.31 & 0.159 & 0.107 & 0.063 & 0.091 & 0.053 & 0.031 \\
K--O & 209 & 893 & 0.57 & 0.56 & 0.25 & 0.199 & 0.160 & 0.057 & 0.101 & 0.091 & 0.033 \\
Ca--O & 157 & 529 & 0.54 & 0.53 & 0.38 & 0.101 & 0.065 & 0.022 & 0.072 & 0.035 & 0.012 \\
Ba--O & 205 & 622 & 0.58 & 0.59 & 0.32 & 0.143 & 0.099 & 0.042 & 0.090 & 0.058 & 0.024 \\
Na--F & 56 & 160 & 0.46 & 0.52 & 0.46 & 0.097 & 0.060 & 0.042 & 0.064 & 0.028 & 0.019 \\
Al--S & 20 & 75 & 0.41 & 0.42 & 0.42 & 0.042 & 0.023 & 0.022 & 0.028 & 0.009 & 0.009 \\
K--S & 57 & 295 & 0.71 & 0.70 & 0.36 & 0.135 & 0.093 & 0.058 & 0.088 & 0.066 & 0.041 \\
Ag--S & 76 & 360 & 0.60 & 0.62 & 0.25 & 0.227 & 0.189 & 0.040 & 0.114 & 0.113 & 0.024 \\
K--As & 31 & 173 & 0.94 & 0.98 & 0.14 & 0.218 & 0.182 & 0.048 & 0.116 & 0.170 & 0.045 \\
\bottomrule
\end{tabular}
}
\end{table*}

\clearpage
{\footnotesize\setlength{\tabcolsep}{2.4pt}%
\begin{longtable}{lrrcccccccccc}
\caption{\captiontransfercomp}\label{tab:transfercomp}\\
\toprule
 & & & \multicolumn{3}{c}{$B$ (\AA)} & \multicolumn{4}{c}{scatter of $\ln s$} & \multicolumn{2}{c}{spread of $R_0$ (\AA)} & \\
\cmidrule(lr){4-6}\cmidrule(lr){7-10}\cmidrule(lr){11-12}
family & hosts & sites & within & fit & $B^{*}$ & total & host & site & face & host & site & cv \\
\midrule
\endfirsthead
\multicolumn{13}{c}{\tablename~\thetable{} (continued)}\\
\toprule
 & & & \multicolumn{3}{c}{$B$ (\AA)} & \multicolumn{4}{c}{scatter of $\ln s$} & \multicolumn{2}{c}{spread of $R_0$ (\AA)} & \\
\cmidrule(lr){4-6}\cmidrule(lr){7-10}\cmidrule(lr){11-12}
family & hosts & sites & within & fit & $B^{*}$ & total & host & site & face & host & site & cv \\
\midrule
\endhead
\bottomrule
\endlastfoot
\multicolumn{13}{l}{\emph{F$^{-}$} (8 families)} \\
Zr--F & 11 & 29 & 0.48 & 0.46 & 0.36 & 0.027 & 0.019 & 0.000 & 0.021 & 0.009 & 0.000 & 1.11 \\
Ba--F & 17 & 57 & 0.55 & 0.55 & 0.41 & 0.092 & 0.022 & 0.028 & 0.086 & 0.012 & 0.015 & 1.03 \\
Ca--F & 22 & 55 & 0.51 & 0.48 & 0.36 & 0.056 & 0.037 & 0.006 & 0.042 & 0.019 & 0.003 & 1.08 \\
Li--F & 46 & 141 & 0.44 & 0.44 & 0.36 & 0.074 & 0.051 & 0.026 & 0.048 & 0.022 & 0.011 & 1.04 \\
Na--F & 56 & 160 & 0.46 & 0.52 & 0.46 & 0.097 & 0.060 & 0.042 & 0.064 & 0.028 & 0.019 & 1.04 \\
Cs--F & 22 & 67 & 0.46 & 0.50 & 0.32 & 0.155 & 0.093 & 0.059 & 0.110 & 0.043 & 0.027 & 1.08 \\
K--F & 86 & 249 & 0.51 & 0.57 & 0.39 & 0.149 & 0.098 & 0.060 & 0.091 & 0.050 & 0.031 & 1.03 \\
Rb--F & 22 & 60 & 0.49 & 0.77 & 0.25 & 0.174 & 0.138 & 0.031 & 0.110 & 0.067 & 0.015 & 1.16 \\
\midrule
\multicolumn{13}{l}{\emph{Cl$^{-}$} (8 families)} \\
Ba--Cl & 9 & 38 & 0.66 & 0.66 & 0.57 & 0.083 & 0.014 & 0.031 & 0.074 & 0.009 & 0.020 & --- \\
Na--Cl & 17 & 65 & 0.47 & 0.52 & 0.50 & 0.094 & 0.054 & 0.043 & 0.062 & 0.025 & 0.020 & 1.07 \\
K--Cl & 27 & 132 & 0.64 & 0.71 & 0.74 & 0.120 & 0.049 & 0.059 & 0.090 & 0.031 & 0.038 & 1.05 \\
Li--Cl & 15 & 52 & 0.51 & 0.61 & 0.59 & 0.071 & 0.062 & 0.014 & 0.039 & 0.032 & 0.007 & 1.16 \\
Rb--Cl & 22 & 76 & 0.72 & 0.73 & 0.53 & 0.107 & 0.081 & 0.047 & 0.053 & 0.058 & 0.034 & 1.12 \\
Cs--Cl & 49 & 165 & 0.67 & 0.76 & 0.44 & 0.146 & 0.104 & 0.048 & 0.092 & 0.069 & 0.032 & 1.07 \\
Pd--Cl & 15 & 41 & 0.61 & 0.60 & $-$0.18 & 0.176 & 0.135 & 0.043 & 0.115 & 0.083 & 0.026 & 1.16 \\
Au--Cl & 13 & 41 & 0.75 & 0.73 & 0.09 & 0.297 & 0.527 & 0.048 & 0.150 & 0.394 & 0.036 & 1.04 \\
\midrule
\multicolumn{13}{l}{\emph{O$^{2-}$} (35 families)} \\
B--O & 183 & 858 & 0.38 & 0.36 & 0.34 & 0.024 & 0.017 & 0.007 & 0.015 & 0.007 & 0.002 & 1.01 \\
Si--O & 260 & 1205 & 0.33 & 0.38 & 0.35 & 0.039 & 0.027 & 0.007 & 0.026 & 0.009 & 0.002 & 1.02 \\
Al--O & 123 & 502 & 0.42 & 0.42 & 0.40 & 0.045 & 0.029 & 0.016 & 0.030 & 0.012 & 0.007 & 1.02 \\
N--O & 16 & 84 & 0.75 & 0.74 & $-$0.70 & 0.067 & 0.023 & 0.021 & 0.055 & 0.017 & 0.016 & 2.68 \\
As$^{5+}$--O & 53 & 192 & 0.52 & 0.49 & 0.30 & 0.049 & 0.034 & 0.017 & 0.022 & 0.018 & 0.009 & 1.37 \\
Be--O & 34 & 128 & 0.49 & 0.46 & 0.33 & 0.044 & 0.038 & 0.006 & 0.025 & 0.018 & 0.003 & 1.11 \\
Nb--O & 112 & 387 & 0.62 & 0.62 & 0.28 & 0.050 & 0.032 & 0.014 & 0.035 & 0.020 & 0.009 & 1.03 \\
Ta$^{5+}$--O & 79 & 231 & 0.60 & 0.58 & 0.27 & 0.046 & 0.033 & 0.012 & 0.028 & 0.020 & 0.007 & 1.03 \\
Ge--O & 116 & 400 & 0.51 & 0.45 & 0.32 & 0.060 & 0.044 & 0.014 & 0.031 & 0.022 & 0.007 & 1.10 \\
Hf--O & 11 & 34 & 0.59 & 0.53 & 0.40 & 0.057 & 0.039 & 0.015 & 0.039 & 0.023 & 0.009 & 1.18 \\
Y--O & 60 & 205 & 0.54 & 0.51 & 0.38 & 0.074 & 0.043 & 0.014 & 0.057 & 0.023 & 0.008 & 1.02 \\
Ga--O & 63 & 212 & 0.52 & 0.47 & 0.35 & 0.064 & 0.052 & 0.014 & 0.033 & 0.027 & 0.007 & 1.10 \\
Mg--O & 95 & 304 & 0.52 & 0.50 & 0.37 & 0.064 & 0.053 & 0.020 & 0.037 & 0.027 & 0.010 & 1.05 \\
Zr--O & 52 & 157 & 0.54 & 0.49 & 0.34 & 0.056 & 0.051 & 0.007 & 0.029 & 0.027 & 0.004 & 1.07 \\
Li--O & 214 & 1052 & 0.47 & 0.52 & 0.43 & 0.091 & 0.060 & 0.034 & 0.055 & 0.028 & 0.016 & 0.99 \\
Lu--O & 17 & 58 & 0.60 & 0.56 & 0.37 & 0.071 & 0.048 & 0.019 & 0.049 & 0.029 & 0.011 & 1.14 \\
Ti--O & 117 & 409 & 0.59 & 0.57 & 0.31 & 0.061 & 0.053 & 0.013 & 0.035 & 0.031 & 0.008 & 1.00 \\
La--O & 74 & 227 & 0.55 & 0.52 & 0.39 & 0.102 & 0.058 & 0.019 & 0.079 & 0.032 & 0.010 & 1.01 \\
Ca--O & 157 & 529 & 0.54 & 0.53 & 0.38 & 0.101 & 0.065 & 0.022 & 0.072 & 0.035 & 0.012 & 1.03 \\
Sr--O & 175 & 590 & 0.53 & 0.53 & 0.41 & 0.110 & 0.067 & 0.028 & 0.078 & 0.035 & 0.015 & 1.01 \\
Se$^{4+}$--O & 25 & 150 & 0.63 & 0.62 & 0.18 & 0.118 & 0.058 & 0.024 & 0.083 & 0.037 & 0.015 & 1.05 \\
In--O & 47 & 130 & 0.46 & 0.42 & 0.25 & 0.099 & 0.095 & 0.020 & 0.046 & 0.044 & 0.009 & 1.06 \\
Zn--O & 110 & 424 & 0.52 & 0.48 & 0.32 & 0.097 & 0.090 & 0.020 & 0.045 & 0.047 & 0.011 & 1.02 \\
Te$^{4+}$--O & 24 & 132 & 0.72 & 0.72 & 0.15 & 0.117 & 0.069 & 0.033 & 0.092 & 0.050 & 0.023 & 1.02 \\
Na--O & 276 & 1223 & 0.49 & 0.49 & 0.31 & 0.159 & 0.107 & 0.063 & 0.091 & 0.053 & 0.031 & 1.03 \\
Pb--O & 25 & 60 & 0.57 & 0.55 & 0.25 & 0.099 & 0.094 & 0.010 & 0.044 & 0.054 & 0.006 & 1.29 \\
Ba--O & 205 & 622 & 0.58 & 0.59 & 0.32 & 0.143 & 0.099 & 0.042 & 0.090 & 0.058 & 0.024 & 1.02 \\
Cd--O & 40 & 137 & 0.53 & 0.50 & 0.33 & 0.128 & 0.125 & 0.022 & 0.057 & 0.066 & 0.012 & 1.13 \\
Pd--O & 22 & 59 & 0.55 & 0.52 & $-$0.25 & 0.158 & 0.141 & 0.020 & 0.065 & 0.078 & 0.011 & 1.10 \\
Ag--O & 32 & 195 & 0.50 & 0.45 & 0.34 & 0.193 & 0.157 & 0.092 & 0.104 & 0.079 & 0.047 & 1.03 \\
Rb--O & 65 & 319 & 0.53 & 0.53 & 0.32 & 0.212 & 0.158 & 0.051 & 0.131 & 0.084 & 0.027 & 1.03 \\
K--O & 209 & 893 & 0.57 & 0.56 & 0.25 & 0.199 & 0.160 & 0.057 & 0.101 & 0.091 & 0.033 & 1.00 \\
Cs--O & 65 & 294 & 0.54 & 0.51 & 0.32 & 0.229 & 0.181 & 0.043 & 0.132 & 0.097 & 0.023 & 1.03 \\
Tl--O & 11 & 42 & 0.76 & 0.72 & 0.44 & 0.138 & 0.143 & 0.060 & 0.032 & 0.109 & 0.046 & 1.34 \\
Hg$^{2+}$--O & 15 & 37 & 0.66 & 0.63 & 0.44 & 0.164 & 0.196 & 0.003 & 0.068 & 0.130 & 0.002 & 1.19 \\
\midrule
\multicolumn{13}{l}{\emph{S$^{2-}$} (21 families)} \\
As$^{2+}$--S & 5 & 64 & 0.76 & 0.76 & 0.19 & 0.264 & 0.000 & 0.241 & 0.103 & 0.000 & 0.182 & --- \\
Al--S & 20 & 75 & 0.41 & 0.42 & 0.42 & 0.042 & 0.023 & 0.022 & 0.028 & 0.009 & 0.009 & 1.05 \\
B--S & 18 & 83 & 0.26 & 0.38 & 0.39 & 0.044 & 0.037 & 0.005 & 0.027 & 0.009 & 0.001 & 1.06 \\
Lu--S & 16 & 85 & 0.55 & 0.52 & 0.40 & 0.055 & 0.030 & 0.020 & 0.041 & 0.016 & 0.011 & 1.05 \\
Ca--S & 10 & 47 & 0.51 & 0.50 & 0.43 & 0.058 & 0.033 & 0.012 & 0.047 & 0.017 & 0.006 & 1.35 \\
Sn--S & 33 & 105 & 0.55 & 0.53 & 0.32 & 0.061 & 0.032 & 0.004 & 0.046 & 0.018 & 0.002 & 1.06 \\
Ta$^{5+}$--S & 15 & 44 & 0.69 & 0.46 & 0.40 & 0.072 & 0.036 & 0.012 & 0.056 & 0.025 & 0.008 & 1.19 \\
Y--S & 23 & 121 & 0.68 & 0.58 & 0.41 & 0.066 & 0.040 & 0.020 & 0.048 & 0.027 & 0.013 & 1.03 \\
Cd--S & 24 & 68 & 0.65 & 0.65 & 0.27 & 0.086 & 0.046 & 0.010 & 0.067 & 0.030 & 0.006 & 1.32 \\
La--S & 26 & 130 & 0.66 & 0.65 & 0.33 & 0.072 & 0.046 & 0.015 & 0.054 & 0.030 & 0.010 & 1.03 \\
In--S & 37 & 170 & 0.69 & 0.57 & 0.36 & 0.077 & 0.055 & 0.030 & 0.044 & 0.038 & 0.021 & 1.05 \\
Mg--S & 11 & 40 & 0.67 & 0.49 & 0.35 & 0.068 & 0.058 & 0.027 & 0.039 & 0.039 & 0.018 & 1.44 \\
Li--S & 37 & 165 & 0.42 & 0.50 & 0.43 & 0.118 & 0.099 & 0.027 & 0.057 & 0.042 & 0.011 & 1.12 \\
Ba--S & 49 & 226 & 0.69 & 0.70 & 0.43 & 0.117 & 0.061 & 0.040 & 0.089 & 0.043 & 0.028 & 1.00 \\
Na--S & 50 & 286 & 0.52 & 0.56 & 0.35 & 0.141 & 0.089 & 0.061 & 0.085 & 0.047 & 0.032 & 0.99 \\
K--S & 57 & 295 & 0.71 & 0.70 & 0.36 & 0.135 & 0.093 & 0.058 & 0.088 & 0.066 & 0.041 & 1.00 \\
As$^{3+}$--S & 15 & 60 & 0.63 & 0.63 & 0.06 & 0.168 & 0.115 & 0.014 & 0.114 & 0.073 & 0.009 & 1.18 \\
Rb--S & 40 & 211 & 0.73 & 0.71 & 0.33 & 0.166 & 0.136 & 0.046 & 0.101 & 0.099 & 0.033 & 1.00 \\
Cs--S & 28 & 116 & 0.74 & 0.72 & 0.32 & 0.171 & 0.153 & 0.042 & 0.101 & 0.113 & 0.031 & 1.03 \\
Ag--S & 76 & 360 & 0.60 & 0.62 & 0.25 & 0.227 & 0.189 & 0.040 & 0.114 & 0.113 & 0.024 & 1.06 \\
Hg$^{2+}$--S & 24 & 92 & 0.85 & 0.88 & 0.16 & 0.230 & 0.147 & 0.035 & 0.171 & 0.125 & 0.030 & 1.08 \\
\midrule
\multicolumn{13}{l}{\emph{Se$^{2-}$} (13 families)} \\
Al--Se & 17 & 62 & 0.52 & 0.44 & 0.42 & 0.033 & 0.029 & 0.004 & 0.018 & 0.015 & 0.002 & 1.41 \\
Sn--Se & 26 & 79 & 0.61 & 0.59 & 0.29 & 0.052 & 0.026 & 0.003 & 0.042 & 0.016 & 0.002 & 1.49 \\
Li--Se & 16 & 64 & 0.45 & 0.59 & 0.54 & 0.092 & 0.078 & 0.016 & 0.052 & 0.035 & 0.007 & 1.24 \\
In--Se & 22 & 79 & 0.77 & 0.63 & 0.36 & 0.067 & 0.046 & 0.002 & 0.042 & 0.035 & 0.002 & 1.37 \\
Ba--Se & 23 & 89 & 0.63 & 0.66 & 0.56 & 0.098 & 0.057 & 0.028 & 0.071 & 0.036 & 0.018 & 1.02 \\
La--Se & 11 & 52 & 0.70 & 0.67 & 0.40 & 0.072 & 0.056 & 0.007 & 0.051 & 0.039 & 0.005 & 1.12 \\
Na--Se & 38 & 253 & 0.53 & 0.57 & 0.25 & 0.139 & 0.094 & 0.060 & 0.081 & 0.050 & 0.032 & 1.05 \\
Mg--Se & 13 & 41 & 0.83 & 0.53 & 0.40 & 0.069 & 0.061 & 0.023 & 0.038 & 0.050 & 0.019 & 1.16 \\
K--Se & 40 & 210 & 0.68 & 0.68 & 0.36 & 0.152 & 0.099 & 0.074 & 0.089 & 0.068 & 0.051 & 1.03 \\
Rb--Se & 20 & 96 & 0.67 & 0.63 & 0.34 & 0.161 & 0.123 & 0.050 & 0.104 & 0.083 & 0.034 & 1.11 \\
As$^{3+}$--Se & 14 & 76 & 0.65 & 0.63 & 0.34 & 0.263 & 0.163 & 0.054 & 0.191 & 0.105 & 0.035 & 1.11 \\
Cs--Se & 20 & 106 & 0.68 & 0.67 & 0.29 & 0.196 & 0.157 & 0.048 & 0.130 & 0.106 & 0.032 & 1.05 \\
Ag--Se & 36 & 149 & 0.59 & 0.57 & 0.24 & 0.213 & 0.196 & 0.039 & 0.089 & 0.117 & 0.023 & 1.05 \\
\midrule
\multicolumn{13}{l}{\emph{Te$^{2-}$} (3 families)} \\
Na--Te & 9 & 46 & 0.58 & 0.61 & 0.53 & 0.112 & 0.042 & 0.072 & 0.078 & 0.024 & 0.042 & --- \\
K--Te & 16 & 88 & 0.77 & 0.75 & 0.32 & 0.142 & 0.103 & 0.059 & 0.089 & 0.080 & 0.046 & 1.04 \\
Ag--Te & 9 & 50 & 0.66 & 0.67 & 0.23 & 0.220 & 0.194 & 0.088 & 0.112 & 0.128 & 0.058 & --- \\
\midrule
\multicolumn{13}{l}{\emph{N$^{3-}$} (7 families)} \\
Mg--N & 16 & 67 & 0.43 & 0.45 & 0.18 & 0.112 & 0.103 & 0.014 & 0.038 & 0.045 & 0.006 & 1.17 \\
Sr--N & 32 & 191 & 0.61 & 0.60 & 0.34 & 0.131 & 0.087 & 0.050 & 0.079 & 0.053 & 0.031 & 1.06 \\
Ba--N & 17 & 94 & 0.58 & 0.57 & 0.35 & 0.164 & 0.104 & 0.045 & 0.111 & 0.060 & 0.026 & 1.05 \\
Ca--N & 32 & 243 & 0.66 & 0.65 & 0.28 & 0.130 & 0.100 & 0.052 & 0.070 & 0.066 & 0.035 & 1.05 \\
Li--N & 38 & 194 & 0.65 & 0.59 & 0.36 & 0.118 & 0.105 & 0.045 & 0.054 & 0.068 & 0.029 & 1.08 \\
Ga--N & 10 & 40 & 0.96 & 0.49 & 0.34 & 0.088 & 0.082 & 0.023 & 0.045 & 0.079 & 0.022 & 1.16 \\
Na--N & 12 & 40 & 0.59 & 0.73 & 0.03 & 0.171 & 0.152 & 0.056 & 0.079 & 0.090 & 0.033 & 1.19 \\
\midrule
\multicolumn{13}{l}{\emph{P$^{3-}$} (4 families)} \\
Na--P & 15 & 100 & 0.65 & 0.73 & 0.45 & 0.114 & 0.092 & 0.040 & 0.063 & 0.060 & 0.026 & 1.08 \\
Ba--P & 16 & 71 & 0.70 & 0.72 & 0.19 & 0.155 & 0.120 & 0.045 & 0.090 & 0.084 & 0.031 & 1.15 \\
Ca--P & 12 & 40 & 0.74 & 0.69 & 0.34 & 0.195 & 0.174 & 0.047 & 0.054 & 0.128 & 0.035 & 1.22 \\
K--P & 17 & 66 & 0.85 & 0.83 & 0.04 & 0.231 & 0.203 & 0.060 & 0.102 & 0.173 & 0.051 & 1.13 \\
\midrule
\multicolumn{13}{l}{\emph{As$^{3-}$} (7 families)} \\
Li--As & 14 & 57 & 0.63 & 0.77 & 0.57 & 0.099 & 0.092 & 0.030 & 0.041 & 0.058 & 0.019 & 1.09 \\
Na--As & 23 & 186 & 0.65 & 0.76 & 0.45 & 0.144 & 0.115 & 0.053 & 0.071 & 0.075 & 0.035 & 1.09 \\
Ca--As & 15 & 48 & 0.75 & 0.72 & 0.42 & 0.180 & 0.155 & 0.054 & 0.054 & 0.116 & 0.040 & 1.16 \\
Ba--As & 18 & 72 & 0.78 & 0.79 & 0.29 & 0.229 & 0.207 & 0.039 & 0.077 & 0.161 & 0.030 & 1.06 \\
K--As & 31 & 173 & 0.94 & 0.98 & 0.14 & 0.218 & 0.182 & 0.048 & 0.116 & 0.170 & 0.045 & 1.06 \\
Cs--As & 9 & 64 & 0.88 & 1.14 & 0.12 & 0.312 & 0.295 & 0.097 & 0.111 & 0.261 & 0.086 & --- \\
Rb--As & 10 & 46 & 0.84 & 1.03 & 0.08 & 0.319 & 0.312 & 0.062 & 0.141 & 0.262 & 0.052 & 1.14 \\
\end{longtable}
}
\clearpage

\begin{figure*}[t]
\includegraphics[width=\textwidth,height=0.8\textheight,keepaspectratio]{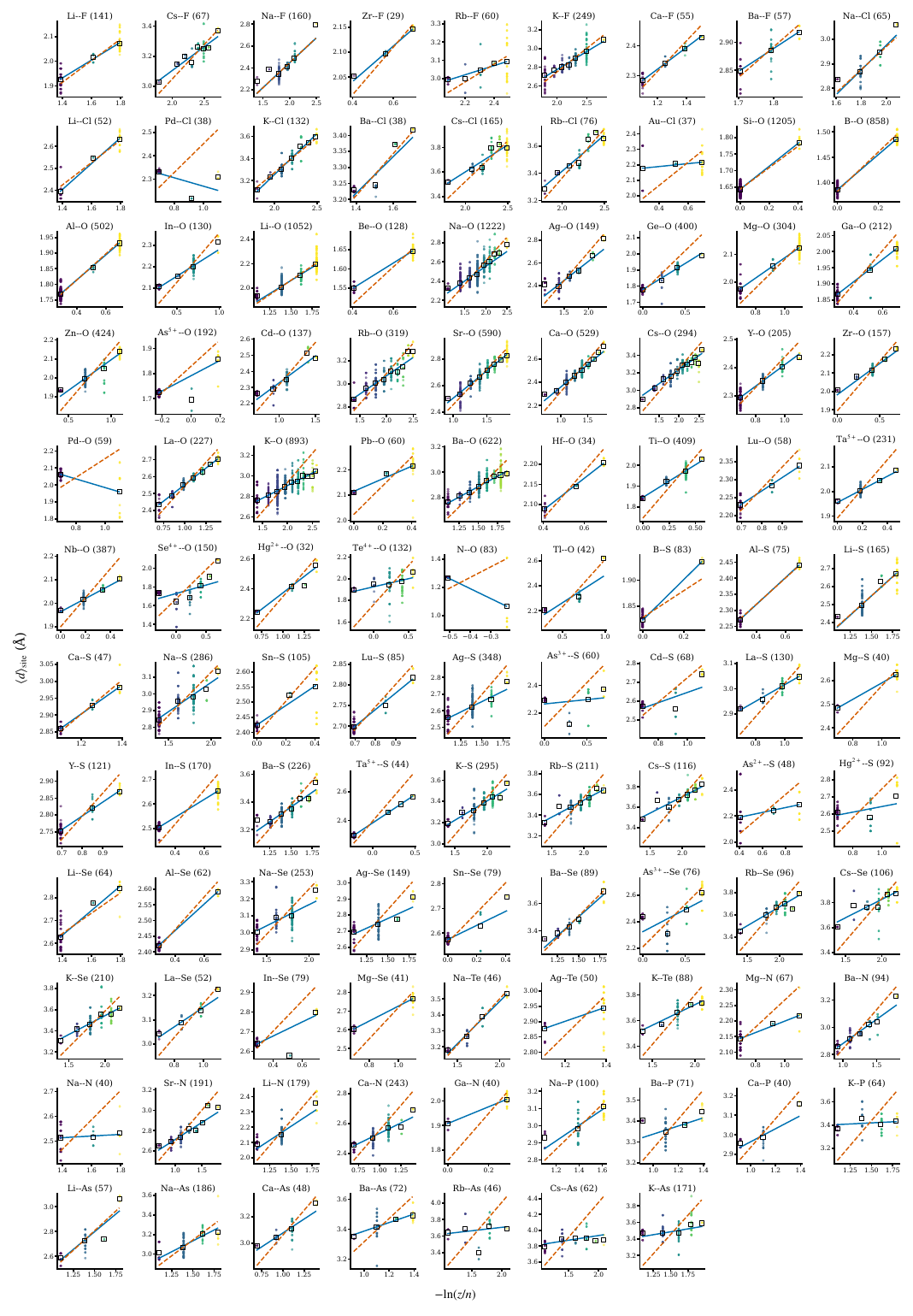}
\caption{The coordination branch of every family of Table~\ref{tab:transfer}, drawn as in Fig.~\ref{fig:branch}, with the number of sites in parentheses.}
\label{fig:branchall}
\end{figure*}

\begin{table*}[t]
\caption{The softness-gap test across the anions. The rows are the cations with eligible families in at least three anions and the columns the anions. Each cell gives $D=B-B^{*}$ in \AA{} with its 68\% half-width in units of the last digit in parentheses, and a check mark where the family passes. A cell is blank where the family does not meet the data requirement.}
\label{tab:anions}
\fitwidth{%
\small
\begin{tabular}{lccccccccc}
\toprule
 & F & Cl & O & S & Se & Te & N & P & As \\
\midrule
Li$^{+}$ & +0.08(4) & $-$0.08(6) & +0.04(2) & $-$0.01(5)\,\checkmark & $-$0.10(7) &  & +0.28(7) &  & +0.06(10)\,\checkmark \\
Na$^{+}$ & $0.00$(5)\,\checkmark & $-$0.04(12)\,\checkmark & +0.18(2) & +0.17(4) & +0.28(5) & +0.05(6)\,\checkmark & +0.56(10) & +0.20(10) & +0.20(8) \\
K$^{+}$ & +0.12(5) & $-$0.11(6) & +0.32(2) & +0.34(4) & +0.33(5) & +0.45(6) &  & +0.81(8) & +0.79(8) \\
Rb$^{+}$ & +0.24(10) & +0.19(6) & +0.21(3) & +0.40(4) & +0.34(6) &  &  &  & +0.76(14) \\
Ag$^{+}$ &  &  & +0.17(6) & +0.34(4) & +0.35(5) & +0.43(13) &  &  &  \\
Cs$^{+}$ & +0.14(6) & +0.22(6) & +0.21(2) & +0.42(7) & +0.38(10) &  &  &  & +0.77(11) \\
Mg$^{2+}$ &  &  & +0.15(2) & +0.32(9) & +0.43(8) &  & +0.25(8) &  &  \\
Ca$^{2+}$ & +0.15(4) &  & +0.15(2) & +0.08(5) &  &  & +0.39(4) & +0.40(12) & +0.33(8) \\
Ba$^{2+}$ & +0.14(9) & +0.09(9)\,\checkmark & +0.26(2) & +0.27(5) & +0.07(8)\,\checkmark &  & +0.22(7) & +0.50(12) & +0.49(7) \\
Al$^{3+}$ &  &  & +0.02(1)\,\checkmark & $-$0.01(5)\,\checkmark & +0.10(3) &  &  &  &  \\
In$^{3+}$ &  &  & +0.21(8) & +0.33(2) & +0.41(6) &  &  &  &  \\
La$^{3+}$ &  &  & +0.16(2) & +0.33(5) & +0.30(9) &  &  &  &  \\
\bottomrule
\end{tabular}
}
\end{table*}

\addtolength{\textheight}{18pt}
\makeatletter\@maxdepth=4pt\maxdepth=4pt\makeatother
\clearpage
\twocolumngrid
 
\end{document}